\documentclass[12pt,a4paper]{article}

\usepackage{amsmath}
\usepackage{amssymb}
\usepackage{graphicx}
\numberwithin{equation}{section} 
\allowdisplaybreaks
\begin{document}
\begin{flushright}
\texttt{OU-HET/460}\\
\texttt{hep-th/0312260}\\
December 2003
\end{flushright}
\bigskip
\bigskip
\begin{center}
{\Large \textbf{Non-Linear Field Equation}}
\end{center}
\begin{center}
{\Large \textbf{from Boundary State Formalism}}
\end{center}
\bigskip
\bigskip
\renewcommand{\thefootnote}{\fnsymbol{footnote}}
\begin{center}
Takashi Maeda\footnote{E-mail: \texttt{maeda@het.phys.sci.osaka-u.ac.jp}}, 
Toshio Nakatsu\footnote{E-mail: \texttt{nakatsu@het.phys.sci.osaka-u.ac.jp}} 
and 
Taku Oonishi\footnote{E-mail: \texttt{ohnishi@het.phys.sci.osaka-u.ac.jp}}\\
\bigskip
{\small 
\textit{Department of Physics, Graduate School of Science, 
Osaka University,\\
Toyonaka, Osaka 560-0043, Japan}}
\end{center}
\bigskip
\bigskip
\renewcommand{\thefootnote}{\arabic{footnote}}
\begin{abstract}
Boundary interactions of closed-string with 
open-strings are examined intended to 
provide a constructive formulation of boundary string field theory.
As an illustration, 
we consider the BPS $D$-brane of the type II superstring 
in a constant NS-NS two-form background, and study 
the boundary interaction with arbitrary configurations 
of gauge field on the brane. 
The boundary interaction is presented, within the world-sheet cut-off
theories, as an off-shell boundary state in 
the closed-string Hilbert space. 
It is regarded as a closed-string theoretical counterpart of 
the Wilson loop in the world-volume gauge theory. 
We show that the action of the closed-string BRST operator 
on the boundary state is translated into the non-linear BRST 
transformation of the open-string fields on the world-volume. 
In particular, the BRST invariance condition 
at the $\alpha'$-order becomes the non-linear equation 
of motion for the non-commutative gauge theory. 
The action of the closed-string BRST operator 
on the boundary state is also shown to be identified with 
the beta functions of the world-sheet renormalization group flow. 
\end{abstract}

\setcounter{footnote}{0}
\newpage
\section{Introduction and Summary}
\label{section1}
Boundary string field theory (BSFT) \cite{Witten1} has been proposed as 
a background independent formulation of open-string field theory 
and exploited in 
\cite{Witten2},\cite{LiWitten},\cite{Shata},\cite{BSSFT}. 
The configuration space of open-string fields 
is identified with the space of two-dimensional field theories 
with arbitrary interactions on the boundary 
and a fixed conformal invariant world-sheet action in the bulk.  
An odd-symplectic structure on the space of the open-string fields 
is given by the two-point functions of the world-sheet theories with 
the boundary interactions. 
A nilpotent fermionic vector field is introduced 
as a certain limit of the bulk BRST operator. 
The string field action is determined from 
the above vector field and the odd-symplectic structure 
by using the Batalin-Vilkovisky formalism. 
However, the original construction of \cite{Witten1} is purely formal. 
Due to the short-distance divergences of the world-sheet theories, 
an introduction of the space of the boundary interactions 
and the odd-symplectic structure requires some regularization procedure. 
There is also an ambiguity in the definition of the nilpotent vector field 
and the notion of the classical equation of motion in the BSFT 
is still not clear.

In this paper, we will clarify these issues  
and make the first-step towards a constructive formulation of the BSFT. 
The aim of the present work are 
to provide a concrete construction of the boundary interactions 
within the cut-off theory, 
and to illustrate how the classical open-string interactions 
are extracted from the boundary interactions 
which are described in the world-sheet theory terminology. 
In order to be explicit, we will consider the BPS $D$-branes 
of the type II superstring theory in a constant NS-NS two-form 
background, and investigate the boundary interactions with the 
off-shell gauge fields on the branes.

The first quantization of open-string becomes intractable 
when the superconformal invariance of the world-sheet theory 
is broken on the boundary. However, 
if not broken in the bulk, the closed-string theory is quantized 
still in the conformal field theoretical method, and 
the boundary state formalism \cite{BoundaryState} 
developed in \cite{MN} will provide 
an appropriate framework for the description of the boundary 
interactions. 
The relevant boundary states are the states which do not 
satisfy the Ishibashi conditions (more precisely, the closed-string 
BRST invariance). In section 2 we construct the boundary states 
with arbitrary numbers of gluons.  
The resultant boundary states correctly 
include the open-string interactions, and reproduce 
the corresponding open-string one-loop amplitudes. 
Our construction in this paper 
is limited to the closed-string NS-NS sector. 
This is sufficient to see the classical physics of 
the open-string NS sector (disk amplitudes of gluons), 
which is dealt with by the NS-NS sector.

The boundary interaction of closed-string 
with arbitrary configurations of the gauge 
field on the brane is obtained from the above boundary states. 
In section 3, the boundary interaction is constructed in the closed-string 
Hilbert space by integrating the gluon boundary states 
over the moduli spaces of the path-ordered points on the boundary 
and summing them over the number of the gluons. 
The UV divergences of the open-string interactions arise 
when the gluon vertices collide with one another on the boundary. 
We will therefore make a point-splitting regularization 
by restricting the integrations over the moduli spaces 
so that intervals between any two gluon vertices on the boundary 
can not be less than a small cut-off parameter $\epsilon$. 
We also modify the path-ordering procedure 
to respect the global world-sheet supersymmetry, 
following to \cite{AT}.  
It is observed in \cite{AT} that the global world-sheet supersymmetry 
controls the short-distance divergences of the world-sheet theories. 
The resultant boundary state is denoted by 
$\bigl|W_{\epsilon}[A];\eta\bigr\rangle_{tot}$. 
It represents the boundary interaction with the off-shell 
gauge field on the brane. 
For another approach to the boundary interaction, 
see \cite{Hashimoto}.

In the loop space approach to string field theory \cite{Z}, 
the interactions are introduced 
by adding the string field vertices to the string field action. 
While, in BSFT, 
the interactions of the open-string fields 
should be extracted from the world-sheet boundary interaction. 
In section 4, we will overview the mechanism.
We investigate the action of the closed-string BRST operator $Q_c$ 
on $\bigl|W_{\epsilon}[A];\eta\bigr\rangle_{tot}$. 
The use of the closed-string BRST operator 
is required by the background independence. 
Since the closed-string BRST operator 
does not include any coupling of the open-strings, 
the BRST invariance condition formally 
leads the linearized equation of motion for the gluons.  
However, the action of the closed-string BRST operator 
on the gluon vertices includes total-derivative terms. 
With the above point-splitting regularization, 
the total-derivative terms do not vanish in the integrals  
and contribute as the boundary terms of the regularized moduli space. 
These boundary terms turn out to provide 
the non-linear BRST transformation of the open-string fields.  
The zero of this non-linear BRST transformation 
gives the non-linear equation of motion. 
In this course, the leading contribution of each boundary term 
becomes too singular at $\epsilon\rightarrow0$. 
Typically it behaves as $\sim 1/\epsilon$. 
The origin is the propagation of the open-string unphysical particles 
\cite{GS}. 
These unpleasant terms turn out to cancel 
with the contact terms which are brought about by  
the supersymmetric path-ordering procedure. 
The leading contributions are cancelled, 
and the next-to-leading contributions, 
which describe the propagation of the gluons,  
become the non-linear corrections to 
the BRST transformation and the equation of motion.

In the constant NS-NS two-form background, 
it is known \cite{SW} that 
the gauge field $A_{\mu}(x)$ on the brane acquires 
the non-commutativity and that the equation of motion 
for the gauge field becomes non-linear:
\begin{equation}
G^{\mu\nu}\nabla_{\mu}F_{\nu\rho}(x)=0\;,
\label{intro1}
\end{equation} 
where $\nabla_{\mu}$ and $F_{\mu \nu}$ are  
the covariant derivative and the field strength 
of the non-commutative gauge theory. 
In section 5, we compute the BRST invariance condition 
of the world-sheet boundary interaction at the $\alpha'$-order 
and confirm that the condition 
$Q_{c}\bigl|W_{\epsilon}[A];\eta\bigr\rangle_{tot}=0$ 
leads to the non-linear equation of motion (\ref{intro1}). 
We would notice that the above mechanism to extract 
the non-linear equation of motion is not restricted 
to the present non-commutative case. 
It can be easily generalized to the  non-abelian Yang-Mills 
equation by associating the Chan-Paton factors with 
the gluon vertices.

In the original attempts to formulate 
string field theory as a theory of 
``all two-dimensional quantum field theories'', 
it was suggested in \cite{BM}  
that the classical equation of motion 
for string fields is given by 
the renormalization group equation 
of the world-sheet theories.  
In section 6, 
the boundary interaction is examined 
from the perspective of the world-sheet renormalization group. 
In our boundary state formulation, 
the beta functions of the boundary couplings $A_{\mu}(k)$ 
are defined by the following condition:
\begin{equation}
\int{d}^{p+1}k\;\beta_{A_{\mu}(k)}
\frac{\delta}{\delta{A}_{\mu}(k)}\;
\Bigl|\;W_{\epsilon}[A];\eta\;\Bigr\rangle_{tot}
=-\epsilon\frac{\partial}{\partial\epsilon}\;
\Bigl|\;W_{\epsilon}[A];\eta\;\Bigr\rangle_{tot}\;.\label{intro2}
\end{equation}
They become linear 
if the $\epsilon$-dependence of the regularized moduli spaces 
is neglected. By taking account of the $\epsilon$-dependence 
of the regularized moduli spaces, the beta functions  
turn to acquire the non-linearity:
\begin{equation}
\beta_{A_{\mu}(k)}
=\alpha^{\prime}\int\frac{d^{p+1}x}{(2\pi)^{\frac{p+1}{2}}}
e^{-ik\cdot{x}}G^{\nu\rho}\nabla_{\nu}F_{\rho\mu}(x)
+{\cal O}(\alpha'^{2})\;.
\label{intro3}
\end{equation}
These beta functions coincide with 
the non-linear BRST transformations of the 
open-string fields. 
Therefore the action of $Q_c$ on 
$\bigl|W_{\epsilon}[A];\eta\bigr\rangle_{tot}$ 
is identified with the beta functions.   
This provides a concrete definition of the nilpotent vector field 
in the Batalin-Vilkovisky formulation of the BSFT.  
Gauge structure of the BSFT can be also studied from our approach. 
It will be reported elsewhere \cite{toappear}.

Finally, 
we comment on a possible extension of 
our formulation of the classical BSFT 
to the quantum BSFT. 
The quantum open-string theory inevitably 
contains the closed-string degrees of freedom. 
The resultant string field theory 
is an open-closed string field theory 
in which the closed-string interactions are encoded in 
the closed-string field vertices while 
the open-string interactions are encoded in the boundary states 
$\bigl|W_{\epsilon}[A];\eta\bigr\rangle_{tot}$. 
The equation of motion for $A_{\mu}(x)$ is 
deformed from 
$Q_{c}\bigl|W_{\epsilon}[A];\eta\bigr\rangle_{tot}=0$ 
by the closed-string interaction, and becomes 
to contain the interaction between 
$\bigl|W_{\epsilon}[A];\eta\bigr\rangle_{tot}$. 
From the space-time viewpoint, 
the boundary states $\bigl|W_{\epsilon}[A];\eta\bigr\rangle_{tot}$ 
correspond to the Wilson loops of the gauge theory on the branes.  
The deformed equation of motion may be the string theoretical realization of 
the loop equation of the gauge theory, 
and the closed-string BRST operator $Q_{c}$ 
corresponds to the loop Laplacian \cite{Polyakov}. 
We hope that the progress along this line 
will shed light on 
the issues of the open-closed string duality.

\section{Gluons in Closed-String Theory}
\label{section2}
In the present section, 
we make a description of the open-string degrees of freedom 
on a non-commutative D-brane in terms of the boundary state formalism. 
Our study is restricted to the massless particles in the open-string theory 
- gluons, and the closed-string NS-NS sector.

This section has two parts. 
As a set-up, boundary states without any open-string leg are briefly reviewed 
in the first subsection
\footnote{For reviews, see \cite{Review} and references therein.}. 
In the second subsection, we construct boundary states with 
arbitrary numbers of gluons, with the method developed in \cite{MN}. 
The resultant boundary states by themselves capture the open-string 
tree interactions. 
The tree propagation between the boundary states 
correctly reproduce the open-string 
one-loop amplitudes. 
The BRST invariance of the boundary state with one-gluon leads to 
the linearized equation of motion for the gluon.

\subsection{A brief review of boundary states without open-string vertices}
Let us consider type II superstring theory 
with a non-commutative D$p$-brane. 
The case of even $p$ corresponds to the type IIA theory, 
while odd $p$ corresponds to the type IIB theory. 
We will use $\mu,\nu=0,1,2,\dotsc,p$ 
to denote the world-volume directions of the D$p$-brane, 
$i,j=p+1,p+2,\dotsc,9$ to denote the transverse directions of the brane 
and $M,N=0,1,2,\dotsc,9$ for the whole space-time indices. 
Closed strings couple to a constant space-time metric $g_{MN}$ 
with $g_{\mu{i}}=0$ and a constant 
two-form gauge field $B_{\mu\nu}$ in the bulk. 
The world-sheet action is given by
\begin{equation}
 \begin{split}
S=\frac{i}{2\pi}\int_{\Sigma} dw\wedge d\bar{w}
\Bigl[\;&\frac{1}{\alpha^{\prime}}\bigl\{
g_{MN}\partial_{w}X^{M}\partial_{\bar{w}}X^{M}
-2\pi\alpha^{\prime}B_{\mu\nu}
\partial_{w}X^{\mu}\partial_{\bar{w}}X^{\nu}\bigr\}\\
&+\frac{1}{2}g_{MN}\psi^{M}\partial_{\bar{w}}\psi^{N}
+\frac{1}{2}g_{MN}\bar{\psi}^{M}\partial_{w}\bar{\psi}^{N}
\;\Bigr]\;.
 \end{split}
\end{equation}
Here we use the cylinder coordinates $w=\tau+i\sigma,\bar{w}=\tau-i\sigma$ 
($\tau\!\geq\!0,\;0\!\leq\!\sigma\!\leq\!2\pi$) 
and the world-sheet $\Sigma$ has a boundary at $\tau=0$. 
We impose the boundary conditions 
\begin{equation}
 \begin{split}
&E_{\mu\nu}\partial_{w}X^{\nu}
+E_{\mu\nu}^{T}\partial_{\bar{w}}X^{\nu}\bigr|_{\tau=0}=0\;,\quad
X^{i}\bigr|_{\tau=0}=x_{0}^{i}\;,\\
&E_{\mu\nu}\psi^{\nu}-i{\eta}E_{\mu\nu}^{T}\bar{\psi}^{\nu}
\bigr|_{\tau=0}=0\;,\quad
\psi^{i}+i\eta\bar{\psi}^{i}\bigr|_{\tau=0}=0\qquad (\eta=\pm1)\;,
 \end{split}\label{bc}
\end{equation}
where $E_{\mu\nu}=g_{\mu\nu}+2\pi\alpha^{\prime}B_{\mu\nu}$ and 
$E_{\mu\nu}^{T}=g_{\mu\nu}-2\pi\alpha^{\prime}B_{\mu\nu}$.

In the NS-NS sector, the world-sheet variables 
$X^{M}$, $\psi^{M}$ and $\bar{\psi}^{M}$ have 
the following mode-expansions:
\begin{equation}
X^{M}(\tau,\sigma)=\hat{x}^{M}
-i\alpha^{\prime}\hat{p}^{M}\tau
+i\sqrt{\frac{\alpha^{\prime}}{2}}\sum_{n\neq{0}}
\Bigl\{\;\frac{\alpha^{M}_{n}}{n}e^{-n(\tau+i\sigma)}
+\frac{\bar{\alpha}^{M}_{n}}{n}e^{-n(\tau-i\sigma)}\;\Bigr\}\;,
\end{equation}
\begin{equation}
\psi^{M}(\tau,\sigma)=\sum_{r\in\mathbb{Z}+\frac{1}{2}}
\psi^{M}_{r}e^{-r(\tau+i\sigma)} ,\quad
\bar{\psi}^{M}(\tau,\sigma)=\sum_{r\in\mathbb{Z}+\frac{1}{2}}
\bar{\psi}^{M}_{r}e^{-r(\tau-i\sigma)}\;.
\end{equation}
The first quantization of the closed-string is achieved by imposing 
the following commutation relations:
\begin{equation}
 \begin{split}
&[x^{M},p^{N}]=ig^{MN},\quad
[\alpha_{m}^{M},\alpha_{n}^{N}]
=[\bar{\alpha}^{M}_{m},\bar{\alpha}^{N}_{n}]
=g^{MN}m\delta_{m,-n}\;,\\
&\{\psi^{M}_{r},\psi_{s}^{N}\}
=\{\bar{\psi}^{M}_{r},\bar{\psi}^{N}_{s}\}=g^{MN}\delta_{r,-s}\;,
 \end{split}
\end{equation}
with all the other commutators equal to zero.

In the closed-string channel, the boundary conditions (\ref{bc}) 
are encoded in the boundary state 
$|B;\eta\rangle_{m}$ , which is defined by the following overlap conditions:
\begin{equation}
 \begin{split}
&\hat{p}^{\mu}|B;\eta\rangle_{m}
=(\hat{x}^{i}-x_{0}^{i})|B;\eta\rangle_{m}=0\;,\\
&(E_{\mu\nu}\alpha_{n}^{\nu}+E^{T}_{\mu\nu}\bar{\alpha}_{-n}^{\nu})
|B;\eta\rangle_{m}
=(\alpha_{n}^{i}-\bar{\alpha}_{-n}^{i})|B;\eta\rangle_{m}=0\;,
\qquad\mathrm{for}\quad n\neq0\;,\\
&(E_{\mu\nu}\psi^{\nu}_{r}-i{\eta}E_{\mu\nu}^{T}\bar{\psi}_{-r})
|B;\eta\rangle_{m}
=(\psi^{i}_{r}+i\eta\bar{\psi}^{i}_{-r})|B;\eta\rangle_{m}=0\;,
\qquad\mathrm{for}\quad r\in\mathbb{Z}+\frac{1}{2}\;.
 \end{split}\label{overlap}
\end{equation}
By using the above overlap conditions, 
we can easily check that
the boundary state $|B;\eta\rangle_{m}$ satisfies 
the Ishibashi conditions
\begin{equation}
\bigl(L_{n}^{m}-\bar{L}_{-n}^{m}\bigr)
|B;\eta\rangle_{m}=0\;,\quad
\bigl(G_{r}^{m}+i{\eta}\bar{G}_{-r}^{m}\bigr)
|B;\eta\rangle_{m}=0\;,\label{Ishibashi}
\end{equation}
where $L_{n}^{m}$, $\bar{L}_{n}^{m}$ and 
$G_{r}^{m}$, $\bar{G}_{r}^{m}$ are the Virasoro and super Virasoro 
generators of the ($X$,$\psi$,$\bar{\psi}$) CFT respectively. 
An explicit form of $|B;\eta\rangle_{m}$ is given by 
$|B;\eta\rangle_{m}=Cg_{\eta}|x_{0}^{i}\rangle_{m}$, where
\begin{equation}
C=\biggl\{\frac{\det(E^{T}g^{-1}E)_{\mu\nu}}{(2\alpha^{\prime})^{p+1}}
\biggr\}^{\frac{1}{4}}
\biggl\{\frac{(2\pi^{2}\alpha^{\prime})^{9-p}}{\det{g_{ij}}}
\biggr\}^{\frac{1}{4}}\;,
\end{equation}
\begin{equation}
 g_{\eta}= \prod_{n=1}^{\infty}\exp\Bigl\{
-\frac{1}{n}N_{\mu\nu}\alpha_{-n}^{\mu}\bar{\alpha}^{\nu}_{-n}
+\frac{1}{n}g_{ij}\alpha_{-n}^{i}\bar{\alpha}^{j}_{-n}
\Bigr\}        
\prod_{r=1/2}^{\infty}\exp\Bigl\{
i{\eta}N_{\mu\nu}\psi_{-r}^{\mu}\bar{\psi}_{-r}^{\nu}
-i{\eta}g_{ij}\psi^{i}_{-r}\bar{\psi}^{j}_{-r}
\Bigr\}\;,  \label{BTgen}          
\end{equation}
\begin{equation}
|x_{0}^{i}\rangle_{m}=\int\frac{d^{9-p}k}{\sqrt{(2\pi)^{9-p}}}
e^{ik_{i}(\hat{x}^{i}-x_{0}^{i})}|0\rangle_{m}\;.
\end{equation}
Here $N_{\mu\nu}=\bigl(g\frac{1}{E}E^{T}\bigr)_{\mu\nu}$ and 
$|0\rangle_{m}$ is the $SL(2,\mathbb{C})$ invariant vacuum 
of the ($X$, $\psi$, $\bar{\psi}$) CFT. 
The normalization factor $C$ is determined by 
factorizing open-string annulus partition function 
in the closed-string channel.

The complete boundary state is obtained by tensoring 
$|B;\eta\rangle_{m}$ with states for the conformal ghosts 
$b,c,\bar{b},\bar{c}$ and for the superconformal ghost 
$\beta,\gamma,\bar{\beta},\bar{\gamma}$ :
\begin{equation}
|B;\eta\rangle_{tot}=
|B;\eta\rangle_{m}\otimes|B\rangle_{gh}
\otimes|B;\eta\rangle_{sgh}\;.\label{complete}
\end{equation}
The boundary states $|B\rangle_{gh}$ and 
$|B;\eta\rangle_{sgh}$ are given by 
\begin{align}
|B\rangle_{gh}&=\prod_{n=1}^{\infty}\exp\Bigl\{
c_{-n}\bar{b}_{-n}+\bar{c}_{-n}b_{-n}\Bigr\}
\Bigl(\frac{c_{0}+\bar{c}_{0}}{2}\Bigr)c_{1}\bar{c}_{1}
|0\rangle_{gh}\;,\label{ghpart}\\
|B;\eta\rangle_{sgh}&=\prod_{r=1/2}^{\infty}\exp\Bigl\{
i\eta\bigl(\gamma_{-r}\bar{\beta}_{-r}-\bar{\gamma}_{-r}\beta_{-r}\bigr)
\Bigr\}|-1,-1\rangle_{sgh}\;,\label{sghpart}
\end{align}
where $|0\rangle_{gh}$ is the $SL(2,\mathbb{C})$ invariant vacuum of 
the conformal ghost system and $|-1,-1\rangle_{sgh}$ is the ground state of 
the superconformal ghost system in the ($-1$,$-1$) picture. 
The mode operators of the ghost and superghost systems 
satisfy the following commutation relations:
\begin{equation}
 \begin{split}
\{c_{m},b_{n}\}
&=\{\bar{c}_{m},\bar{b}_{n}\}
=\delta_{m,-n}\;,\\
[\gamma_{r},\beta_{s}]
&=[\bar{\gamma}_{r},\bar{\beta}_{s}]
=\delta_{r,-s}\;,
 \end{split}
\end{equation}
with all the other commutators equal to zero. 
The boundary states $|B\rangle_{gh}$ and 
$|B;\eta\rangle_{sgh}$ enjoy the overlap conditions
\begin{equation}
 \begin{split}
(c_{n}+\bar{c}_{-n})|B\rangle_{gh}
&=(b_{n}-\bar{b}_{-n})|B\rangle_{gh}=0\;,\\
(\gamma_{r}+i\eta\bar{\gamma}_{-r})|B;\eta\rangle_{sgh}
&=(\beta_{r}+i\eta\bar{\beta}_{-r})|B;\eta\rangle_{sgh}=0\;.
 \end{split}\label{ghostoverlap}
\end{equation}
Combining these conditions with the Ishibashi conditions (\ref{Ishibashi}), 
one can easily check that the complete boundary state $|B;\eta\rangle_{tot}$ 
is annihilated by the closed-string BRST operator 
\begin{equation}
 \begin{split}
Q_{c}=&-\sum_{n\in\mathbb{Z}}c_{-n}L^{m}_{n}
+\frac{1}{2}\sum_{r\in\mathbb{Z}+1/2}\gamma_{-r}G^{m}_{r}
+\frac{1}{2}\sum_{m,n\in\mathbb{Z}}(m-n):c_{-m}c_{-n}b_{m+n}:\\
&\quad
-\sum_{n\in\mathbb{Z}}\sum_{r\in\mathbb{Z}+1/2}
\bigl(\frac{n}{2}-r\bigr):c_{-n}\gamma_{-r}\beta_{n+r}:
+\frac{1}{4}\sum_{r,s\in\mathbb{Z}}
:\gamma_{-r}\gamma_{-s}b_{r+s}:\\
&\quad\quad+
\Bigl\{\mathrm{anti-chiral{\;\;}part}\Bigr\}\;.
 \end{split}\label{BRST}
\end{equation}

The boundary state (\ref{complete}) 
depends on the value of $\eta$. 
The GSO projection of the closed-string theory 
selects a specific combination of the two values of $\eta=\pm1$. 
In the NS-NS sector, the GSO projection operator is defined by
\begin{equation}
P_{GSO}=\frac{1+(-1)^{N_{\psi}+G_{sgh}}}{2}
\frac{1+(-1)^{\bar{N}_{\psi}+\bar{G}_{sgh}}}{2}\;,\label{GSO}
\end{equation}
where $N_{\psi}$ and $G_{sgh}$ are the fermion number operator 
and the superghost number operator:
\begin{equation}
N_{\psi}=\sum_{r=1/2}^{\infty}g_{MN}\psi^{M}\psi^{N}\;,\quad
G_{sgh}=-\sum_{r=1/2}^{\infty}
(\gamma_{-r}\beta_{r}+\beta_{-r}\gamma_{r})-1\;,
\end{equation}
and similarly for $\bar{N}_{\psi}$ and $\bar{G}_{sgh}$ 
of the anti-chiral sector. 
The GSO projection operator (\ref{GSO}) 
acts on the boundary state $|B;\eta\rangle_{tot}$ as follows:
\begin{equation}
P_{GSO}|B;\eta\rangle_{tot}=
\frac{1}{2}\bigl\{\;|B;\eta\rangle_{tot}
-|B;-\eta\rangle_{tot}\bigr\}\;.
\end{equation}
In the absence of any open-string leg, 
boundary states in the R-R sector are also constructed in \cite{Review}. 
\subsection{Boundary states with gluons}
We now construct the boundary states which describe the gauge field living 
on the D-brane. 
For a while, our attention will be concentrated on the matter part. 
It is convenient to introduce the one-dimensional superspace notation. 
Let $\theta$ be an anti-commuting coordinate on the world-sheet boundary. 
The one-dimensional superspace consists of the pair ($\sigma$, $\theta$). 
The superfield and the superderivative are defined by 
\begin{equation}
\mathbf{X}^{\mu}=X^{\mu}-i\sqrt{\frac{\alpha^{\prime}}{2}}\theta
\bigl(\psi^{\mu}+i\eta\bar{\psi}^{\mu}\bigr)\;,\quad
D_{\theta}=\frac{\partial}{\partial\theta}
-i\theta\frac{\partial}{\partial\sigma}\;.
\end{equation}
Let $k_{\mu}$ and $A_{\mu}(k)$ be the momentum and polarization vectors. 
We begin with the following auxiliary operator:
\begin{equation}
:\exp\Bigl\{ik_{\mu}\mathbf{X}^{\mu}
+{\lambda}A_{\mu}(k)D_{\theta}\mathbf{X}^{\mu}\Bigr\}:\;,
\label{2a}
\end{equation}
where $\lambda$ is an anti-commuting auxiliary parameter \cite{IM}. 
By expanding the auxiliary operator (\ref{2a}) 
in powers of $\lambda$, 
we get analogues of the open-string tachyon and gluon vertex operators 
as follows:
\begin{equation}
:\exp\Bigl\{ik_{\mu}\mathbf{X}^{\mu}
+{\lambda}A_{\mu}(k)D_{\theta}\mathbf{X}^{\mu}\Bigr\}:
\;=\;:e^{ik_{\mu}\mathbf{X}^{\mu}}:
-i\lambda\Bigl\{\;
i:A_{\mu}(k)D_{\theta}\mathbf{X}^{\mu}
e^{ik_{\mu}\mathbf{X}^{\mu}}:\Bigr\}\;.
\label{2b}
\end{equation}

Let us overview how the operators in the closed-string theory 
describe the open-string interactions, 
with the help of the boundary state 
$|B;\eta\rangle_{m}$. 
The overlap conditions (\ref{overlap}) say that the annihilation operators 
with respect to the boundary state $|B;\eta\rangle_{m}$ are 
\begin{equation}
 \begin{split}
E_{\mu\nu}\alpha_{n}^{\nu}
+E^{T}_{\mu\nu}\bar{\alpha}^{\nu}_{-n}\;,&\quad
\alpha^{i}_{n}-\bar{\alpha}^{i}_{-n}\;,\quad 
n\in\mathbb{Z}\;,\\
E_{\mu\nu}\psi^{\nu}_{r}
-i{\eta}E^{T}_{\mu\nu}\bar{\psi}^{\nu}_{-r}\;,&\quad
\psi_{r}^{i}+i{\eta}\bar{\psi}^{i}_{-r}\;,\quad
r\in\mathbb{Z}+\frac{1}{2}\;,
 \end{split}\label{annihilation}
\end{equation}
rather than the lowering operators 
$\alpha^{M}_{n}$, $\bar{\alpha^{M}_{n}}$ ($n\geq1$) and 
$\psi^{M}_{r}$, $\bar{\psi}^{M}_{r}$ ($r\geq1/2$). 
Thus, for the operators acting on $|B;\eta\rangle_{m}$, 
the relevant normal-ordering operation is provided by 
placing all of the operators (\ref{annihilation}) 
to the right, and the relevant operator product expansion (OPE) 
is obtained by using this modified normal-ordering prescription. 
The modified normal-ordering and OPE reproduce 
the open-string interactions. 
For instance, the modified OPE between $\mathbf{X}^{\mu}$ becomes 
\begin{equation}
 \begin{split}
&\mathbf{X}^{\mu}(0,\sigma_{1},\theta_{1})
\mathbf{X}^{\nu}(0,\sigma_{2},\theta_{2})\\
&\sim
\alpha^{\prime}G^{\mu\nu}
\mathbf{G}(\sigma_{1},\theta_{1};\sigma_{2},\theta_{2})
-\frac{i}{2}\theta^{\mu\nu}\mathrm{sign}(\sigma_{1}-\sigma_{2})
+\frac{i}{2\pi}\theta^{\mu\nu}(\sigma_{1}-\sigma_{2})\;,
 \end{split}
\end{equation}
where the tensors
\begin{equation}
G^{\mu\nu}=\frac{1}{2}
\Bigl(\frac{1}{E}+\frac{1}{E^{T}}\Bigr)^{\mu\nu}\;,\quad
\theta^{\mu\nu}=
\pi\alpha^{\prime}\Bigl(\frac{1}{E}-\frac{1}{E^{T}}\Bigr)^{\mu\nu}
\label{tensors}
\end{equation}
are the open-string metric 
and the non-commutative parameter respectively \cite{SW}, 
and Green's function on the superspace is defined by
\begin{equation}
\mathbf{G}(\sigma_{1},\theta_{1};\sigma_{2},\theta_{2})
=-\ln|e^{i\sigma_{1}}-e^{i\sigma_{2}}|^{2}
-i\theta_{1}\theta_{2}
\frac{1}{\;\sin\bigl({\frac{\sigma_{1}-\sigma_{2}}{2}}\bigr)\;}\;.\label{Green}
\end{equation}

For dealing with the normal-ordering and the OPE 
with respect to $|B;\eta\rangle_{m}$, 
it becomes convenient to consider the Bogolubov transformation 
\begin{equation}
O\quad\longmapsto\quad
\widehat{O}=g_{\eta}^{-1}Og_{\eta}\;,\label{BT}
\end{equation}
which maps the operators (\ref{annihilation}) to 
the lowering operators. 
Therefore, the normal-ordering of $O$ with respect to 
the boundary state $|B;\eta\rangle_{m}$ is translated to 
the normal-ordering of $\widehat{O}$ 
with respect to the $SL(2,\mathbb{C})$ invariant vacuum $|0\rangle_{m}$. 
The OPE between $O$ with respect to $|B;\eta\rangle_{m}$ is translated to 
the OPE between $\widehat{O}$ with respect to $|0\rangle_{m}$. 
We employ this description in the subsequent part of this paper.

We now return to the operator (\ref{2a}). 
Under the Bogolubov transformation (\ref{BT}), 
the auxiliary operator (\ref{2a}) transforms as
\begin{equation}
 \begin{split}
g_{\eta}^{-1}: \exp\Bigl\{ik_{\mu}\mathbf{X}^{\mu}
&+{\lambda}A_{\mu}(k)D_{\theta}\mathbf{X}^{\mu}\Bigr\}:g_{\eta}\\
=R_{A}(\tau,\theta,\lambda;k)
&\prod_{n=1}^{\infty}\exp\Bigl\{
\sqrt{\frac{\alpha^{\prime}}{2}}\frac{1}{n}
\bigl(k_{\mu}-i\lambda{\theta}nA_{\mu}(k)\bigr)
(g^{-1}N^{T})^{\mu}_{\;\;\nu}\alpha_{-n}^{\nu}
e^{-n(\tau-i\sigma)}\Bigr\}\\
\times&\prod_{n=1}^{\infty}\exp\Bigl\{
\sqrt{\frac{\alpha^{\prime}}{2}}\frac{1}{n}
\bigl(k_{\mu}+i\lambda{\theta}nA_{\mu}(k)\bigr)
(g^{-1}N)^{\mu}_{\;\;\nu}\bar{\alpha}^{\nu}_{-n}
e^{-n(\tau+i\sigma)}\Bigl\}\\
\times&\prod_{r=1/2}^{\infty}\exp\Bigl\{
\sqrt{\frac{\alpha^{\prime}}{2}}
\bigl({\theta}k_{\mu}-i{\lambda}A_{\mu}(k)\bigr)
(g^{-1}N^{T})^{\mu}_{\;\;\nu}\psi_{-r}^{\nu}
e^{-r(\tau-i\sigma)}\Bigr\}\\
\times&\prod_{r=1/2}^{\infty}\exp\Bigl\{
i\eta\sqrt{\frac{\alpha^{\prime}}{2}}
\bigl({\theta}k_{\mu}-i{\lambda}A_{\mu}(k)\bigr)
(g^{-1}N)_{\;\;\nu}^{\mu}\bar{\psi}^{\nu}_{-r}
e^{-r(\tau+i\sigma)}\Bigr\}\\
\times& : \exp\Bigl\{ik_{\mu}\mathbf{X}^{\mu}
+{\lambda}A_{\mu}(k)D_{\theta}\mathbf{X}^{\mu}\Bigr\}:\;.
 \end{split}
\end{equation}
The explicit form of the factor  $R_{A}(\tau,\theta,\lambda;k)$ is given by
\begin{equation}
 \begin{split}
R_{A}(\tau,\theta,\lambda;k)
=&\Bigl(1-e^{-2\tau}\Bigr)
^{\frac{\alpha^{\prime}}{2}k_{\mu}(g^{-1}Ng^{-1})^{\mu\nu}k_{\nu}}\\
&\times\exp\Bigl\{-\frac{\alpha^{\prime}}{2}\lambda\theta
f_{\mu\nu}(k)(g^{-1}Ng^{-1})^{\mu\nu}\frac{1}{1-e^{2\tau}}\Bigr\}\\
&\times\exp\Bigl\{+\frac{\alpha^{\prime}}{2}\lambda\theta
f_{\mu\nu}(k)(g^{-1}Ng^{-1})^{\mu\nu}\frac{e^{\tau}}{1-e^{2\tau}}\Bigr\}\;,
 \end{split}\label{singular}
\end{equation}
where $f_{\mu\nu}(k)=ik_{\mu}A_{\nu}(k)-ik_{\nu}A_{\mu}(k)$ 
is the Fourier transform of the field strength 
in the commutative gauge theory. 
The factor (\ref{singular}) has a singular behavior in the limit 
$\tau\rightarrow+0$. 
Thus, the action of the operator (\ref{2a}) on the boundary state 
$|B;\eta\rangle_{m}$ is singular at the world-sheet boundary. 
This is the consequence of the fact that operators have interactions with 
their mirror images through the existence of $|B;\eta\rangle_{m}$.

By subtracting the singular factor (\ref{singular}) from (\ref{2a}), 
we define the renormalized operator
\begin{equation}
 \begin{split}
W_{A}(\tau,\sigma,\theta,\lambda;k)
=R_{A}^{-1}(\tau,\theta,\lambda;k)
&\exp\Bigl\{-\frac{\alpha^{\prime}}{4}{\lambda\theta}
f_{\mu\nu}(k)(g^{-1}Ng^{-1})^{\mu\nu}\Bigr\}\\
\times&:\exp\Bigl\{ik_{\mu}\mathbf{X}^{\mu}
+\lambda{A_{\mu}(k)}D_{\theta}\mathbf{X}^{\mu}\Bigr\}:\;.
 \end{split}\label{renormalized}
\end{equation}
The renormalized operators 
$\mathbf{V}_{tac}(\tau,\sigma,\theta;k)$ and 
$\mathbf{V}_{gl}^{\mu}(\tau,\sigma,\theta;k)$, 
corresponding to the open-string tachyon and gluon, 
are obtained by the following expansion of $W_{A}$ :
\begin{equation}
W_{A}(\tau,\sigma,\theta,\lambda;k)
=\mathbf{V}_{tac}(\tau,\sigma,\theta;k)
-i\lambda\;A_{\mu}(k)\mathbf{V}_{gl}^{\mu}(\tau,\sigma,\theta;k)\;.
\label{ex1}
\end{equation}
An additional finite subtraction 
has been made in (\ref{renormalized}) by multiplying 
$\exp\bigl\{-\frac{\alpha^{\prime}}{4}\lambda\theta
f_{\mu\nu}(g^{-1}Ng^{-1})^{\mu\nu}\bigr\}$. 
Similarly to the bosonic string case \cite{MN}, 
this additional subtraction is necessary 
to reproduce the correct open-string one-loop amplitudes 
and the standard BRST action on the boundary state 
with one-gluon.

As a consequence of the fact that 
the action of (\ref{renormalized}) on the boundary state 
$|B;\eta\rangle_{m}$ is no longer singular 
at the world-sheet boundary, the operator
\begin{equation}
\widehat{W}_{A}(\sigma,\theta,\lambda;k)
=\lim_{\tau\rightarrow+0}g_{\eta}^{-1}
W_{A}(\tau,\sigma,\theta,\lambda;k)g_{\eta}\;,\label{hatW}
\end{equation}
is well-defined. 
In powers of the auxiliary parameter $\lambda$, 
the operator $\widehat{W}_{A}$ is expanded as 
\begin{equation}
\widehat{W}_{A}(\sigma,\theta,\lambda;k)
=\widehat{\mathbf{V}}_{tac}(\sigma,\theta;k)
-i\lambda\;A_{\mu}(k)
\widehat{\mathbf{V}}_{gl}^{\mu}(\sigma,\theta;k)\;.
\label{ex2}
\end{equation}
The components $\widehat{\mathbf{V}}_{tac}(\sigma,\theta;k)$, 
$\widehat{\mathbf{V}}_{gl}^{\mu}(\sigma,\theta;k)$ 
are also obtained by making the Bogolubov transformation of 
$\mathbf{V}_{tac}(\tau,\sigma,\theta;k)$, 
$\mathbf{V}_{gl}^{\mu}(\tau,\sigma,\theta;k)$, 
and then taking the limit 
$\tau\rightarrow+0$. 
The OPE between $\widehat{W}_{A}$ takes the following form :
\begin{equation}
 \begin{split}
&\widehat{W}_{A}(\sigma_{1},\theta_{1},\lambda_{1};k^{(1)})\;
\widehat{W}_{A}(\sigma_{2},\theta_{2},\lambda_{2};k^{(2)})\\
&=\exp\Bigl\{\;\frac{i}{2}k^{(1)}{\times}k^{(2)}
\mathrm{sign}(\sigma_{1}-\sigma_{2})\Bigr\}\\
&\quad\times\exp\Bigl\{-\frac{i}{2\pi}
k^{(1)}{\times}k^{(2)}(\sigma_{1}-\sigma_{2})\Bigr\}\\
&\quad\times\exp\Bigl\{\;
\frac{i}{2\pi}\lambda_{1}\theta_{1}A(k^{(1)}){\times}k_{\nu}^{(2)}
+\frac{i}{2\pi}\lambda_{2}\theta_{2}A(k^{(2)}){\times}k_{\nu}^{(1)}
\Bigr\}\\
&\quad\times\exp\Bigl[\;\alpha^{\prime}G^{\mu\nu}
\bigl\{ik^{(1)}_{\mu}+\lambda_{1}A_{\mu}(k^{(1)})D_{\theta_{1}}\bigr\}
\bigl\{ik^{(2)}_{\nu}+\lambda_{2}A_{\nu}(k^{(2)})D_{\theta_{2}}\bigr\}
\mathbf{G}(\sigma_{1},\theta_{1};\sigma_{2},\theta_{2})\Bigr]\\
&\;\quad\times:\widehat{W}_{A}(\sigma_{1},\theta_{1},\lambda_{1};k^{(1)})\;
\widehat{W}_{A}(\sigma_{2},\theta_{2},\lambda_{2};k^{(2)}):\;,
 \end{split}\label{OPE}
\end{equation}
where we have employed the abbreviated notation : 
$k^{(\alpha)}{\times}k^{(\beta)}
=k^{(\alpha)}_{\mu}\theta^{\mu\nu}k^{(\beta)}_{\nu}$, 
$k^{(\alpha)}{\times}A(k^{(\beta)})
=k^{(\alpha)}_{\mu}\theta^{\mu\nu}A_{\nu}(k^{(\beta)})$. 
The appearance of the open-string metric $G^{\mu\nu}$ 
and the non-commutativity factor 
\begin{equation}
\exp\bigl\{\;\frac{i}{2}k^{(1)}\times{k}^{(2)}
\mathrm{sign}(\sigma_{1}-\sigma_{2})\bigr\} \label{NC}
\end{equation} 
in the OPE (\ref{OPE}), implies that the operator 
(\ref{hatW}) captures the open-string interactions \cite{SW}. 
From (\ref{OPE}), we can check that the operators 
$\widehat{\mathbf{V}}_{tac}(\sigma,\theta;k)$ 
and $\widehat{\mathbf{V}}_{gl}^{\mu}(\sigma,\theta;k)$ 
in (\ref{ex2}) satisfy the analogues of the OPEs between 
the tachyon and gluon vertices in the open-string theory 
with the constant $B_{\mu\nu}$ background. 
The factor 
\begin{equation}
 \begin{split}
&\exp\Bigl\{-\frac{i}{2\pi}
k^{(1)}{\times}k^{(2)}(\sigma_{1}-\sigma_{2})\Bigr\}\\
&\times\exp\Bigl\{
\;\frac{i}{2\pi}\lambda_{1}\theta_{1}A(k^{(1)}){\times}k^{(2)}
+\frac{i}{2\pi}\lambda_{2}\theta_{2}A(k^{(2)}){\times}k^{(1)}
\Bigr\}
 \end{split}\label{strange}
\end{equation}
in (\ref{OPE}) looks like strange in the open-string interpretation. 
However, in the presence of this factor, we can 
correctly reproduce the open-string one-loop amplitudes of gluons. 
The factor (\ref{strange}) also plays an important role in the derivation of 
the supergravity couplings of non-commutative D-branes 
from the boundary state formalism \cite{MN,toappear}.

By expanding the superfield operators 
$\mathbf{V}_{tac}(\tau,\sigma,\theta;k)$ and 
$\mathbf{V}_{gl}^{\mu}(\tau,\sigma,\theta;k)$ 
in powers of $\theta$, 
we obtain two kinds of renormalized operators for 
the open-string tachyon and gluon respectively:
\begin{align}
\mathbf{V}_{tac}(\tau,\sigma,\theta;k)
&=V^{(-1)}_{tac}(\tau,\sigma;k)
+{\theta}V^{(0)}_{tac}(\tau,\sigma;k)\;,\label{s1}\\
\mathbf{V}_{gl}^{\mu}(\tau,\sigma,\theta;k)
&=V^{(-1)\mu}_{gl}(\tau,\sigma;k)
+{\theta}V^{(0)\mu}_{gl}(\tau,\sigma;k)\;.\label{s2}
\end{align}
The operators $V_{tac}^{(-1)}(\tau,\sigma;k)$ 
and $V^{(-1)\mu}_{gl}(\tau,\sigma;k)$ correspond to 
the $-1$ picture vertex operators in the open-string theory, 
while $V_{tac}^{(0)}(\tau,\sigma;k)$ 
and $V^{(0)\mu}_{gl}(\tau,\sigma;k)$ correspond to 
the $0$ picture vertex operators. 
We employ the operators $V_{tac}^{(0)}(\tau,\sigma;k)$ 
and $V^{(0)\mu}_{gl}(\tau,\sigma;k)$ 
and integrate all positions of the vertices on the boundary. 
So we do not need to add any contribution of the ghosts 
and the superghosts to the vertices. 
This is relevant to compute the amplitudes for the annulus, 
in which the volume of the conformal Killing group is finite and 
the super-conformal Killing groups are absent.

The boundary state with a single open-string tachyon 
and the boundary state with a single gluon are obtained 
by acting $V_{tac}^{(0)}(\tau,\sigma;k)$ 
and $V^{(0)\mu}_{gl}(\tau,\sigma;k)$ on $|B;\eta\rangle_{m}$ 
respectively, and then by taking the limit $\tau\rightarrow+0$ :
\begin{equation}
 \begin{split}
|B_{tac};(\sigma;k);\eta\rangle_{tot}
&\equiv\lim_{\tau\rightarrow+0}V_{tac}^{(0)}
(\tau,\sigma;k)|B;\eta\rangle_{tot}\\
&=Cg_{\eta}\widehat{V}_{tac}^{(0)}(\sigma;k)
|x^{i}_{0}\rangle_{m}\otimes|B\rangle_{gh}\otimes|B;\eta\rangle_{sgh}\;,
 \end{split}\label{singletachyon}
\end{equation}
\begin{equation}
 \begin{split}
|B[A];(\sigma;k);\eta\rangle_{tot}
&\equiv\lim_{\tau\rightarrow+0}
A_{\mu}(k)V_{gl}^{(0)\mu}(\tau,\sigma;k)|B;\eta\rangle_{tot}\\
&=Cg_{\eta}A_{\mu}(k)\widehat{V}_{gl}^{(0)\mu}(\sigma;k)
|x^{i}_{0}\rangle_{m}\otimes|B\rangle_{gh}\otimes|B;\eta\rangle_{sgh}\;,
 \end{split}\label{singlegluon}
\end{equation}
where the operators 
$\widehat{V}^{(0)}_{tac}(\sigma;k)$ and 
$\widehat{V}^{(0)\mu}_{gl}(\sigma;k)$ are the components of 
the superfield operators 
$\widehat{\mathbf{V}}_{tac}(\sigma,\theta;k)$ and 
$\widehat{\mathbf{V}}^{\mu}_{gl}(\sigma,\theta;k)$ :
\begin{align}
\widehat{\mathbf{V}}_{tac}(\sigma,\theta;k)
&=\widehat{V}^{(-1)}_{tac}(\sigma;k)
+\theta\widehat{V}^{(0)}_{tac}(\sigma;k)\;,\label{hats1}\\
\widehat{\mathbf{V}}_{gl}^{\mu}(\sigma,\theta;k)
&=\widehat{V}^{(-1)\mu}_{gl}(\sigma;k)
+\theta\widehat{V}^{(0)\mu}_{gl}(\sigma;k)\;.\label{hats2}
\end{align}
In (\ref{singletachyon}) and (\ref{singlegluon}), 
we include the ghost part $|B\rangle_{gh}$ 
and the superghost part $|B;\eta\rangle_{sgh}$. 
The states (\ref{singletachyon}) and (\ref{singlegluon}) 
depend on the value of $\eta$, 
through both the state $|B;\eta\rangle_{tot}$ 
and the operators $V_{tac}^{(0)}$, $V^{(0)\mu}_{gl}$. 
The closed-string GSO projection operator (\ref{GSO}) 
projects out the one-tachyon state (\ref{singletachyon}) :
\begin{equation}
P_{GSO}|B_{tac};(\sigma;k);\eta\rangle_{tot}=0\;,
\end{equation}
and selects a specific combination of the one-gluon state 
$|B[A];(\sigma;k);\eta=\pm1\rangle_{tot}$ :
\begin{equation}
P_{GSO}|B[A];(\sigma;k);\eta\rangle_{tot}
=\frac{1}{2}\bigl\{\;
|B[A];(\sigma;k);\eta\rangle_{tot}
-|B[A];(\sigma;k);-\eta\rangle_{tot}\bigr\}\;,
\end{equation}
respectively.

The boundary state with $N$ gluons is obtained by 
taking the same step as for $N=1$ :
\begin{equation}
 \begin{split}
&\bigl|B[A];(\sigma_{1},k^{(1)}),
(\sigma_{2},k^{(2)}),\dotsc,(\sigma_{N},k^{(N)})
;\eta\bigr\rangle_{tot}\\
&\equiv\lim_{\tau_{\alpha}\rightarrow+0}
V_{A}^{(0)}(\tau_{1},\sigma_{1};k^{(1)})
V_{A}^{(0)}(\tau_{2},\sigma_{2};k^{(2)})\dots
V_{A}^{(0)}(\tau_{N},\sigma_{N};k^{(N)})|B;\eta\rangle_{tot}\\
&\quad=C\;g_{\eta}\widehat{V}^{(0)}_{A}(\sigma_{1};k^{(1)})
\widehat{V}^{(0)}_{A}(\sigma_{2};k^{(2)})\dots
\widehat{V}^{(0)}_{A}(\sigma_{N};k^{(N)})|x_{0}^{i}\rangle_{m}
\otimes|B\rangle_{gh}\otimes|B;\eta\rangle_{sgh}\;,
 \end{split}\label{Ngluon}
\end{equation}
where we adopt an abbreviated notation :
\begin{align}
\mathbf{V}_{A}(\tau,\sigma,\theta;k)
&={V}_{A}^{(-1)}(\tau,\sigma;k)
+\theta{V}_{A}^{(0)}(\tau,\sigma;k)
\equiv{A}_{\mu}(k)\mathbf{V}_{gl}^{\mu}(\tau,\sigma,\theta;k)\;,
\label{abb1}\\
\widehat{\mathbf{V}}_{A}(\sigma,\theta;k)
&=\widehat{{V}}_{A}^{(-1)}(\sigma;k)
+\theta\widehat{{V}}_{A}^{(0)}(\tau,\sigma;k)
\equiv{A}_{\mu}(k)\widehat{\mathbf{V}}_{gl}^{\mu}(\sigma,\theta;k)\;.
\label{abb2}
\end{align}
From the OPE (\ref{OPE}), 
the $N$-gluons boundary state (\ref{Ngluon}) by itself 
describes the open-string $N$-points tree amplitude. 
The GSO projection of the $N$-gluon boundary state (\ref{Ngluon}) 
is given by the following combination :
\begin{equation}
 \begin{split}
\frac{1}{2}\Bigl\{\;&\bigr|B[A];(\sigma_{1},k^{(1)}),
(\sigma_{2},k^{(2)}),\dotsc,(\sigma_{N},k^{(N)});
\eta=+1\bigr\rangle_{tot}\\
&\quad\quad-\bigr|B[A];(\sigma_{1},k^{(1)}),(\sigma_{2},k^{(2)}),
\dotsc,(\sigma_{N},k^{(N)})
;\eta=-1\bigl\rangle_{tot}\Bigr\}\;.
\end{split}\label{GSOgluon}
\end{equation}

The closed-string cylinder amplitudes in the NS-NS sector 
should reproduce the open-string one-loop amplitudes 
with the anti-periodic boundary condition 
in the time direction of the world-sheet. 
The dual boundary state with $N$ gluons 
${}_{tot}\langle{B[A]};(\sigma_{1};k^{(1)}),\dotsc,(\sigma_{N};k^{(N)});\eta|$ 
is obtained from (\ref{Ngluon}) by taking 
the Hermitian conjugation of it and flipping the sign of $\eta$. 
The tree propagations between the boundary states are given by
\begin{equation}
 \begin{split}
{}_{tot}\bigl\langle{B[A]};(\sigma_{N+1};k^{(N+1)})&,\dotsc,
(\sigma_{N+M};k^{(N+M)});\eta^{\prime}\bigr|(c_{0}-\bar{c}_{0})\\
&\times(b_{0}+\bar{b}_{0})e^{-\tau_{(c)}(L_{0}^{tot}+\bar{L}_{0}^{tot})}
\bigl|{B[A]};(\sigma_{1};k^{(1)}),\dotsc,
(\sigma_{N};k^{(N)});\eta\bigr\rangle_{tot}\;,
 \end{split}\label{sw}
\end{equation}
where $L_{0}^{tot}$ and $\bar{L}_{0}^{tot}$ are the Virasoro zero modes 
including the contributions of the ghosts and the superghosts, 
and $\tau_{(c)}$ is the length of the cylinder. 
The insertion $(c_{0}-\bar{c}_{0})$ is associated with the rotation of 
the cylinder, while the insertion $(b_{0}+\bar{b}_{0})$ is associated 
with the moduli of the cylinder $\tau_{(c)}$. 
By integrating (\ref{sw}) 
over $\sigma_{1},\dotsc,\sigma_{M+N}$ and $\tau_{(c)}$, 
we obtain the open-string $M+N$ annulus amplitude of the Ramond sector 
when $\eta\eta^{\prime}=+1$, 
while we obtain the open-string amplitudes of the NS sector 
when $\eta\eta^{\prime}=-1$, as we expect. 
The GSO-projected combination (\ref{GSOgluon}) gives the correct combination 
of the open-string one-loop amplitudes.

\subsubsection*{Linearized equation of motion from BRST invariance}
In the closed-string field theory, an arbitrary string field 
is a vector in the closed-string Hilbert space, 
and the BRST invariance condition for it leads to the (linearized) 
equation of motion of the closed-string field theory. 
The gluon boundary states (\ref{singlegluon}), (\ref{Ngluon}) are also 
elements of the closed-string Hilbert space. 
Therefore, it is reasonable to expect that the BRST invariance condition 
for the boundary states 
with gluons leads to the equation of motion for the gluons.

Let us consider the BRST invariance 
condition for the one-gluon boundary state \\
$\oint{d\sigma}|B[A];(\sigma,k);\eta\rangle_{tot}$. 
When the vertex operators do not contain the ghosts and the superghosts, 
the overlap conditions for $|B\rangle_{gh}$ 
and $|B;\eta\rangle_{sgh}$ (\ref{ghostoverlap}) 
implies that the action of the closed-string BRST operator 
(\ref{BRST}) reduces to 
\begin{equation}
Q_{c}\quad\sim\quad
-\frac{1}{\;2\;}
\sum_{n\in\mathbb{Z}}(c_{-n}-\bar{c}_{n})(L_{n}^{m}-\bar{L}_{-n}^{m})
+\frac{1}{\;4\;}\!\!\sum_{r\in\mathbb{Z}+1/2}\!
(\gamma_{-r}-i\eta\bar{\gamma}_{r})
(G_{r}^{m}+i\eta\bar{G}_{-r}^{m})\;.\label{reduce}
\end{equation}
Thus, the BRST transformation of 
the one-gluon boundary state becomes as follows:
\begin{equation}
 \begin{split}
&Q_{c}\biggl[\int_{0}^{2\pi}d\sigma
|B[A];(\sigma,k);\eta\rangle_{tot}\biggr]\\
&=\frac{C}{2}\int_{0}^{2\pi}d\sigma\;g_{\eta}\;
\biggl\{\;
-\sum_{n\in\mathbb{Z}}(c_{-n}-\bar{c}_{n})
\bigl[\mathcal{L}_{n},\widehat{V}_{A}^{(0)}(\sigma;k)\bigr]\\
&\qquad\qquad\qquad\qquad+
\frac{1}{\;2\;}\!\!
\sum_{r\in\mathbb{Z}+1/2}
(\gamma_{-r}-i\eta\bar{\gamma}_{r})
\bigl[\mathcal{G}_{r},\widehat{V}_{A}^{(0)}(\sigma;k)\bigr]
\;\biggr\}                
|x^{i}_{0}\rangle_{m}\otimes
|B\rangle_{gh}\otimes|B;\eta\rangle_{sgh}\;,
 \end{split}\label{BRST1}
\end{equation} 
where $\mathcal{L}_{n}$, $\mathcal{G}_{r}$ are the Bogolubov transformations 
of the generators of the super-diffeomorphism on the boundary:
\begin{equation}
\mathcal{L}_{n}
=g_{\eta}^{-1}(L_{n}^{m}-\bar{L}_{-n}^{m})g_{\eta}\;,\quad
\mathcal{G}_{r}
=g_{\eta}^{-1}(G_{r}^{m}+i\eta\bar{G}_{-r}^{m})g_{\eta}\;.
\label{superdiff}
\end{equation}
The commutation relations between the generators in (\ref{superdiff}) 
and the gluon vertex operators in (\ref{abb2}) are as follows:
\begin{equation}
 \begin{split}
\;\bigl[\mathcal{L}_{n},\widehat{V}^{(0)}_{A}(\sigma;k)\bigr]
&=-i\frac{d}{d\sigma}\bigl\{
e^{in\sigma}\widehat{V}_{A}^{(0)}(\sigma;k)\bigr\}
 +\alpha^{\prime}\bigl(k\cdot{k}\bigr)ne^{in\sigma}
\widehat{V}_{A}^{(0)}(\sigma;k)\\
&\qquad\qquad\qquad\qquad\qquad\qquad
 +\alpha^{\prime}\bigl(k\cdot{A}(k)\bigr)n^{2}e^{in\sigma}
\widehat{V}_{tac}^{(-1)}(\sigma;k)\;,
 \end{split}\label{v0}
\end{equation}
\begin{equation}
 \begin{split}
\bigl[\mathcal{G}_{r},\widehat{V}^{(0)}_{A}(\sigma;k)\bigr]
&=-i\frac{d}{d\sigma}\bigl\{
e^{ir\sigma}\widehat{V}_{A}^{(-1)}(\sigma;k)\bigr\}
 +2\alpha^{\prime}\bigl(k\cdot{k}\bigr)re^{ir\sigma}
\widehat{V}_{A}^{(-1)}(\sigma;k)\\
&\qquad\qquad\qquad\qquad\qquad\qquad
 -2\alpha^{\prime}\bigl(k\cdot{A}(k)\bigr)re^{ir\sigma}
\widehat{V}_{tac}^{(0)}(\sigma;k)\;,
 \end{split}\label{sv0}
\end{equation}
\begin{align}
\bigl[\mathcal{L}_{n},\widehat{V}^{(-1)}_{A}(\sigma;k)\bigr]
&=-i\frac{d}{d\sigma}\bigl\{
e^{in\sigma}\widehat{V}_{A}^{(-1)}(\sigma;k)\bigr\}
+\bigl(\;\alpha^{\prime}k\cdot{k}-\frac{1}{2}\;\bigr)
ne^{in\sigma}\widehat{V}_{A}^{(-1)}(\sigma;k)\;,
\label{v1}\\
\bigl\{\mathcal{G}_{r},\widehat{V}^{(-1)}_{A}(\sigma;k)\bigr\}
&=e^{ir\sigma}\widehat{V}_{A}^{(0)}(\sigma;k)
+2\alpha^{\prime}\bigl(k\cdot{A}(k)\bigr)re^{ir\sigma}
\widehat{V}_{tac}^{(-1)}(\sigma;k)\;,\label{sv1}
\end{align}
where we have abbreviated as 
$k\cdot{k}
=k_{\mu}G^{\mu\nu}k_{\nu}$, 
$k\cdot{A}(k)
=k_{\mu}G^{\mu\nu}A_{\nu}(k)$. 
By plugging (\ref{v0}) and (\ref{sv0}) into (\ref{BRST1}), 
we obtain
\begin{equation}
 \begin{split}
&Q_{c}\biggl[\int_{0}^{2\pi}d\sigma
|B[A];(\sigma,k);\eta\rangle_{tot}\biggr]\\
&=-\frac{iC}{2}\int^{2\pi}_{0}d\sigma\;\frac{d}{d\sigma}
\Bigl\{-\sum_{n\in\mathbb{Z}}(c_{-n}-\bar{c}_{n})e^{in\sigma}
\widehat{V}_{A}^{(0)}(\sigma;k)
+\frac{1}{\;2\;}\!\!\sum_{r\in\mathbb{Z}+1/2}\!
(\gamma_{-r}-i\eta\bar{\gamma}_{r})e^{ir\sigma}
\widehat{V}_{A}^{(-1)}(\sigma;k)\Bigr\}\\
&\quad-C\int^{2\pi}_{0}d\sigma\int{d\theta}\;g_{\eta}
\biggl[\;-\alpha^{\prime}\bigl(k\cdot{k}\bigr)A_{\mu}(k)
D_{\theta}^{\;2}\mathbf{C}(\sigma,\theta)
\widehat{\mathbf{V}}_{gl}^{\mu}(\sigma,\theta;k)\\
&\qquad\qquad\qquad\qquad\qquad\quad
+\alpha^{\prime}\bigl(k\cdot{A}(k)\bigr)
D^{\;3}_{\theta}\mathbf{C}(\sigma,\theta)
\widehat{\mathbf{V}}_{tac}(\sigma,\theta;k)\;
\biggr]\;
|x^{i}_{0}\rangle_{m}\otimes
|B\rangle_{gh}\otimes|B;\eta\rangle_{sgh}\;,
 \end{split}\label{BRST2}
\end{equation}
where we have combined the ghosts and the superghosts 
into the superfield on the boundary 
\begin{equation}
\mathbf{C}(\sigma,\theta)
=\frac{\;1\;}{2}\Bigl\{\;
\sum_{n\in\mathbb{Z}}\;(c_{-n}-\bar{c}_{n})\;e^{in\sigma}
\;+\;\theta\!\!\sum_{r\in\mathbb{Z}+1/2}\!
(\gamma_{-r}-i\eta\bar{\gamma}_{r})\;e^{ir\sigma}\;\Bigr\}\;.
\end{equation}
Since the the total derivative term in (\ref{BRST2}) vanishes, 
the BRST transformation law of the single-gluon boundary state 
becomes
\begin{equation}
 \begin{split}
&Q_{c}\biggl[\int_{0}^{2\pi}d\sigma
|B[A];(\sigma,k);\eta\rangle_{tot}\biggr]\\
&=\alpha^{\prime}\bigl(k\cdot{k}\bigr)A_{\mu}(k)\;
C\int^{2\pi}_{0}\!d\sigma\int{d\theta}\;g_{\eta}\;
D_{\theta}^{\;2}\mathbf{C}(\sigma,\theta)
\widehat{\mathbf{V}}_{gl}^{\mu}(\sigma,\theta;k)
|x^{i}_{0}\rangle_{m}\otimes
|B\rangle_{gh}\otimes|B;\eta\rangle_{sgh}\\
&\quad-\alpha^{\prime}\bigl(k\cdot{A}(k)\bigr)\;
C\int^{2\pi}_{0}\!d\sigma\int{d\theta}\;g_{\eta}\;
D^{\;3}_{\theta}\mathbf{C}(\sigma,\theta)
\widehat{\mathbf{V}}_{tac}(\sigma,\theta;k)
|x^{i}_{0}\rangle_{m}\otimes
|B\rangle_{gh}\otimes|B;\eta\rangle_{sgh}\;.
 \end{split}\label{BRST3}
\end{equation}
Therefore, the BRST invariance condition 
of the single-gluon boundary state 
$\oint{d\sigma}|B[A];(\sigma;k);\eta\rangle_{tot}$ requires 
\begin{equation}
 \bigl(k\cdot{k}\bigr)A_{\mu}(k)=0\;,\quad
 k\cdot{A}(k)=0\;.
\end{equation}
These are nothing but the linearized equation of motion and 
the Lorentz gauge condition for the gluon with the open-string metric 
$G^{\mu\nu}$.

It is enlightening to rewrite (\ref{BRST3}) 
from the string field theory viewpoint. 
In the string field theory, 
the world-sheet BRST transformation law 
for the vertex operators 
is translated into 
the target space BRST transformation law 
for the string fields. 
The equation (\ref{BRST3}) says that 
the world-sheet BRST transformation of the single-gluon boundary state 
becomes the linear combination of the boundary states with the vertices  
\begin{equation}
\widehat{\mathbf{V}}^{\ast\mu}_{gl}(\sigma,\theta;k)
=\Bigl[\;(D_{\theta}^{\;2}\mathbf{C})\widehat{\mathbf{V}}^{\mu}_{gl}
\;\Bigr](\sigma,\theta;k)\;,\quad
\widehat{\mathbf{V}}_{a.g}(\sigma,\theta;k)
=\Bigl[\;(D_{\theta}^{\;3}\mathbf{C})\widehat{\mathbf{V}}_{tac}
\;\Bigr](\sigma,\theta;k)\;.
\label{anti}
\end{equation}
Each coefficient in (\ref{BRST3}) is identified with 
the target space BRST transformation of the string fields 
which are associated with the vertices 
$\widehat{\mathbf{V}}^{\ast\mu}_{gl}(\sigma,\theta;k)$ 
and $\widehat{\mathbf{V}}_{a.g}(\sigma,\theta;k)$.

The open-string field theory in itself is equipped 
\cite{TB} with the structure 
of the Batalin-Vilkovisky formulation \cite{BV}. 
Due to the ghost-number anomaly, 
the inner-product in the open-string Hilbert space 
becomes the odd-symplectic form, 
and the anti-bracket is obtained from this odd-symplectic form. 
Each string field is paired with its antifield 
in accord with the anti-bracket. 
States which correspond to the vertices 
$\widehat{\mathbf{V}}^{\ast\mu}_{gl}(\sigma,\theta;k)$ 
and $\widehat{\mathbf{V}}_{a.g}(\sigma,\theta;k)$ 
can be found in the open-string Hilbert space. 
Since $D_{\theta}^{\;2}\mathbf{C}(\sigma,\theta)$ 
corresponds to the ghost zero-mode of the open-string theory, 
the state which corresponds to the vertex 
$\widehat{\mathbf{V}}^{\ast\mu}_{gl}(\sigma,\theta;k)$ 
must be paired with the state of the gluon vertex 
$\widehat{\mathbf{V}}^{\mu}_{gl}(\sigma,\theta;k)$. 
Therefore, the string field associated with the vertex operator 
$\widehat{\mathbf{V}}^{\ast\mu}_{gl}(\sigma,\theta;k)$ 
is the antifield for the gluon $A_{\mu}(k)$ . 
We denote it by $A_{\mu}^{\ast}(k)$. 
The string field which is associated with 
$\widehat{\mathbf{V}}_{a.g}(\sigma,\theta;k)$ 
is the \textit{target space} antighost $\mathcal{B}(k)$. 
Both $A_{\mu}^{\ast}(k)$ and $\mathcal{B}(k)$ 
are the Grassmann-odd fields 
and have the ghost number $-1$. 
Each coefficient in (\ref{BRST3}) is identified with 
the \textit{target space} BRST transformation of 
$A_{\mu}^{\ast}(k)$ and $\mathcal{B}(k)$ respectively. 
Therefore, (\ref{BRST3}) is rewritten as follows:
\begin{equation}
 \begin{split}
&Q_{c}\;\biggl[\;\int_{0}^{2\pi}\!d\sigma\int{d}\theta\;
g_{\eta}\;A_{\mu}(k)
\widehat{\mathbf{V}}_{gl}(\sigma,\theta;k)\;\biggr]\;
|x^{i}_{0}\rangle_{m}\otimes
|B\rangle_{gh}\otimes|B;\eta\rangle_{sgh}\\
&=-\int^{2\pi}_{0}d\sigma\int{d\theta}\;g_{\eta}
\biggl[\;\delta_{\mathbf{B}}^{0}A_{\mu}^{\ast}(k)
\widehat{\mathbf{V}}^{\ast\mu}_{gl}(\sigma,\theta;k)
+\delta_{\mathbf{B}}^{0}\mathcal{B}(k)
\widehat{\mathbf{V}}_{a.g}(\sigma,\theta;k)\;
\biggr]\\
&\qquad\qquad\qquad\qquad\qquad
\qquad\qquad\qquad\qquad\qquad
\times
|x^{i}_{0}\rangle_{m}\otimes
|B\rangle_{gh}\otimes|B;\eta\rangle_{sgh}\;,
 \end{split}\label{BRST4}
\end{equation}
where the target space BRST transformations are given by
\footnote{Strictly speaking, 
the transformation laws are given by 
$\delta_{\mathbf{B}}^{0}A_{\mu}^{\ast}(k)
=\alpha^{\prime}\bigl\{\bigl(k\cdot{k}\bigr)A_{\mu}(k)
-k_{\mu}\bigl(k\cdot{A}(k)\bigr)\bigr\}
+i\alpha^{\prime}\pi(k)$ and 
$\delta_{\mathbf{B}}^{0}\mathcal{B}(k)
=-i\alpha^{\prime}\pi(k)$, where $\pi(k)$ is the string field 
corresponding to the Nakanishi-Lautrup field. 
We put $\pi(k)=i\bigl(k\cdot{A}(k)\bigr)$ and obtain (\ref{linear}).}
\begin{equation}
\delta_{\mathbf{B}}^{0}A_{\mu}^{\ast}(k)
=-\alpha^{\prime}\bigl(k\cdot{k}\bigr)A_{\mu}(k)\;,\quad
\delta_{\mathbf{B}}^{0}\mathcal{B}(k)
=\alpha^{\prime}\bigl(k\cdot{A}(k)\bigr)\;.
\label{linear}
\end{equation}

\section{Boundary Interaction of Closed-String}
\label{section3}
In the previous section, we have constructed the boundary states with 
arbitrary numbers of the gluons. 
They reproduce the open-string amplitudes correctly. 
In this section, by using these boundary states, 
we describe the interaction with external gauge field 
as an element of the closed-string Hilbert space.

The boundary state with $N$-gluons (\ref{Ngluon}) 
can be rewritten by means of the supersymmetric gluon vertices 
(\ref{abb2}) as follows : 
\begin{equation}
 \begin{split}
&\bigl|B[A];(\sigma_{1},k^{(1)}),\dotsc,
(\sigma_{N},k^{(N)});\eta\bigr\rangle_{tot}\\
&=C\;g_{\eta}
\int{d}\theta_{1}\widehat{\mathbf{V}}_{A}(\sigma_{1},\theta_{1};k^{(1)})
\dots
\int{d}\theta_{N}\widehat{\mathbf{V}}_{A}(\sigma_{N},\theta_{N};k^{(N)})
|x^{i}_{0}\rangle_{m}\otimes|B\rangle_{gh}\otimes|B;\eta\rangle_{sgh}\;.
 \end{split}\label{30}
\end{equation}
One may expect that the closed-string state which represents 
the boundary interaction with the external gauge field 
\begin{equation}
A_{\mu}(x)=\int\frac{d^{p+1}k}{(2\pi)^{\frac{p+1}{2}}}A_{\mu}(k)e^{ik\cdot{x}}
\end{equation}
is obtained by integrating (\ref{30}) over the momenta 
$k^{(1)},\dotsc,k^{(N)}$ and the position of the gluons 
$\sigma_{1},\dotsc,\sigma_{N}$, 
and then by summing it over $N=0,1,2,\dotsc,\infty$ :
\begin{equation}
 \begin{split}
&C\;g_{\eta}\sum_{N=0}^{\infty}i^{N}
\int_{0}^{2\pi}\!d\sigma_{1}d\theta_{1}
\widehat{\mathbf{V}}_{A}\bigl(\mathbf{X}(\sigma_{1},\theta_{1})\bigr)
\cdots
\int_{0}^{2\pi}\!d\sigma_{N}d\theta_{N}
\widehat{\mathbf{V}}_{A}\bigl(\mathbf{X}(\sigma_{N},\theta_{N})\bigr)\\
&\qquad\qquad\qquad\qquad\qquad\qquad\times
\prod_{\alpha=1}^{N-1}\Theta(\sigma_{\alpha+1}-\sigma_{\alpha})\;
|x^{i}_{0}\rangle_{m}\otimes|B\rangle_{gh}\otimes|B;\eta\rangle_{sgh}\;,
 \end{split}\label{3a}
\end{equation}
where $\Theta(\sigma)$ is the step function defined as 
$\Theta(\sigma)=1$ for $\sigma>0$ and 
$\Theta(\sigma)=0$ for $\sigma<0$. 
In (\ref{3a}), we have introduced the notation 
\begin{equation}
\widehat{\mathbf{V}}_{A}\bigl(\mathbf{X}(\sigma,\theta)\bigr)
=\int\frac{d^{p+1}k}{(2\pi)^{\frac{p+1}{2}}}
\widehat{\mathbf{V}}_{A}(\sigma,\theta;k)\;.
\label{conf}
\end{equation}

However, due to the short-distance divergence of the 
boundary correlation functions (\ref{Green}), 
the state (\ref{3a}) is not a well-defined object, 
and needs some regularization procedure. 
We make a point-splitting regularization at the world-sheet boundary, 
by introducing a small positive cut-off parameter $\epsilon$ and 
replacing the step-functions $\Theta(\sigma_{\alpha+1}-\sigma_{\alpha})$ 
in (\ref{3a}) with the regularized ones 
$\Theta(\sigma_{\alpha+1}-\sigma_{\alpha}-\epsilon)$. 
In the cut-off theory, 
it is convenient to deal with the dimensionless couplings. 
The equation (\ref{v0}) says that the gluon vertex operator 
$\int{d\sigma}{d\theta}\widehat{\mathbf{V}}_{gl}^{\mu}(\sigma,\theta;k)$ 
has the scaling dimension 
$\alpha^{\prime}(k{\cdot}k)$. 
To make the couplings $A_{\mu}(k)$ dimensionless, 
we make the redefinition of the couplings as 
\begin{equation}
A_{\mu}(k)\quad\longmapsto\quad
\epsilon^{\alpha^{\prime}k\cdot{k}}A_{\mu}(k)\;,
\end{equation}
and the vertex operators 
$\widehat{\mathbf{V}}_{A}(\sigma,\theta;k)$ and 
$\widehat{\mathbf{V}}_{A}\bigl(\mathbf{X}(\sigma,\theta)\bigr)$ 
are replaced with 
\begin{equation}
\widehat{\mathbf{V}}_{A,\epsilon}(\sigma,\theta;k)
=\epsilon^{\alpha^{\prime}k\cdot{k}}
\widehat{\mathbf{V}}_{A}(\sigma,\theta;k)\;,\quad
\widehat{\mathbf{V}}_{A,\epsilon}\bigl(\mathbf{X}(\sigma,\theta)\bigr)
=\int\frac{d^{p+1}k}{(2\pi)^{\frac{p+1}{2}}}
\widehat{\mathbf{V}}_{A,\epsilon}(\sigma,\theta;k)\;,
\label{rvertex}
\end{equation}
respectively. 
Instead of (\ref{3a}), we thus obtain
\begin{equation}
 \begin{split}
&C\;g_{\eta}\sum_{N=0}^{\infty}\frac{i^{N}}{N}
\int_{0}^{2\pi}\!d\sigma_{1}d\theta_{1}
\widehat{\mathbf{V}}_{A,\epsilon}\bigl(\mathbf{X}(\sigma_{1},\theta_{1})\bigr)
\int_{-\infty}^{+\infty}\!\!d\sigma_{2}d\theta_{2}
\widehat{\mathbf{V}}_{A,\epsilon}
\bigl(\mathbf{X}(\sigma_{2},\theta_{2})\bigr)\\
&\qquad\qquad\qquad\qquad\qquad\qquad
\times\cdots\times
\int_{-\infty}^{+\infty}\!\!d\sigma_{N}d\theta_{N}
\widehat{\mathbf{V}}_{A,\epsilon}
\bigl(\mathbf{X}(\sigma_{N},\theta_{N})\bigr)\\
&\qquad\qquad\qquad\qquad\qquad\qquad\qquad\times
\prod_{\alpha=1}^{N}\Theta(\sigma_{\alpha+1}-\sigma_{\alpha}-\epsilon)\;
|x^{i}_{0}\rangle_{m}\otimes|B\rangle_{gh}\otimes|B;\eta\rangle_{sgh}\;.
 \end{split}\label{3b}
\end{equation}
In the above expression (\ref{3b}), 
we formally extend the range of the coordinates 
$\sigma_{\alpha}$ ($\alpha=2,\dotsc,N$) 
to $-\infty\leq\sigma_{\alpha}\leq+\infty$ 
by using the $2\pi$ periodicity of the boundary state. 
We put $\sigma_{N+1}=\sigma_{1}+2\pi$. 
The singularity which arises when the two gluons 
at $\sigma_{1}$ and $\sigma_{N}$ collide 
has been regularized by inserting 
$\Theta(\sigma_{N+1}-\sigma_{N}-\epsilon)
=\Theta(\sigma_{1}+2\pi-\sigma_{N}-\epsilon)$.

The state (\ref{3b}) is not invariant 
under the global world-sheet supersymmetry on the boundary :
$\delta_{\xi}\sigma=i\xi\theta$, $\delta_{\xi}\theta=\xi$. 
The obstruction is the use of the step functions 
$\Theta(\sigma_{\alpha+1}-\sigma_{\alpha}-\epsilon)$ 
in the path-ordering procedure. 
Following to \cite{AT}, we recover this global world-sheet supersymmetry 
by replacing $\sigma_{\alpha+1}-\sigma_{\alpha}$, 
the argument of the step functions in (\ref{3b}), with 
the supersymmetric invariant distances 
$\sigma_{\alpha+1}-\sigma_{\alpha}
-i\theta_{\alpha}\theta_{\alpha+1}$. 
The supersymmetrized step functions 
\begin{equation}
 \begin{split}
&\Theta(\sigma_{\alpha+1}-\sigma_{\alpha}
       -i\theta_{\alpha}\theta_{\alpha+1}-\epsilon)\\
&=\Theta(\sigma_{\alpha+1}-\sigma_{\alpha}-\epsilon)
-i\theta_{\alpha}\theta_{\alpha+1}
\delta(\sigma_{\alpha+1}-\sigma_{\alpha}-\epsilon)\;,
 \end{split}
\end{equation}
introduce the contact terms in (\ref{3b}). 
It is observed in \cite{AT} that 
these contact terms exclude unpleasant divergences in 
the superstring amplitudes, 
and the global world-sheet supersymmetry on the boundary 
is also associated with the non-abelian gauge invariance of 
the effective theory.

With the point-splitting regularization and 
the supersymmetric path-ordering procedure, 
we finally obtain
\begin{align}
&\Bigl|\;W_{\epsilon}[A];\eta\;\Bigr\rangle\notag\\*
&=\mathrm{SP}_{\epsilon}\biggl[\;C\;g_{\eta}
\exp\Bigl\{\;i\oint{d\sigma}\int{d\theta}\;
\widehat{\mathbf{V}}_{A,\epsilon}\bigl(\mathbf{X}(\sigma,\theta)\bigr)
\;\Bigr\}\;
\biggr]
|x^{i}_{0}\rangle_{m}\otimes|B\rangle_{gh}\otimes|B;\eta\rangle_{sgh}\notag\\
&=
C\;g_{\eta}\sum_{N=0}^{\infty}\frac{i^{N}}{N}
\int^{2\pi}_{0}\!\!d\sigma_{1}{d}\theta_{1}
\widehat{\mathbf{V}}_{A,\epsilon}\bigl(\mathbf{X}(\sigma_{1},\theta_{1})\bigr)
\int^{+\infty}_{-\infty}\!\!d\sigma_{2}{d}\theta_{2}
\widehat{\mathbf{V}}_{A,\epsilon}
\bigl(\mathbf{X}(\sigma_{2},\theta_{2})\bigr)\notag\\*
&\qquad\qquad\qquad\qquad\qquad
\times\cdots\cdots\times
\int^{+\infty}_{-\infty}\!\!d\sigma_{N}{d}\theta_{N}
\widehat{\mathbf{V}}_{A,\epsilon}
\bigl(\mathbf{X}(\sigma_{N},\theta_{N})\bigr)\notag\\*
&\qquad\qquad\qquad\qquad\qquad\times
\prod_{\alpha=1}^{N}
\Theta(\sigma_{\alpha+1}-\sigma_{\alpha}
-i\theta_{\alpha}\theta_{\alpha+1}-\epsilon)\;
|x^{i}_{0}\rangle_{m}\otimes|B\rangle_{gh}\otimes|B;\eta\rangle_{sgh}\notag\\
&=C\;\sum_{N=0}^{\infty}\frac{i^{N^{2}}}{N}
\int^{2\pi}_{0}\!\!{d}\sigma_{1}d\theta_{1}
\int^{+\infty}_{-\infty}\!\!{d}\sigma_{2}d\theta_{2}
\cdots\int^{+\infty}_{-\infty}\!\!{d}\sigma_{N}d\theta_{N}
\prod_{\alpha=1}^{N}
\Theta(\sigma_{\alpha+1}-\sigma_{\alpha}
-i\theta_{\alpha}\theta_{\alpha+1}-\epsilon)\notag\\*
&\qquad\qquad\times{g}_{\eta}\;
\widehat{\mathbf{V}}_{A,\epsilon}\bigl(\mathbf{X}(\sigma_{1},\theta_{1})\bigr)
\widehat{\mathbf{V}}_{A,\epsilon}\bigl(\mathbf{X}(\sigma_{2},\theta_{2})\bigr)
\cdots
\widehat{\mathbf{V}}_{A,\epsilon}\bigl(\mathbf{X}(\sigma_{N},\theta_{N})\bigr)
|x^{i}_{0}\rangle_{m}\otimes|B\rangle_{gh}\otimes|B;\eta\rangle_{sgh}
\label{SWilson}
\end{align}
where the symbol $\mathrm{SP}_{\epsilon}$ denotes the supersymmetric 
path-ordering operation with the cut-off parameter $\epsilon$. 
We put $(\sigma_{N+1},\theta_{N+1})=(\sigma_{1}+2\pi,-\theta_{1})$. 
We argue that the state (\ref{SWilson}) is the correct closed-string state 
which describes the boundary interaction with the gauge field. 
In section~4, we will confirm that the invariance condition of 
(\ref{SWilson}) under the closed-string BRST transformation leads 
the equation of motion for the non-commutative gauge theory. 
In particular, we will see that the supersymmetric path-ordering procedure 
controls the short-distance divergence of the world-sheet theory 
in the exquisite way.

From the space-time point of view, the state 
$\bigl|W_{\epsilon}[A]\bigr\rangle_{tot}$ 
is the closed-string theoretical counterpart of the Wilson loop. 
In the gauge theory, 
the basic gauge-invariant observables are provided by the Wilson loops 
\footnote{In the non-commutative gauge theory, 
the Wilson loops do not form closed loops in the target space, and 
are called open Wilson lines. 
One can check that the coupling of (\ref{SWilson}) 
with the closed-string states 
reproduces the open Wilson line 
in the $\alpha^{\prime}\rightarrow{0}$ limit \cite{toappear}.}.  
Therefore, the state 
$\bigl|W_{\epsilon}[A]\bigr\rangle_{tot}$ 
becomes the basic ingredient of the open-closed string field theory. 
In the sequel, we call the state (\ref{SWilson}) the Wilson loop.

\section{Non-linear Equation of Motion from BRST Invariance}\label{section4}
In this section 
we describe the BRST invariance condition 
of the Wilson loop. 
Let us first explain the origin of the non-linearity 
of the BRST transformation.

\begin{figure}[t]
\begin{center}
\includegraphics[scale=0.8]{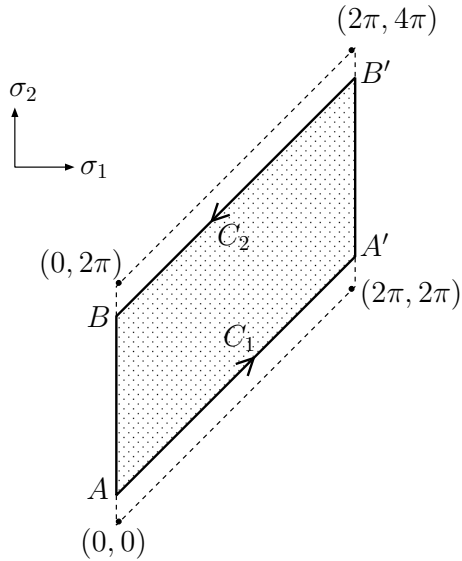}
\caption{\small 
Configuration space 
$\mathcal{M}_{2,\epsilon}$ is the shaded parallelogram 
$AA'B'B$, where the two lines 
$\overline{AB}$ and $\overline{A'B'}$
are identified with each other. 
$A=(0,\epsilon)$, $B=(0,2\pi-\epsilon)$, 
$A'=(2\pi,2\pi+\epsilon)$, $B'=(2\pi,4\pi-\epsilon)$. 
The boundary of $\mathcal{M}_{2,\epsilon}$ 
consists of the two lines 
$C_1$ and $C_2$ . 
The parallelogram drawn by the dotted lines is 
the configuration space of the unregularized theory 
($\epsilon=0$). 
}
\label{Fig:2-point}
\end{center}
\end{figure}

\begin{figure}[t]
\begin{center}
\includegraphics[scale=0.8]{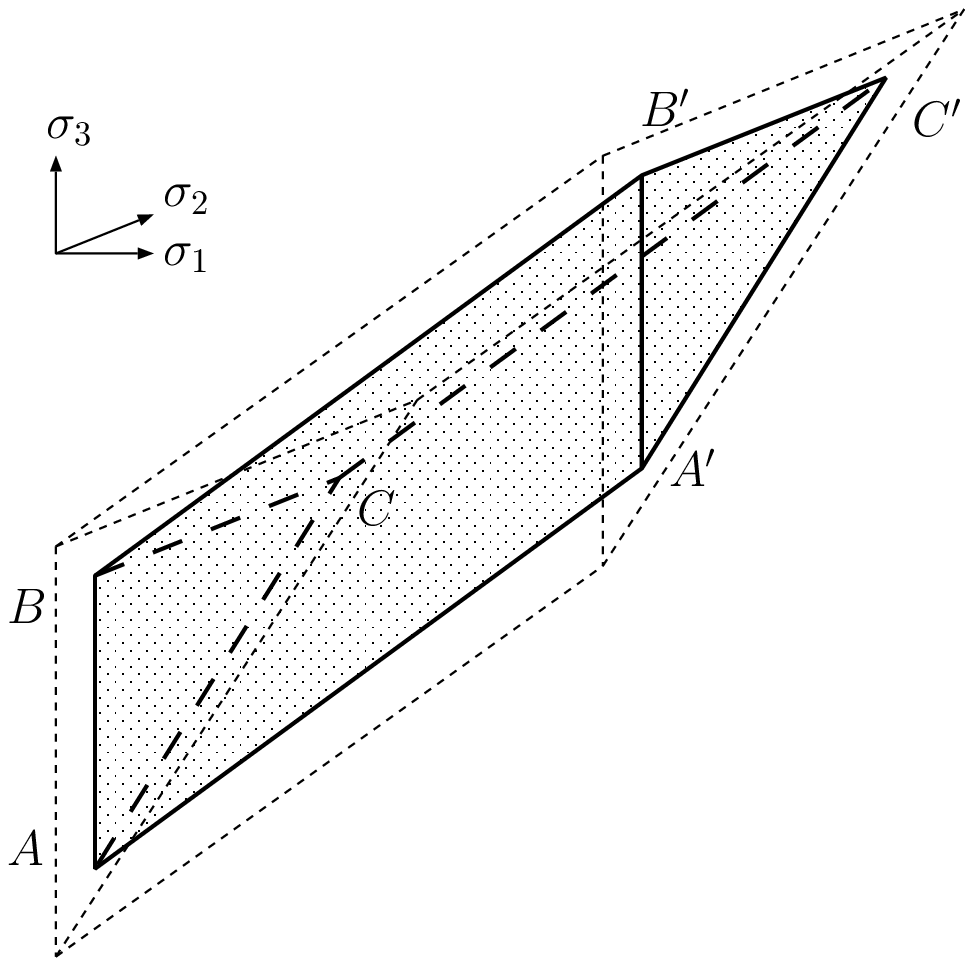}
\caption{\small 
Configuration space 
$\mathcal{M}_{3,\epsilon}$ 
is the shaded triangular prism, 
where the two sides $ABC$ and $A'B'C'$ 
are identified with each other. 
$A=(0,\epsilon,2\epsilon),\ 
B=(0,\epsilon,2\pi-\epsilon),\ 
C=(0,2\pi-2\epsilon,2\pi-\epsilon),\ 
A'=(2\pi,2\pi+\epsilon,2\pi+2\epsilon),\ 
B'=(2\pi,2\pi+\epsilon,4\pi-\epsilon),\ 
C'=(2\pi,4\pi-2\epsilon,4\pi-\epsilon)
$
The boundary of $\mathcal{M}_{3,\epsilon}$ 
consists of the three parallelograms 
$AA'B'B,\ CC'A'A,$ and $BB'C'C$. 
The triangular prism drawn by the dotted lines 
is the configuration space of the unregularized theory 
($\epsilon=0$). 
}
\label{Fig:3-point}
\end{center}
\end{figure}

The Wilson loop consists of the boundary states 
of several numbers of gluons, where the gluon's positions 
$(\sigma_s,\theta_s)$ are integrated over according 
to the supersymmetric path-ordering procedure (\ref{SWilson}). 
The action of the closed-string BRST operator on the Wilson loop 
is obtained from the collection of the actions 
on these boundary states of gluons.
For the multi-gluon boundary states, 
computations of the actions of the closed-string 
BRST operator become more complicated than the single gluon boundary state. 
Let us consider the following state 
which appears in the non-supersymmetric path-ordered Wilson loop 
(\ref{3b}):
\begin{eqnarray}
&&
\frac{i^N}{N}C
\int
\!\prod_{\alpha=1}^N 
\frac{d^{p+1}k^{(\alpha)}}{(2\pi)^{\frac{p+1}{2}}}
~g_{\eta}
\int_{{\cal M}_{N,\epsilon}}
\!\!\! 
d\sigma_1 \wedge \cdots \wedge d\sigma_N 
\prod_{\beta=1}^N 
\widehat{V}_{A,\epsilon}^{(0)}
(\sigma_{\beta};k^{(\beta)})
\nonumber \\*
&&
~~~~~~~~~~~~~~~~~~~~~~~~~
\times 
|x_0^i\rangle_m 
\otimes 
|B \rangle_{gh}
\otimes 
|B;\eta \rangle_{sgh}~.  
\label{test 1}
\end{eqnarray}
In the supersymmetric path-ordered Wilson loop (\ref{SWilson}), 
the above state (\ref{test 1}) arises as an element 
of the $N$-gluon boundary state.
It is obtained from the $N$-gluon boundary state 
by integrating out all the anti-commuting coordinates 
$\theta_{\alpha}$ of the superfield operators 
$\widehat{{\bf V}}_{A,\epsilon}({\bf X}(\sigma_{\alpha},\theta_{\alpha}))$. 
The supersymmetrized step functions in the path-ordering procedure 
just contribute as the step functions 
$\Theta(\sigma_{\alpha+1}-\sigma_{\alpha}-\epsilon)$. 
The integration region ${\cal M}_{N,\epsilon}$ of the above integral is 
the configuration space of $\epsilon$-separated ordered 
$N$ points on the boundary circle.  
It consists of the $N$ points $(\sigma_1,\cdots,\sigma_N)$ 
which satisfy the conditions 
\begin{eqnarray}
0 \leq \sigma_1 <2\pi~,~~~~~~~
\sigma_{\alpha+1}-\sigma_{\alpha} \geq \epsilon~~~~
(1 \leq \alpha \leq N)~.
\label{MN-epsilon}
\end{eqnarray}
${\cal M}_{2,\epsilon}$ and ${\cal M}_{3,\epsilon}$
are depicted in 
Figures \ref{Fig:2-point} and \ref{Fig:3-point}. 

\begin{figure}[t]
\begin{center}
\includegraphics[scale=0.8]{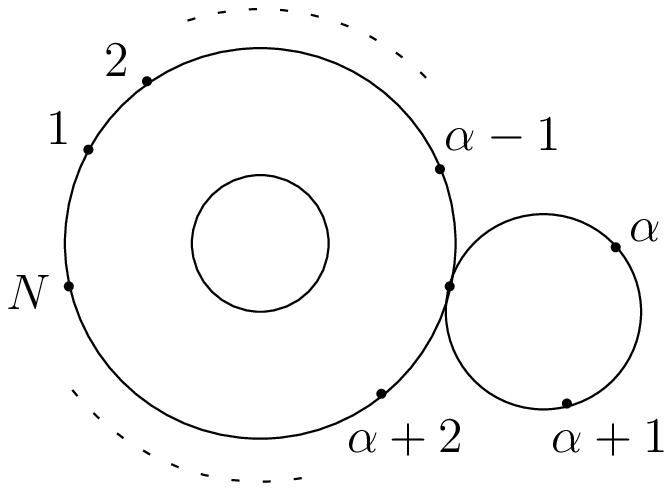}
\caption{\small 
Typical configuration of $\mathcal{N}^{(\alpha)}_{N,\epsilon}$.
}
\label{Fig:boundarydisc}
\end{center}
\end{figure}

The action of the superconformal generators
on the $N$-gluon boundary state  
can be computed by using the commutation relations 
(\ref{v0})-(\ref{sv1}) 
as in the case of the single gluon boundary state. 
However, the vanishings of the total derivative terms 
such as occurred for the single
gluon boundary state do not happen in the present case 
since the boundary of the configuration space 
is not empty. For instance, let us consider the Virasoro 
action on the above state (\ref{test 1}). 
The computation by using the commutation relation 
(\ref{v0}) 
gives rise to the following boundary integral: 
\begin{eqnarray}
&&
\frac{i^{(N-1)}}{N}\int_{{\cal M}_{N,\epsilon}}
d\sigma_1 \wedge \cdots \wedge d\sigma_N 
\sum_{\alpha=1}^N
\frac{d}{d\sigma_{\alpha}}
\left\{ 
e^{in\sigma_{\alpha}}
\prod_{\beta=1}^N
\widehat{V}_{A,\epsilon}^{(0)}
(\sigma_{\beta};k^{(\beta)})
\right\}
\nonumber \\
&&
=
\frac{i^{(N-1)}}{N}\int_{\partial {\cal M}_{N,\epsilon}}
\sum_{\alpha=1}^N(-)^{\alpha-1}
e^{in\sigma_{\alpha}}
d\sigma_1 \wedge \cdots 
\stackrel{\alpha}{\stackrel{\vee}{\cdot}}
\cdots \wedge d\sigma_N 
\prod_{\beta=1}^N
\widehat{V}_{A,\epsilon}^{(0)}
(\sigma_{\beta};k^{(\beta)})~.
\label{Virasoro test 1}
\end{eqnarray}
The boundary $\partial {\cal M}_{N,\epsilon}$ has several 
components. It can be written formally in the form 
$\partial {\cal M}_{N,\epsilon}=
\cup_{\alpha=1}^N {\cal N}_{N,\epsilon}^{(\alpha)}$, where 
${\cal N}_{N,\epsilon}^{(\alpha)}$ is the boundary component 
characterized by the condition 
$\sigma_{\alpha+1}=\sigma_{\alpha}+\epsilon$. 
For the case of $N=2$,  
the boundary components ${\cal N}_{2,\epsilon}^{(1)}$ 
and ${\cal N}_{2,\epsilon}^{(2)}$ are respectively 
the lines $C_1$ and $C_2$ in Figure \ref{Fig:2-point}. 
For the case of $N=3$, the boundary components 
${\cal N}_{3,\epsilon}^{(1)}$, ${\cal N}_{3,\epsilon}^{(2)}$ and 
${\cal N}_{3,\epsilon}^{(3)}$ are respectively the parallelograms 
$AA'B'B$, $CC'A'A$ and $BB'C'C$ 
in Figure \ref{Fig:3-point}. 
The integration in (\ref{Virasoro test 1}) becomes 
the sum of the integrations over the components 
${\cal N}_{N,\epsilon}^{(\alpha)}$ of 
$\partial {\cal M}_{N,\epsilon}$. 
A typical configuration of the $N$ points on the boundary component 
${\cal N}_{N,\epsilon}^{(\alpha)}$ 
is depicted in Figure \ref{Fig:boundarydisc}. 
On this component, two gluons marked by $\alpha$ and 
$\alpha+1$ in the integral (\ref{Virasoro test 1}) 
are close to each other. 
They are thought to be in the open-string tree channel 
and their propagating modes are described by open-string 
vertex operators. Therefore the integral over this component 
becomes effectively the integral of the other gluons and 
the open-string vertex operator which describes the 
lightest propagating mode of the above two gluons. 
This indicates that boundary states with $N-1$ legs 
are brought about from the $N$-gluon boundary state 
by the action of the superconformal generators. 
This happens also for the action of the closed-string 
BRST operator and becomes the origin of the non-linearity. 
The configuration space ${\cal M}_{N,\epsilon}$ has corners 
when $N \geq 3$. The boundary components 
${\cal N}_{N,\epsilon}^{(\alpha)}$ are patched together 
along the corners. For instance, 
${\cal N}_{3,\epsilon}^{(\alpha)}$ are patched together 
along the lines $\overline{AA'}$,$\overline{BB'}$ and 
$\overline{CC'}$ in Figure \ref{Fig:3-point}. 
On the tubular neighbourhoods of the corners, 
more than two gluons are thought to be in the open-string 
channel. Suitable modifications of the above argument 
are required there.  

It becomes important to study the boundary states 
which are brought about effectively from the boundary 
integrals such as (\ref{Virasoro test 1}). 
In particular, we need to know the open-string vertex 
operator which describes the lightest propagating mode 
of the two gluons. This can be read from the expansion 
of the product of the $0$ picture gluon vertex operators. 
By using the OPE (\ref{OPE}) we obtain 
\begin{eqnarray}
&&
\widehat{V}^{(0)}_{A,\epsilon}(\sigma;k_1)
\widehat{V}^{(0)}_{A,\epsilon}(\sigma+\epsilon;k_2)
\nonumber \\
&&
=
2\alpha'
\epsilon^{\alpha'(k_1+k_2)^2-2} 
(2\alpha'
k_1
\cdot 
k_2-1)
e^{-\frac{i}{2}k_1\times k_2}
A(k_1)
\cdot 
A(k_2)
\widehat{V}_{tac}^{(-1)}(\sigma;k_1+k_2)
\nonumber \\
&&
~~~~
+
{\cal O}(\epsilon^{\alpha'(k_1+k_2)^2-1})~.
\label{OPE between 0 gluon vertices}
\end{eqnarray}
The operator $\widehat{V}_{tac}^{(-1)}$ 
in (\ref{OPE between 0 gluon vertices}) is \textit{not} 
the vertex operator of the open-string tachyon 
which is excluded by the GSO projection.
It is rather the vertex operator of an \textit{unphysical} 
open-string particle with 
$\alpha'(\mathrm{mass})^2 = -1$.
The equation (\ref{OPE between 0 gluon vertices}) says that 
the unphysical open-string particles appear as 
the dominant contribution in the intermediate open-string channels.
This means that the boundary integral (\ref{Virasoro test 1}) 
yields the boundary states with less legs 
which include the unphysical open-string particles.

However, the multi-gluon boundary states of the Wilson loop 
include states other than (\ref{test 1}). These states are brought about by  
the supersymmetric path-ordering. 
They are induced from the $\delta$-functions 
$-i\theta_{\alpha}\theta_{\alpha+1}
\delta(\sigma_{\alpha+1}-\sigma_{\alpha}-\epsilon)$ 
of the supersymmetric step functions  
and give the contact interactions between 
the $-1$ picture gluon vertex operators. 
For instance, the following states appear in the $N$-gluon 
boundary state:  
\begin{eqnarray}
&&
\frac{i^{(N-1)}}{N}C
\int
\!\prod_{\beta=1}^N 
\frac{d^{p+1}k^{(\beta)}}{(2\pi)^{\frac{p+1}{2}}}
~g_{\eta}
\int_{{\cal N}_{N,\epsilon}^{(\alpha)}}
\!
d\sigma_1 \wedge \cdots 
\!\!\!\stackrel{\alpha+1}{\stackrel{\vee}{\cdot}}\!\!\! 
\cdots \wedge d\sigma_N 
\widehat{V}_{A,\epsilon}^{(-1)}(\sigma_{\alpha};k^{(\alpha)})
\widehat{V}_{A,\epsilon}^{(-1)}(\sigma_{\alpha}+\epsilon;k^{(\alpha+1)})
\nonumber \\*
&&
~~~~~~~~~~~~~~~~~~~~~~
\times 
\prod_{\beta \neq \alpha,\alpha+1}^N 
\widehat{V}_{A,\epsilon}^{(0)}
(\sigma_{\beta};k^{(\beta)})
|x_0^i\rangle_m 
\otimes 
|B \rangle_{gh}
\otimes 
|B;\eta \rangle_{sgh}~,   
\label{test 2}
\end{eqnarray}
for $1 \leq \alpha \leq N$.

The action of the closed-string BRST operator on 
the multi-gluon boundary states are obtained 
by piecing together all the actions on their elements. 
The contact interactions of the $-1$ picture operators 
turn out to play important roles under this incorporation. 
For instance, 
the Virasoro action on the contact interaction of the state 
(\ref{test 2}) modifies the previous description 
of the nearby two gluons 
on the boundary component ${\cal N}_{N,\epsilon}^{(\alpha)}$. 
We will see in section \ref{section5} 
that these two gluons are described by the following 
combination of the gluon vertex operators: 
\begin{eqnarray}
&&
\widehat{V}_{A,\epsilon}^{(0)}(\sigma;k_1)
\widehat{V}_{A,\epsilon}^{(0)}(\sigma+\epsilon;k_2)
+
\frac{i}{2}
\widehat{V}_{A,\epsilon}^{(-1)}(\sigma;k_1)
\partial_{\sigma}\widehat{V}_{A,\epsilon}^{(-1)}(\sigma+\epsilon;k_2) 
\nonumber \\*
&&
~~
-
\frac{i}{2}
\partial_{\sigma} \widehat{V}_{A,\epsilon}^{(-1)}(\sigma;k_1)
\widehat{V}_{A,\epsilon}^{(-1)}(\sigma+\epsilon;k_2)~. 
\label{two-gluon contact term test}
\end{eqnarray}
The above combination is named the contact term 
$\Delta_{A,\epsilon}^{(0)}(\sigma;k_1,k_2)$ 
in section \ref{section5}. The $-1$ picture operators 
in (\ref{two-gluon contact term test})
come from the contact interactions which are required 
by the supersymmetric path-ordering. 
The role of these operators could be stressed in comparison with 
(\ref{OPE between 0 gluon vertices}). 
The expansion into the 
power series of $\epsilon$ can be calculated by making use of 
the OPE 
(\ref{OPE}). 
It can be written in the following form: 
\begin{eqnarray}
\Delta_{A,\epsilon}^{(0)}(\sigma;k_1,k_2)
=
i\epsilon^{\alpha' (k_1+k_2)^2-1}
\Biggl\{
I_{\mu}(k_1,k_2)
\widehat{V}_{gl}^{(0)\mu}(\sigma;k_1+k_2)
+
{\cal O}(\epsilon)
\Biggr\}
~, 
\label{expansion of two-gluon contact term 0}
\end{eqnarray}
where the tensor $I_{\mu}$ is given by  
\begin{eqnarray}
&&
I_{\mu}(k_1,k_2)
\nonumber \\*
&&
=
\alpha'e^{-\frac{i}{2}k_1 \times k_2}
\Biggl\{
(k_{1\mu}-k_{2\mu})
A(k_1)
\cdot 
A(k_2)
-2k_1
\cdot 
A(k_2)A_{\mu}(k_1)
+2k_2
\cdot   
A(k_1)A_{\mu}(k_2)
\Biggr\} .
\end{eqnarray}
By comparing the expansion 
(\ref{expansion of two-gluon contact term 0}) 
with (\ref{OPE between 0 gluon vertices}), 
we see that the unphysical open-string particle, 
which appears in the open-string channel of the 
product  
$\widehat{V}_{A,\epsilon}^{(0)}
\widehat{V}_{A,\epsilon}^{(0)}$,   
is actually cancelled out 
with the leading term of the expansion 
of the $-1$ picture operators.  
The next-to-leading terms of these expansions are combined 
and form the $0$ picture gluon vertex operator 
in (\ref{expansion of two-gluon contact term 0}).   
$\mathcal{O}(\epsilon)$ term in 
(\ref{expansion of two-gluon contact term 0}) 
represents the massive modes of the open-string.
These massive modes become irrelevant at the short distance 
$\epsilon\sim 0$.

The cancellations of the unphysical open-string particle such as 
in the above turn to occur when the superconformal generators act 
on the multi-gluon boundary states. 
Owing to these cancellations the gluon boundary states 
with less legs are brought about. 
These happen also for the action of the closed-string BRST 
operator. 
This means that the actions of the closed-string BRST operator 
on the multi-gluon boundary states 
change the actions on the gluon boundary states 
with less legs effectively. 
The modifications can be interpreted as corrections to 
the (linear) BRST transformation $\delta^0_{\mathbf{B}} A^*_{\mu}$. 
The first correction $\delta^1_{\mathbf{B}} A^*_{\mu}$, 
which is quadratic with respect to $A_{\mu}$, 
turns out to be given by the tensor $I_{\mu}$. 
It can be written by using the Moyal $\star$-product
\footnote{
$
f \star g(x) \equiv 
\lim_{y \rightarrow x}
\exp \left\{\frac{i}{2}\theta^{\mu \nu}
\frac{\partial}{\partial x^{\mu}}
\frac{\partial}{\partial y^{\nu}} \right\}
f(x)g(y)
$.}
as follows:
\begin{eqnarray}
&&
\delta_{{\bf B}}^1
A^*_{\mu}(k)
=
i \alpha'
\int 
\!
\frac{d^{p+1}x}{(2\pi)^{\frac{p+1}{2}}}
\, 
e^{-ik_{\mu}x^{\mu}}
G^{\nu \rho}
\Bigl(
\partial_{\mu}A_{\nu}\star A_{\rho}
-
A_{\nu}\star\partial_{\mu}A_{\rho}
\nonumber \\*
&&
~~~~~~~~~~~~~~~~~~~~~~~~~~~~~~~~~~~~~~~~~~~~~~~
+
2A_{\nu}\star\partial_{\rho}A_{\mu}
-
2
\partial_{\nu}A_{\mu}\star A_{\rho}
\Bigr)(x)~,
\end{eqnarray}
where  
$A_{\mu}(x)=
\int\frac{d^{p+1}k}{(2\pi)^{\frac{p+1}{2}}}e^{ik\cdot x}A_{\mu}(k)$ 
is the gauge field on the non-commutative world-volume. 
The non-linear BRST transformation, 
which we call $\delta_{\mathbf{B}}$, 
is obtained by putting together these corrections; 
$\delta_{\mathbf{B}}=\sum_{L=0}^{\infty}\delta_{\mathbf{B}}^L$. 
The $L$-th correction 
$\delta_{\mathbf{B}}^L\ (L\ge 1)$ 
generates polynomials of $A_{\mu}$ with the degree equal to $L+1$. 
We will compute the first two corrections 
$\delta_{\mathbf{B}}^1$ and $\delta_{\mathbf{B}}^2$ 
in section~\ref{section5}. 
Together with $\delta_{\mathbf{B}}^0$, 
they  
become the leading term of the $\alpha'$-expansion 
of the non-linear transformation $\delta_{\mathbf{B}}$.

Let us describe the result: 
The action of the closed-string BRST operator 
on the Wilson loop has the form,  
\begin{eqnarray}
\lefteqn{
Q_c \times \textrm{SP}_\epsilon
\Biggl[\, C\,  g_{\eta} 
\exp \left\{i\oint d\sigma d\theta \>
\widehat{\mathbf{V}}_{A,\epsilon}
\bigl(\mathbf{X}(\sigma,\theta)\bigr)
\Biggr\} \right]
|x^i_0\rangle_m \otimes |B\rangle_{gh}\otimes |B;\eta\rangle_{sgh}
}
\nonumber
\\
&&=
- \textrm{SP}_{\epsilon}
\Biggl[\, C\, g_{\eta} \ 
i\oint d\sigma d\theta
\int\!\frac{d^{p+1}k}{(2\pi)^{\frac{p+1}{2}}}\>
\epsilon^{\alpha'k\cdot k}
\biggl\{\delta_{\mathbf{B}} A^*_{\mu}(k)\,
\widehat{\mathbf{V}}_{gl}^{* \mu}(\sigma,\theta;k)
+
\delta_{\mathbf{B}} \mathcal{B}(k) \,
\widehat{\mathbf{V}}_{a.g}(\sigma,\theta;k)
\biggr\}
\nonumber
\\*
&& ~~~~~~~~~~~~~~~~~~~~~
\times 
\exp \Biggl\{i\oint d\sigma d\theta \>
\widehat{\mathbf{V}}_{A,\epsilon}
\bigl(\mathbf{X}(\sigma,\theta)\bigr)
\Biggr\}
\Biggr]
|x^i_0\rangle_m \otimes |B\rangle_{gh}\otimes |B;\eta\rangle_{sgh},
\label{Qc on Wilsonloop}
\end{eqnarray}
where 
\begin{eqnarray}
\delta_{\mathbf{B}} A^*_{\mu}(k)
&=&
\alpha'\int\frac{d^{p+1}x}{(2\pi)^{\frac{p+1}{2}}}\>
e^{-ik\cdot x}
G^{\nu \rho}\nabla_{\rho}F_{\nu\mu}\Big|_{\partial\cdot A=0}(x)
+ \mathcal{O}(\alpha'^{\,2}),
\label{non-linear A*}
\\
\delta_{\mathbf{B}}\mathcal{B}(k)
&=&
\alpha' \, k\cdot A(k) 
~~~~ 
\biggl( =\delta^0_{\mathbf{B}}\mathcal{B}(k) \biggr).
\label{non-linear Lorentz}
\end{eqnarray}
Here 
$\nabla_{\mu} = \partial_{\mu}+iA_{\mu}$ 
is the covariant derivative of the non-commutative 
gauge theory and 
$F_{\mu\nu}=\partial_{\mu}A_{\nu}-\partial_{\nu}A_{\mu}
+iA_{\mu}\star A_{\nu}-iA_{\nu}\star A_{\mu}$ 
is the field strength.
Therefore the BRST invariance of the Wilson loop 
requires the conditions, 
$\delta_{\mathbf{B}} A^*_{\mu}(k)=0$ 
and 
$\delta_{\mathbf{B}} \mathcal{B}(k) = 0. $ 
These are respectively the equation of motion 
for the non-commutative gauge theory in the Lorentz gauge 
and the Lorentz gauge condition.

\section{Derivation of the Result}\label{section5}
In this section we provide a proof of the result stated in 
section \ref{section4}.  
It is convenient to use the language of the superspace formalism.   
Let $S=(\sigma,\theta)$ be the superspace coordinate. 
The supersymmetric step function is denoted by 
$\Theta(S,S')=\Theta(\sigma'-\sigma-i\theta \theta')$. 
We write the regularized multiple-ordered integrals 
in the Wilson loop 
(\ref{SWilson}) 
as follows: 
\begin{eqnarray}
\int_{{\cal I}_N(\epsilon)}
\bigl[ d^NS \bigr]=
\int_0^{2\pi}\!\! d\sigma_1 d\theta_1 
\int_{-\infty}^{+\infty}\!\! d\sigma_2 d\theta_2 
\cdots 
\int_{-\infty}^{+\infty}\!\! d\sigma_N d\theta_N  
\prod_{\alpha=1}^N
\Theta(S_{\alpha}+\epsilon,S_{\alpha+1})~,
\label{super multiple-ordered integral}
\end{eqnarray}
where 
$\bigl[ d^NS \bigr]=dS_1 \cdots dS_N$  
with the $(1,1)$-superforms  
$dS_{\alpha}=d\sigma_{\alpha}d\theta_{\alpha}$. 
In the above we have introduced the supermoduli 
${\cal I}_{N}(\epsilon)$ to represent the integration region of 
the RHS. We express the multi-gluon boundary states 
of the Wilson loop in the following form: 
\begin{eqnarray}
\frac{i^{N^2}}{N}C 
\int_{{\cal I}_{N}(\epsilon)}
\bigl[ d^NS \bigr]
~g_{\eta} 
\widehat{{\bf V}}_{A,\epsilon}({\bf X}(S_1))
\cdots 
\widehat{{\bf V}}_{A,\epsilon}({\bf X}(S_N)) 
|x_0^i \rangle_m \otimes 
|B \rangle_{gh} \otimes 
|B;\eta \rangle_{sgh}~.
\label{N-gluon BS}
\end{eqnarray}
The action of the closed-string BRST operator on the Wilson loop 
is the collection of the actions on the multi-gluon boundary states. 
These actions can be written as 
\begin{eqnarray}
&&
Q_c \times 
\frac{i^{N^2}}{N}C 
\int_{{\cal I}_{N}(\epsilon)}
\bigl[ d^NS \bigr]
~g_{\eta} 
\widehat{{\bf V}}_{A,\epsilon}({\bf X}(S_1))
\cdots 
\widehat{{\bf V}}_{A,\epsilon}({\bf X}(S_N)) 
|x_0^i \rangle_m \otimes 
|B \rangle_{gh} \otimes 
|B;\eta \rangle_{sgh} 
\nonumber \\
&&
=
\sum_{\alpha=1}^N
\pm
\frac{i^{N^2}}{N}C 
\int_{{\cal I}_{N}(\epsilon)}
\bigl[ d^NS \bigr]
~g_{\eta} 
\widehat{{\bf V}}_{A,\epsilon}({\bf X}(S_1))
\cdots
\delta
\widehat{{\bf V}}_{A,\epsilon}({\bf X}(S_{\alpha}))
\cdots
\widehat{{\bf V}}_{A,\epsilon}({\bf X}(S_N))
\nonumber \\*
&&
~~~~~~~~~~~~~~~~~~~~~~~~~~~~~~~~~~~~~~~~~
\times 
|x_0^i \rangle_m \otimes 
|B \rangle_{gh} \otimes 
|B;\eta \rangle_{sgh}~, 
\label{Qc N-gluon BS 1}
\end{eqnarray}
where 
$\delta \widehat{{\bf V}}_{A,\epsilon}({\bf X}(S))
=\int \frac{d^{p+1}k}{(2\pi)^{\frac{p+1}{2}}}
\delta \widehat{{\bf V}}_{A,\epsilon}(S;k)$ 
is the BRST transformation of the gluon vertex operator. 
It follows from (\ref{reduce}) as 
\begin{eqnarray}
&&
\delta
\widehat{\bf{V}}_{A,\epsilon}(S;k)
\nonumber \\*
&&
=
-\frac{1}{2}\sum_{n \in Z}
(c_{-n}-\bar{c}_{n})
\left[ {\cal L}_n, \widehat{{\bf V}}_{A,\epsilon}(S;k) \right]
+\frac{1}{4}
\sum_{r \in Z+\frac{1}{2}}
(\gamma_{-r}-i\eta \bar{\gamma}_r)
\left\{ {\cal G}_r, \widehat{{\bf V}}_{A,\epsilon}(S;k) \right\}~, 
\label{BRS gluon vertex}
\end{eqnarray}
where the commutation relations with the superconformal generators 
are given in 
(\ref{v0})-(\ref{sv1}). 
The actions (\ref{Qc N-gluon BS 1}) are obtainable  
from the actions of the superconformal generators 
by taking account of the ghost $c, \bar{c}$ 
and the superghosts $\gamma, \bar{\gamma}$ as in (\ref{BRST2}).

Our computation is perturbative with respect to the numbers of 
gluons in the Wilson loop and  
the results up to $N=3$ will be described in the below. 
These turn out to be enough 
to give the BRST invariance condition for  
the Wilson loop at the $\alpha'$-order.  

\subsection{Contribution from a single gluon}
For the single gluon boundary state, 
the actions of the superconformal generators 
and the closed-string BRST operator are computed already 
in section \ref{section2}. We summarize them here. 
The action of the superconformal generators is as follows: 
\begin{eqnarray}
&&
(L^m_{n}-\bar{L}_{-n}^m)\times 
iC 
\int_{{\cal I}_1}dS 
~g_{\eta}
\widehat{{\bf V}}_{A,\epsilon}({\bf X}(S))
|x_0^i\rangle_m 
\otimes 
|B \rangle_{gh} 
\otimes 
|B;\eta \rangle_{sgh}
\nonumber \\
&&
~
=-iC 
\int \! 
\frac{d^{p+1}k}{(2\pi)^{\frac{p+1}{2}}}
~g_{\eta}
\int_{0}^{2\pi}
\!\!\! d\sigma
~ne^{in\sigma}\epsilon^{\alpha' k \cdot k}
\left\{
\delta^0_{{\bf B}}A^*_{\mu}(k)
\widehat{V}^{(0)\mu}_{gl}(\sigma;k)
-
n\delta^0_{{\bf B}}{\cal B}(k) 
\widehat{V}^{(-1)}_{tac}(\sigma;k)
\right\}
\nonumber \\*
&&
~~~~~~~~~~~~~~~~~~~~~~~~~~~~~~~~~~~~~
\times
|x_0^i\rangle_m 
\otimes 
|B \rangle_{gh} 
\otimes 
|B;\eta \rangle_{sgh}~, 
\label{one-gluon Virasoro 1}
\\
&&
(G_{r}^m+i\eta \bar{G}_{-r}^m )\times 
iC 
\int_{{\cal I}_1}
dS 
~g_{\eta}
\widehat{{\bf V}}_{A,\epsilon}({\bf X}(S))
|x_0^i\rangle_m 
\otimes 
|B \rangle_{gh} 
\otimes 
|B;\eta \rangle_{sgh}
\nonumber \\
&&
~
=-iC 
\int \! 
\frac{d^{p+1}k}{(2\pi)^{\frac{p+1}{2}}}
~g_{\eta}
\int_{0}^{2\pi}
\!\!\! d\sigma
~2re^{ir\sigma}\epsilon^{\alpha' k \cdot k}
\left\{
\delta^0_{{\bf B}}A^*_{\mu}(k)
\widehat{V}^{(-1)\mu}_{gl}(\sigma;k)
+
\delta^0_{{\bf B}}{\cal B}(k) 
\widehat{V}^{(0)}_{tac}(\sigma;k)
\right\}
\nonumber \\*
&&
~~~~~~~~~~~~~~~~~~~~~~~~~~~~~~~~~~~~~
\times 
|x_0^i\rangle_m 
\otimes 
|B \rangle_{gh} 
\otimes 
|B;\eta \rangle_{sgh}~,
\label{one-gluon supercurrent 1}
\end{eqnarray}
where $\delta_{{\bf B}}^0 A_{\mu}^*(k)$ and 
$\delta^0_{{\bf B}} {\cal B}(k)$ 
are the linearized transformations (\ref{linear}). 
The above actions are combined and yield 
the following action of the closed-string BRST operator: 
\begin{eqnarray}
&&
Q_c \times 
iC
\int_{{\cal I}_1}
\!\! dS 
~g_{\eta}
\widehat{{\bf V}}_{A,\epsilon}({\bf X}(S))
|x_0^i\rangle_m 
\otimes 
|B \rangle_{gh}
\otimes 
|B;\eta \rangle_{sgh}
\nonumber \\
&&
~
=-iC
\int_{{\cal I}_1}
\!\! dS 
~g_{\eta}
\int \frac{d^{p+1}k}{(2\pi)^{\frac{p+1}{2}}}
\epsilon^{\alpha'k\cdot k}
\left\{
\delta_{{\bf B}}^0
A_{\mu}^*(k)
\widehat{{\bf V}}_{gl}^{* \mu}(S;k)
+
\delta^0_{{\bf B}}
{\cal B}(k)
\widehat{{\bf V}}_{a.g}(S;k)
\right\}
\nonumber \\*
&&
~~~~~~~~~~~~~~~~~~~~~~~~~~~~~~~~~~~~~
\times 
|x_0^i\rangle_m 
\otimes 
|B \rangle_{gh}
\otimes 
|B;\eta \rangle_{sgh}~.
\label{Qc on one-gluon BS}
\end{eqnarray}

\subsection{Contribution from two gluons}
We examine the two-gluon boundary state of the Wilson loop 
(\ref{SWilson}). 
The integration of 
the anti-commuting coordinates $\theta_1,\theta_2$ gives 
\begin{eqnarray}
&&
\frac{i^4}{2}
C \int_{{\cal I}_2(\epsilon)}
\bigl[ d^2S \bigr] 
~g_{\eta}
\widehat{{\bf V}}_{A,\epsilon}
({\bf X}(S_1))
\widehat{{\bf V}}_{A,\epsilon}
({\bf X}(S_2))~
|x_0^i\rangle_m 
\otimes 
|B \rangle_{gh}
\otimes 
|B;\eta \rangle_{sgh}
\nonumber \\
&&
=\frac{i^2}{2}C
\int 
\!
\prod_{r=1,2}
\frac{d^{p+1}k^{(r)}}{(2\pi)^{\frac{p+1}{2}}}
~g_{\eta}
\int_0^{2\pi}
\!\! 
d\sigma_1
\widehat{V}^{(0)}_{A,\epsilon}(\sigma_1;k^{(1)}) 
\int_{\sigma_1+\epsilon}^{\sigma_1+2\pi-\epsilon}
\!\!\!
d\sigma_2 
\widehat{V}^{(0)}_{A,\epsilon}(\sigma_2;k^{(2)}) 
\nonumber \\*
&&
~~~~~~~~~~~~
\times 
|x_0^i \rangle_m \otimes 
|B \rangle_{gh} \otimes 
|B;\eta \rangle_{sgh}
\nonumber \\
&&
~~
+\frac{i}{2}C
\int 
\!
\prod_{r=1,2}
\frac{d^{p+1}k^{(r)}}{(2\pi)^{\frac{p+1}{2}}}
~g_{\eta}
\int_0^{2\pi}
\!\!
d\sigma 
\widehat{V}^{(-1)}_{A,\epsilon}(\sigma;k^{(1)}) 
\widehat{V}^{(-1)}_{A,\epsilon}(\sigma+\epsilon;k^{(2)})
\nonumber \\*
&&
~~~~~~~~~~~~
\times  
|x_0^i \rangle_m \otimes 
|B \rangle_{gh} \otimes 
|B;\eta \rangle_{sgh}
\nonumber \\
&&
~~+
\frac{i}{2}C
\int 
\!
\prod_{r=1,2}
\frac{d^{p+1}k^{(r)}}{(2\pi)^{\frac{p+1}{2}}}
~g_{\eta}
\int_0^{2\pi}
\!\!
d\sigma 
\widehat{V}^{(-1)}_{A,\epsilon}(\sigma-\epsilon;k^{(2)}) 
\widehat{V}^{(-1)}_{A,\epsilon}(\sigma;k^{(1)})
\nonumber \\*
&&
~~~~~~~~~~~~
\times 
|x_0^i \rangle_m \otimes 
|B \rangle_{gh} \otimes 
|B;\eta \rangle_{sgh}~.
\label{two-gluon BS}
\end{eqnarray} 
Here the first term is the state (\ref{test 1}) with $N=2$. 
The moduli space ${\cal M}_{2,\epsilon}$ is denoted 
in Figure \ref{Fig:2-point}. 
The supersymmetric step functions in the regularized 
path-ordering give rise to 
the $\delta$-functions, 
$\delta(\sigma_{2}-\sigma_1-\epsilon)$
$=i\int d\theta_2 d\theta_1
\Theta(\sigma_{2}-\sigma_{1}-i\theta_{1}\theta_{2}-\epsilon)$ 
and 
$\delta(\sigma_{1}+2\pi-\sigma_2-\epsilon)=$
$i\int d\theta_2 d\theta_1 
\Theta(\sigma_{1}+2\pi-\sigma_{2}+i\theta_{2}\theta_{1}-\epsilon)$. 
These two $\delta$-functions provide respectively 
the contact interactions of the $-1$ picture operators in 
the second and the third terms of (\ref{two-gluon BS}).

\subsubsection{Action of the superconformal generators}
Let us start with the action of the superconformal generators. 
The computations are made by using the commutation relations 
(\ref{v0})-(\ref{sv1}). It is convenient to rewrite those relations 
by means of $\delta^0_{{\bf B}}A^*_{\mu}$ 
and $\delta^0_{{\bf B}}{\cal B}$: 
\begin{eqnarray}
\Bigl[ 
{\cal L}_n,~ \widehat{V}_{A,\epsilon}^{(0)}(\sigma;k)
\Bigr]
&=&
-i\frac{d}{d\sigma}
\left\{ 
e^{in \sigma}\widehat{V}_{A,\epsilon}^{(0)}(\sigma;k) 
\right\}
\nonumber \\
&&
-ne^{in\sigma}\epsilon^{\alpha' k \cdot k}
\left\{
\delta^0_{{\bf B}}A^*_{\mu}(k)
\widehat{V}^{(0)\mu}_{gl}(\sigma;k)
-
n\delta^0_{{\bf B}}{\cal B}(k) 
\widehat{V}^{(-1)}_{tac}(\sigma;k)
\right\},
\label{CCR 1'} \\
\Bigl[ 
{\cal G}_r,~ \widehat{V}_{A,\epsilon}^{(0)}(\sigma;k) 
\Bigr]
&=&
-i\frac{d}{d\sigma}
\left\{ 
e^{ir \sigma}\widehat{V}_{A,\epsilon}^{(-1)}(\sigma;k) 
\right\}
\nonumber \\
&&
-2re^{ir\sigma}\epsilon^{\alpha' k \cdot k}
\left\{
\delta^0_{{\bf B}}A^*_{\mu}(k)
\widehat{V}^{(-1)\mu}_{gl}(\sigma;k)
+
\delta^0_{{\bf B}}{\cal B}(k) 
\widehat{V}^{(0)}_{tac}(\sigma;k)
\right\}~,
\label{CCR 2'} \\
\Bigl[ 
{\cal L}_n,~ \widehat{V}_{A,\epsilon}^{(-1)}(\sigma;k) 
\Bigr]
&=&
-ie^{in\sigma}\left(
\frac{d}{d\sigma}-\frac{in}{2}
\right)
e^{in \sigma}\widehat{V}_{A,\epsilon}^{(-1)}(\sigma;k) )
\nonumber \\
&&
-ne^{in\sigma}\epsilon^{\alpha' k \cdot k}
\delta^0_{{\bf B}}A^*_{\mu}(k) 
\widehat{V}^{(-1)\mu}_{gl}(\sigma;k),
\label{CCR 3'} \\
\Bigl\{ 
{\cal G}_r,~ \widehat{V}_{A,\epsilon}^{(-1)}(\sigma;k) 
\Bigr\}
&=&
e^{ir \sigma} 
\widehat{V}_{A,\epsilon}^{(0)}(\sigma;k)
+2re^{ir\sigma}\epsilon^{\alpha' k\cdot k}
\delta^0_{{\bf B}}{\cal B}(k)
\widehat{V}^{(-1)}_{tac}(\sigma;k).
\label{CCR 4'}
\end{eqnarray} 
Several operators are brought about  
by these commutation relations.
Among them, the operators 
$\delta_{{\bf B}}^0A_{\mu}^*\widehat{V}_{gl}^{(-1,0)\mu}$ 
and 
$\delta_{{\bf B}}^0{\cal B}\,\widehat{V}_{tac}^{(-1,0)}$ 
will be kept intact under the computations 
because they are always combined and yield  
the operators 
$\delta_{{\bf B}}^0
A_{\mu}^*(k)
\widehat{{\bf V}}_{gl}^{* \mu}
+
\delta^0_{{\bf B}}
{\cal B}(k)
\widehat{{\bf V}}_{a.g}$ 
in the actions of the closed-string BRST operator. 
So we may not write down their contributions 
explicitly in the bellow.   
To denote such omissions 
we use the symbol ``$\sim$''. 
For instance, the commutator (\ref{CCR 1'}) will be expressed 
as $\Bigl[{\cal L}_n,\widehat{V}_{A,\epsilon}^{(0)}\Bigr] 
\sim -i\partial_{\sigma}\bigl\{e^{in\sigma} 
\widehat{V}_{A,\epsilon}^{(0)}\bigr\}$. 

~

\noindent
\underline{Action of the Virasoro generators}:~
We first compute the Virasoro action 
by using the commutation relations 
(\ref{CCR 1'}) and (\ref{CCR 3'}). 
Let us concentrate on the effects of the operators  
$-i\partial_{\sigma}\bigl\{e^{in\sigma} 
V_{A,\epsilon}^{(0)}\bigr\}$ 
and 
$-ie^{in\sigma}(\partial_{\sigma} - \frac{in}{2})
\widehat{V}_{A,\epsilon}^{(-1)}$ 
which are brought about by these commutation relations.  
The first term of (\ref{two-gluon BS}) involves the double 
integral of the $0$ picture operators 
$\frac{i^2}{2}\int_{{\cal M}_{2,\epsilon}}
\widehat{V}_{A,\epsilon}^{(0)}
\widehat{V}_{A,\epsilon}^{(0)}$. 
When the Virasoro generator acts on this term, 
we need to evaluate the integral 
$\Bigl[ {\cal L}_{n},
~\frac{i^2}{2}\int_{{\cal M}_{2,\epsilon}}
\widehat{V}_{A,\epsilon}^{(0)}
\widehat{V}_{A,\epsilon}^{(0)}\Bigr]$. 
By using the commutation relation (\ref{CCR 1'}) 
this becomes  
\begin{eqnarray}
&&
\Bigl[ {\cal L}_{n},
~\frac{i^2}{2}\int_{{\cal M}_{2,\epsilon}}
d\sigma_1 \wedge d\sigma_2
\widehat{V}_{A,\epsilon}^{(0)}(\sigma_1;k^{(1)})
\widehat{V}_{A,\epsilon}^{(0)}(\sigma_2;k^{(2)})
\Bigr]
\nonumber \\
&&
=
\frac{i}{2}
\int_{{\cal M}_{2,\epsilon}}
d \Bigl\{
(-d\sigma_1e^{in\sigma_2}+d\sigma_2e^{in\sigma_1})
\widehat{V}^{(0)}_{A,\epsilon}(\sigma_1;k^{(1)}) 
\widehat{V}^{(0)}_{A,\epsilon}(\sigma_2;k^{(2)}) 
\Bigr\}
\nonumber \\
&&
~~
+ \cdots, 
\label{two-gluon integral 1}
\end{eqnarray}
where the terms proportional to 
$\delta_{{\bf B}}^0A_{\mu}^*$ and $\delta_{{\bf B}}^0{\cal B}$ 
are omitted. 
The above integration  
reduces to the integration over 
$\partial {\cal M}_{2,\epsilon}=C_1+C_2$ (Figure \ref{Fig:2-point}) 
by the Stokes theorem: 
\begin{eqnarray}
&&
\frac{i}{2}
\int_{{\cal M}_{2,\epsilon}}
d \Bigl\{
(-d\sigma_1e^{in\sigma_2}+d\sigma_2e^{in\sigma_1})
\widehat{V}^{(0)}_{A,\epsilon}(\sigma_1;k^{(1)}) 
\widehat{V}^{(0)}_{A,\epsilon}(\sigma_2;k^{(2)}) 
\Bigr\}
\nonumber \\
&&
= 
\frac{i}{2}
\int_{C_1+C_2}
(-d\sigma_1e^{in\sigma_2}+d\sigma_2e^{in\sigma_1})
\widehat{V}^{(0)}_{A,\epsilon}(\sigma_1;k^{(1)}) 
\widehat{V}^{(0)}_{A,\epsilon}(\sigma_2;k^{(2)}) 
~. 
\label{two-gluon integral 2}
\end{eqnarray}
The integration along $C_1$ in (\ref{two-gluon integral 2}) 
becomes 
\begin{eqnarray}
\frac{i}{2}
\int_{0}^{2\pi}
d\sigma
e^{in\sigma}(1-e^{in\epsilon})
\widehat{V}^{(0)}_{A,\epsilon}(\sigma;k^{(1)}) 
\widehat{V}^{(0)}_{A,\epsilon}(\sigma+\epsilon;k^{(2)})~,   
\label{integral C1}
\end{eqnarray}
while the integration along $C_2$ is expressed as 
\begin{eqnarray}
\frac{i}{2}
\int_{0}^{2\pi}
d\sigma
e^{in(\sigma-\epsilon)}(1-e^{in\epsilon})
\widehat{V}^{(0)}_{A,\epsilon}(\sigma-\epsilon;k^{(2)}) 
\widehat{V}^{(0)}_{A,\epsilon}(\sigma;k^{(1)}) ~~~.  
\label{integral C2}
\end{eqnarray}

We move on to the second and the third terms of (\ref{two-gluon BS}). 
They involve the contact 
interactions of the $-1$ picture vertex operators.  
When the Virasoro algebra acts on these terms, 
we need to evaluate the single integrals  
$\Bigl[ {\cal L}_{n},
~\frac{i}{2}\int
\widehat{V}_{A,\epsilon}^{(-1)}
\widehat{V}_{A,\epsilon}^{(-1)}\Bigr]$.  
Let us consider the case of the second term. 
Taking account of the commutation relation 
$\biggl[{\cal L}_n,~\widehat{V}_{A,\epsilon}^{(-1)} \biggr]
\sim -ie^{in\sigma}(\partial_{\sigma} - \frac{in}{2})
\widehat{V}_{A,\epsilon}^{(-1)}$, 
it becomes     
\begin{eqnarray}
&&
\Bigl[ {\cal L}_{n},
~\frac{i}{2}\int_{0}^{2\pi}d\sigma 
\widehat{V}_{A,\epsilon}^{(-1)}(\sigma;k^{(1)})
\widehat{V}_{A,\epsilon}^{(-1)}(\sigma+\epsilon;k^{(2)})\Bigr]
\nonumber \\  
&&
=
\frac{1}{2}
\int_{0}^{2\pi}
\!\!
d\sigma 
\left\{
e^{in\sigma}
\left(
\frac{d}{d\sigma}
-\frac{in}{2}
\right)
\widehat{V}_{A,\epsilon}^{(-1)}(\sigma;k^{(1)})
\widehat{V}_{A,\epsilon}^{(-1)}(\sigma+\epsilon;k^{(2)})
\right.
\nonumber \\
&&
~~~~~~~~~~~~~~~
\left.
+
e^{in(\sigma+\epsilon)}
\widehat{V}_{A,\epsilon}^{(-1)}(\sigma;k^{(1)})
\left(
\frac{d}{d\sigma}
-\frac{in}{2}
\right)
\widehat{V}_{A,\epsilon}^{(-1)}(\sigma+\epsilon;k^{(2)})
\right\}
\nonumber \\
&&
~
+ \cdots~.  
\label{two-gluon integral 3}
\end{eqnarray}
The above single integral can be evaluated by the partial-integrations 
as follows: 
\begin{eqnarray}
&&
\frac{1}{2}
\int_{0}^{2\pi}
\!\!
d\sigma 
\left\{
e^{in\sigma}
\left(
\frac{d}{d\sigma}
-\frac{in}{2}
\right)
\widehat{V}_{A,\epsilon}^{(-1)}(\sigma;k^{(1)})
\widehat{V}_{A,\epsilon}^{(-1)}(\sigma+\epsilon;k^{(2)})
\right.
\nonumber \\
&&
~~~~~~~~~~~~~~~
\left.
+
e^{in(\sigma+\epsilon)}
\widehat{V}_{A,\epsilon}^{(-1)}(\sigma;k^{(1)})
\left(
\frac{d}{d\sigma}
-\frac{in}{2}
\right)
\widehat{V}_{A,\epsilon}^{(-1)}(\sigma+\epsilon;k^{(2)})
\right\}
\nonumber \\
&&
~~
=
\frac{1}{2}
\int_{0}^{2\pi}
\!\!\!
d\sigma 
e^{in\sigma}(1-e^{in\epsilon})
\frac{d\widehat{V}_{A,\epsilon}^{(-1)}(\sigma;k^{(1)})}{d\sigma}
\widehat{V}_{A,\epsilon}^{(-1)}(\sigma+\epsilon;k^{(2)})
\nonumber \\
&&
~~~~~~~~~~~~~
+
\frac{1}{4}
\int_{0}^{2\pi}
\!\!\!
d\sigma
\frac{de^{in\sigma}}{d\sigma}
(1-e^{in\epsilon})
\widehat{V}_{A,\epsilon}^{(-1)}(\sigma;k^{(1)})
\widehat{V}_{A,\epsilon}^{(-1)}(\sigma+\epsilon;k^{(2)})
\nonumber \\
&&
~~
=
\frac{1}{4}
\int_{0}^{2\pi}
\!\!\!d\sigma
e^{in\sigma}(1-e^{in\epsilon})
\nonumber \\*
&&
~~~~
\times 
\left\{
\frac{d\widehat{V}_{A,\epsilon}^{(-1)}(\sigma;k^{(1)})}{d\sigma}
\widehat{V}_{A,\epsilon}^{(-1)}(\sigma+\epsilon;k^{(2)})
-
\widehat{V}_{A,\epsilon}^{(-1)}(\sigma;k^{(1)})
\frac{d\widehat{V}_{A,\epsilon}^{(-1)}(\sigma+\epsilon;k^{(2)})}{d\sigma}
\right\}~. 
\label{two-gluon integral 4}
\end{eqnarray}
The integral analogous to (\ref{two-gluon integral 3}) 
appears from the third term of (\ref{two-gluon BS}). 
This can be computed by following the same step as above.  
Correspondingly to (\ref{two-gluon integral 4}), 
we obtain the following single integral 
from the third term of (\ref{two-gluon BS}):  
\begin{eqnarray}
&&
\frac{1}{4}
\int_{0}^{2\pi}
\!\!\!d\sigma
e^{in(\sigma -\epsilon)}(1-e^{in\epsilon})
\nonumber \\*
&&
~~~
\times 
\left\{
\frac{d\widehat{V}_{A,\epsilon}^{(-1)}(\sigma-\epsilon;k^{(2)})}{d\sigma}
\widehat{V}_{A,\epsilon}^{(-1)}(\sigma;k^{(1)})
-
\widehat{V}_{A,\epsilon}^{(-1)}(\sigma-\epsilon;k^{(2)})
\frac{d\widehat{V}_{A,\epsilon}^{(-1)}(\sigma;k^{(1)})}{d\sigma}
\right\}~. 
\label{two-gluon integral 5}
\end{eqnarray}

In order to describe the collection of the integrals 
(\ref{integral C1}), (\ref{integral C2}), (\ref{two-gluon integral 4}) 
and (\ref{two-gluon integral 5}), 
we introduce the contact term 
$\Delta_{A,\epsilon}^{(0)}(\sigma)$ by  
\begin{eqnarray}
\Delta_{A,\epsilon}^{(0)}(\sigma;k^{(1)},k^{(2)}) 
&=&
\widehat{V}_{A,\epsilon}^{(0)}(\sigma;k^{(1)})
\widehat{V}_{A,\epsilon}^{(0)}(\sigma+\epsilon;k^{(2)})
+
\frac{i}{2}
\widehat{V}_{A,\epsilon}^{(-1)}(\sigma;k^{(1)})
\frac{d \widehat{V}_{A,\epsilon}^{(-1)}(\sigma+\epsilon;k^{(2)})}
{d\sigma}
\nonumber \\*
&&
-
\frac{i}{2}
\frac{d\widehat{V}_{A,\epsilon}^{(-1)}(\sigma;k^{(1)})}
{d\sigma}
\widehat{V}_{A,\epsilon}^{(-1)}(\sigma+\epsilon;k^{(2)})~.
\label{two-gluon contact term 1}
\end{eqnarray}
The integrals (\ref{integral C1}) 
and (\ref{two-gluon integral 4}) are added up to 
the form 
$\frac{i}{2}\int e^{in\sigma}(1-e^{in\epsilon})
\Delta^{(0)}_{A,\epsilon}$. 
It becomes similar for the other two integrals. 
The $-1$ picture operators in the above contact term   
come from the contact interactions of the two-gluon state 
which are required 
by the supersymmetric path-ordering. 
The role of these $-1$ picture operators 
should be stressed in comparison with 
(\ref{OPE between 0 gluon vertices}). 
The expansion into the 
power series of $\epsilon$ can be calculated by making use of 
the OPE 
(\ref{OPE}). 
It can be written in the following form: 
\begin{eqnarray}
&&
\Delta_{A,\epsilon}^{(0)}(\sigma;k^{(1)},k^{(2)})
\nonumber \\*
&&
~~
=
i\epsilon^{\alpha' (k^{(1)}+k^{(2)})^2-1}
I_{\mu}(k^{(1)},k^{(2)})
\widehat{V}_{gl}^{(0)\mu}(\sigma;k^{(1)}+k^{(2)})
+
{\cal O}(\epsilon^{\alpha' (k^{(1)}+k^{(2)})^2})
~. 
\label{expansion of two-gluon contact term 1}
\end{eqnarray}
The tensor $I_{\mu}$ in the above 
is given by  
\begin{eqnarray}
&&
I_{\mu}(k^{(1)},k^{(2)})
\nonumber \\*
&&
=
\alpha'e^{-\frac{i}{2}k^{(1)} \times k^{(2)}}
\nonumber \\*
&&
~
\times
\Biggl\{
(k_{\mu}^{(1)}-k_{\mu}^{(2)})
A(k^{(1)})
\!
\cdot 
\!
A(k^{(2)})
-2k^{(1)}
\!\!
\cdot 
\!
A(k^{(2)})A_{\mu}(k^{(1)})
+2k^{(2)}
\!\!
\cdot 
\!
A(k^{(1)})A_{\mu}(k^{(2)})
\Biggr\}~.
\nonumber \\
\label{I}
\end{eqnarray}
By comparing the expansion 
(\ref{expansion of two-gluon contact term 1}) 
with (\ref{OPE between 0 gluon vertices}), 
we see that the unphysical open-string particle, 
which appears in the open-string channel of the 
product  
$\widehat{V}_{A,\epsilon}^{(0)}
\widehat{V}_{A,\epsilon}^{(0)}$,   
is actually cancelled out 
with the leading term of the expansion 
of the $-1$ picture operators.  
The next-to-leading terms of these expansions are combined 
to form the $0$ picture gluon vertex operator 
in (\ref{expansion of two-gluon contact term 1}).   
We notice that the above $I_{\mu}$ does not satisfy the condition 
$I_{\mu}(k^{(1)},k^{(2)})=$
$-I_{\mu}(k^{(2)},k^{(1)})$ 
due to the presence of the non-commutative phase factor.

Let us write down the Virasoro action on the two-gluon boundary state 
by using the contact term $\Delta_{A,\epsilon}^{(0)}$. 
From the above computations it becomes as follows: 
\begin{eqnarray}
&&
(L_n-\bar{L}_{-n})\times
\frac{i^4}{2}
C
\int_{{\cal I}_2(\epsilon)}
\bigl[ d^2S \bigr]
~g_{\eta}
\widehat{{\bf V}}_{A,\epsilon}
({\bf X}(S_1))
\widehat{{\bf V}}_{A,\epsilon}
({\bf X}(S_2))~
|x_0^i\rangle_m 
\otimes 
|B \rangle_{gh}
\otimes 
|B;\eta \rangle_{sgh}
\nonumber \\
&&
=iC
\int 
\!
\prod_{r=1,2}
\frac{d^{p+1}k^{(r)}}{(2\pi)^{\frac{p+1}{2}}}
g_{\eta}
\int_0^{2\pi}
\!\!d\sigma 
e^{in\sigma}(1-e^{in\epsilon})
\Delta_{A,\epsilon}^{(0)}(\sigma;k^{(1)},k^{(2)}) 
\nonumber \\*
&&
~~~~~~~~~~~~~~~~~~~~~~~~~
\times 
|x_0^i\rangle_m 
\otimes 
|B \rangle_{gh}
\otimes 
|B;\eta \rangle_{sgh}
\nonumber \\
&&
~~
-\frac{i^2}{2}C 
\int 
\!
\prod_{r=1,2}
\frac{d^{p+1}k^{(r)}}{(2\pi)^{\frac{p+1}{2}}}
\epsilon^{\alpha'k^{(1)}\cdot k^{(1)}}
g_{\eta}
\int_{0}^{2\pi}
\!\! 
d\sigma_1
~n e^{in\sigma_1}
\delta_{{\bf B}}^0A_{\mu}^*(k^{(1)})
\widehat{V}_{gl}^{(0)\mu}(\sigma_1;k^{(1)})
\nonumber \\*
&&
~~~~~~~~~~~~~~~~~~~~~~~~~
\times 
\int_{\sigma_1+\epsilon}^{\sigma_1+2\pi-\epsilon}
\!\!\! 
d\sigma_2 
\widehat{V}_{A,\epsilon}^{(0)}(\sigma_2;k^{(2)})
|x_0^i\rangle_m 
\otimes 
|B \rangle_{gh}
\otimes 
|B;\eta \rangle_{sgh}
\nonumber \\
&&
~~-iC 
\int 
\!
\prod_{r=1,2}
\frac{d^{p+1}k^{(r)}}{(2\pi)^{\frac{p+1}{2}}}
\epsilon^{\alpha' k^{(1)} \cdot k^{(1)}}
g_{\eta}
\int_{0}^{2\pi}
\!\!
d\sigma 
~n e^{in\sigma}
\delta_{{\bf B}}^0{\cal A}_{\mu}^*(k^{(1)})
\widehat{V}_{gl}^{(-1)}(\sigma;k^{(1)})
\nonumber \\*
&&
~~~~~~~~~~~~~~~~~~~~~~~~~
\times 
\widehat{V}_{A,\epsilon}^{(-1)\mu}(\sigma+\epsilon;k^{(2)}) 
|x_0^i\rangle_m 
\otimes 
|B \rangle_{gh}
\otimes 
|B;\eta \rangle_{sgh}
\nonumber \\
&&
~~
-\frac{i^2}{2}C 
\int 
\!
\prod_{r=1,2}
\frac{d^{p+1}k^{(r)}}{(2\pi)^{\frac{p+1}{2}}}
\epsilon^{\alpha' k^{(2)}\cdot k^{(2)}}
g_{\eta}
\int_{0}^{2\pi}
\!\! 
d\sigma_1
\widehat{V}_{A,\epsilon}^{(0)}(\sigma_1;k^{(1)})
\nonumber \\*
&&
~~~~~~~~~~~~~~~
\times
\int_{\sigma_1+\epsilon}^{\sigma_1+2\pi-\epsilon}
\!\!\! 
d\sigma_2 
~n e^{in\sigma_2}
\delta_{{\bf B}}^0A_{\mu}^*(k^{(2)})
\widehat{V}_{gl}^{(-1)\mu}(\sigma_2;k^{(2)})
|x_0^i\rangle_m 
\otimes 
|B \rangle_{gh}
\otimes 
|B;\eta \rangle_{sgh}
\nonumber \\
&&
~~
-iC 
\int 
\!
\prod_{r=1,2}
\frac{d^{p+1}k^{(r)}}{(2\pi)^{\frac{p+1}{2}}}
\epsilon^{\alpha' k^{(2)}\cdot k^{(2)}}
\int_{0}^{2\pi}
\!\!
d\sigma 
\widehat{V}_{A,\epsilon}^{(-1)}(\sigma;k^{(1)})
\nonumber \\*
&&
~~~~~~~~~~~~~~~~~~~~~~~~~
\times 
n e^{in\sigma_2}
\delta_{{\bf B}}^0A_{\mu}^*(k^{(2)})
\widehat{V}_{gl}^{(-1)\mu}(\sigma+\epsilon;k^{(2)}) 
|x_0^i\rangle_m 
\otimes 
|B \rangle_{gh}
\otimes 
|B;\eta \rangle_{sgh}
\nonumber \\ 
&&
~~
+ \cdots ~,
\label{Virasoro two-gluon boundary integral}
\end{eqnarray}
where we have omitted the terms proportional 
to $\delta_{{\bf B}}^0{\cal B}$. 

~

\noindent
\underline{Action of the supercurrent}:~
Computation of the action of the supercurrent becomes similar 
to the previous computation of the Virasoro generator. 
Among the operators in the commutation relations 
(\ref{CCR 2'}) and (\ref{CCR 4'}) we concentrate on the 
effects of the operators  
$ -i \partial_{\sigma} 
\left\{e^{ir\sigma}\widehat{V}^{(-1)}_{A,\epsilon} \right\}$
and 
$e^{ir\sigma}\widehat{V}_{A,\epsilon}^{(0)}$. 
When the supercurrent acts on the first term of the two-gluon 
state (\ref{two-gluon BS}), the integral 
$\Bigl[ {\cal G}_{r},
~\frac{i^2}{2}\int_{{\cal M}_{2,\epsilon}}
\widehat{V}_{A,\epsilon}^{(0)}
\widehat{V}_{A,\epsilon}^{(0)}\Bigr]$  need to be evaluated. 
By using the commutation relation 
$\Bigl\{ {\cal G}_r,~\widehat{V}_{A,\epsilon}^{(0)} \Bigr\}
\sim -i \partial_{\sigma} 
\left\{e^{ir\sigma}\widehat{V}^{(-1)}_{A,\epsilon} \right\}$, 
it becomes as follows:  
\begin{eqnarray}
&&
\Bigl[ {\cal G}_{r},
~\frac{i^2}{2}\int_{{\cal M}_{2,\epsilon}}
d\sigma_1 \wedge d\sigma_2 
\widehat{V}_{A,\epsilon}^{(0)}(\sigma_1;k^{(1)})
\widehat{V}_{A,\epsilon}^{(0)}(\sigma_2;k^{(2)})
\Bigr]
\nonumber \\
&&
=
\frac{i}{2}
\int_{{\cal M}_{2,\epsilon}}
d
\Biggl\{
-d\sigma_1e^{ir\sigma_2}
\widehat{V}_{A,\epsilon}^{(0)}(\sigma_1;k^{(1)})
\widehat{V}_{A,\epsilon}^{(-1)}(\sigma_2;k^{(2)})
+d\sigma_2e^{ir\sigma_1}
\widehat{V}_{A,\epsilon}^{(-1)}(\sigma_1;k^{(1)})
\widehat{V}_{A,\epsilon}^{(0)}(\sigma_2;k^{(2)})
\Biggr\} 
\nonumber \\
&&
~~
+ \cdots~, 
\end{eqnarray}
where the terms proportional to 
$\delta_{{\bf B}}^0A_{\mu}^*$ and 
$\delta_{{\bf B}}^0{\cal B}$ are omitted.  
The above integration 
reduces to the integrations over  
the boundaries $C_1$ and $C_2$ by the Stokes theorem.  
The integration along $C_1$ can be written as     
\begin{eqnarray}
\frac{i}{2}
\int_{0}^{2\pi}
d\sigma e^{ir\sigma}
\Biggl\{
\widehat{V}_{A,\epsilon}^{(-1)}(\sigma;k^{(1)})
\widehat{V}_{A,\epsilon}^{(0)}(\sigma+\epsilon;k^{(2)})
-e^{ir\epsilon}
\widehat{V}_{A,\epsilon}^{(0)}(\sigma;k^{(1)})
\widehat{V}_{A,\epsilon}^{(-1)}(\sigma+\epsilon;k^{(2)})
\Biggr\}~, 
\label{s integral C1}
\end{eqnarray}
while the integration along $C_2$ turns out to be  
\begin{eqnarray}
\frac{i}{2}
\int_{0}^{2\pi}
d\sigma e^{ir(\sigma-\epsilon)}
\Biggl\{
\widehat{V}_{A,\epsilon}^{(-1)}(\sigma-\epsilon;k^{(2)})
\widehat{V}_{A,\epsilon}^{(0)}(\sigma;k^{(1)})
-e^{ir\epsilon}
\widehat{V}_{A,\epsilon}^{(0)}(\sigma-\epsilon;k^{(2)})
\widehat{V}_{A,\epsilon}^{(-1)}(\sigma;k^{(1)})
\Biggr\}~.
\label{s integral C2}
\end{eqnarray}
From the viewpoint of the superconformal algebra 
these two integrals are the cousins to the integrals 
(\ref{integral C1}) and (\ref{integral C2}) 
since they are obtained from the same integral 
by the actions of the supercurrent and the Virasoro generator  
respectively.
We move on to the second and the third terms 
of (\ref{two-gluon BS}). 
When the supercurrent acts on these terms, 
the single integrals 
$\Bigl[ {\cal G}_{r},
~\frac{i}{2}\int
\widehat{V}_{A,\epsilon}^{(-1)}
\widehat{V}_{A,\epsilon}^{(-1)}\Bigr]$ 
need to be cared. By using the anti-commutation relation 
$\Bigl\{{\cal G}_r,~\widehat{V}_{A,\epsilon}^{(-1)}\Bigr\}
\sim e^{ir\sigma}\widehat{V}_{A,\epsilon}^{(0)}$, 
the second term of (\ref{two-gluon BS}) becomes as follows: 
\begin{eqnarray}
&&
\Bigl[ {\cal G}_{r},
~\frac{i}{2}\int_{0}^{2\pi}
\widehat{V}_{A,\epsilon}^{(-1)}(\sigma;k^{(1)})
\widehat{V}_{A,\epsilon}^{(-1)}(\sigma+\epsilon;k^{(2)})\Bigr]
\nonumber \\
&&
=
\frac{i}{2}
\int_{0}^{2\pi}
d\sigma e^{ir\sigma}
\Biggl\{
-e^{ir\epsilon}
\widehat{V}_{A,\epsilon}^{(-1)}(\sigma;k^{(1)})
\widehat{V}_{A,\epsilon}^{(0)}(\sigma+\epsilon;k^{(2)})
+
\widehat{V}_{A,\epsilon}^{(0)}(\sigma;k^{(1)})
\widehat{V}_{A,\epsilon}^{(-1)}(\sigma+\epsilon;k^{(2)})
\Biggr\} 
\nonumber \\
&&
~~ 
+ \cdots~. 
\label{s two-gluon integral 1}
\end{eqnarray}
Similarly, the third term becomes 
\begin{eqnarray}
&&
\Bigl[ {\cal G}_{r},
~\frac{i}{2}\int_{0}^{2\pi}
\widehat{V}_{A,\epsilon}^{(-1)}(\sigma-\epsilon;k^{(2)})
\widehat{V}_{A,\epsilon}^{(-1)}(\sigma;k^{(1)})\Bigr]
\nonumber \\
&&
=
\frac{i}{2}
\int_{0}^{2\pi}
d\sigma e^{ir(\sigma -\epsilon)}
\Biggl\{
-e^{ir\epsilon}
\widehat{V}_{A,\epsilon}^{(-1)}(\sigma-\epsilon;k^{(2)})
\widehat{V}_{A,\epsilon}^{(0)}(\sigma;k^{(1)})
+
\widehat{V}_{A,\epsilon}^{(0)}(\sigma-\epsilon;k^{(2)})
\widehat{V}_{A,\epsilon}^{(-1)}(\sigma;k^{(1)})
\Biggr\} 
\nonumber \\
&&
~~ 
+ \cdots~. 
\label{s two-gluon integral 2}
\end{eqnarray}
Two integrals in (\ref{s two-gluon integral 1}) 
and (\ref{s two-gluon integral 2}) 
are the cousins to the integrals 
(\ref{two-gluon integral 4}) and (\ref{two-gluon integral 5}) 
from the viewpoint of the superconformal algebra.

Apart from the terms proportional to 
$\delta_{{\bf B}}^0A_{\mu}^*$ and 
$\delta_{{\bf B}}^0{\cal B}$, 
the action of the supercurrent on the two-gluon boundary state 
is obtained by collecting the above integrals. 
In parallel with the case of the Virasoro algebra,  
the integrals (\ref{s integral C1}) and 
(\ref{s integral C2}) may be paired with 
the integrals in (\ref{s two-gluon integral 1}) 
and (\ref{s two-gluon integral 2}). 
In this way, we are led to introduce another contact term 
$\Delta_{A,\epsilon}^{(-1)}(\sigma)$ of the following form: 
\begin{eqnarray}
\Delta_{A,\epsilon}^{(-1)}(\sigma;k^{(1)},k^{(2)})
=
\widehat{V}_{A,\epsilon}^{(0)}
(\sigma;k^{(1)}) 
\widehat{V}_{A,\epsilon}^{(-1)}
(\sigma+\epsilon;k^{(2)})
+ 
\widehat{V}_{A,\epsilon}^{(-1)}
(\sigma;k^{(1)}) 
\widehat{V}_{A,\epsilon}^{(0)}
(\sigma+\epsilon;k^{(2)})~.
\label{two-gluon contact term 2}
\end{eqnarray}
The integrals 
in (\ref{s integral C1}) and (\ref{s two-gluon integral 1}) 
are added up to the form 
$i\int e^{ir\sigma}(1-e^{ir\epsilon})\Delta_{A,\epsilon}^{(-1)}$. 
It is similar for the other two integrals. 
It can be expected from the viewpoint of the superconformal algebra 
that the contact term 
$\Delta_{A,\epsilon}^{(-1)}(\sigma)$
plays a role of the superpartner of 
$\Delta_{A,\epsilon}^{(0)}(\sigma)$. 
This is confirmed by a comparison between their expansions 
into the power series of $\epsilon$. 
By making use of the OPE (\ref{OPE}) we obtain  
\begin{eqnarray}
&&
\Delta_{A,\epsilon}^{(-1)}(\sigma;k^{(1)},k^{(2)})
\nonumber \\*
&&
~~
=
2i
\epsilon^{\alpha' (k^{(1)}+k^{(2)})^2-1}
I_{\mu}(k^{(1)},k^{(2)})
\widehat{V}_{gl}^{(-1)\mu}(\sigma;k^{(1)}+k^{(2)})
+
{\cal O}(\epsilon^{\alpha' (k^{(1)}+k^{(2)})^2})
~,
\label{expansion of two-gluon contact term 2}
\end{eqnarray}
where the tensor $I_{\mu}$ is given by (\ref{I}).

The action of the supercurrent is obtained from 
the above computations. 
It can be written in the following form by using the contact 
term $\Delta_{A,\epsilon}^{(-1)}$:
\begin{eqnarray}
&&
(G_r+i\eta \bar{G}_{-r})\times
\frac{i^4}{2}
C
\int_{{\cal I}_2(\epsilon)}
\bigl[ d^2S \bigr]
~g_{\eta}
\widehat{{\bf V}}_{A,\epsilon}
({\bf X}(S_1))
\widehat{{\bf V}}_{A,\epsilon}
({\bf X}(S_2))~
|x_0^i\rangle_m 
\otimes 
|B \rangle_{gh}
\otimes 
|B;\eta \rangle_{sgh}
\nonumber \\
&&
= 
iC
\int 
\!
\prod_{r=1,2}
\frac{d^{p+1}k^{(r)}}{(2\pi)^{\frac{p+1}{2}}}
g_{\eta}
\int_0^{2\pi}
\!\!d\sigma 
e^{ir\sigma}(1-e^{ir\epsilon})
\Delta_{A,\epsilon}^{(-1)}(\sigma;k^{(1)},k^{(2)}) 
|x_0^i\rangle_m 
\otimes 
|B \rangle_{gh}
\otimes 
|B;\eta \rangle_{sgh}
\nonumber \\
&& ~~
-\frac{i^2}{2}C 
\int 
\!
\prod_{r=1,2}
\frac{d^{p+1}k^{(r)}}{(2\pi)^{\frac{p+1}{2}}}
g_{\eta}
\int_{0}^{2\pi}
\!\! 
d\sigma_1
~2r e^{ir\sigma_1}
\delta_{{\bf B}}^0A_{\mu}^*(k^{(1)})
\widehat{V}_{gl}^{(-1)\mu}(\sigma_1;k^{(1)})
\nonumber \\*
&&
~~~~~~~~~~~~
\times 
\int_{\sigma_1+\epsilon}^{\sigma_1+2\pi-\epsilon}
\!\!\! 
d\sigma_2 
\widehat{V}_{A,\epsilon}^{(0)}(\sigma_2;k^{(2)})
|x_0^i\rangle_m 
\otimes 
|B \rangle_{gh}
\otimes 
|B;\eta \rangle_{sgh}
\nonumber \\
&& ~~
-\frac{i^2}{2}C 
\int 
\!
\prod_{r=1,2}
\frac{d^{p+1}k^{(r)}}{(2\pi)^{\frac{p+1}{2}}}
g_{\eta}
\int_{0}^{2\pi}
\!\! 
d\sigma_1
\widehat{V}_{A,\epsilon}^{(0)}(\sigma_1;k^{(1)})
\nonumber \\*
&&
~~~~~~~~~~~~
\times 
\int_{\sigma_1+\epsilon}^{\sigma_1+2\pi-\epsilon}
\!\!\! 
d\sigma_2 
~2r e^{ir\sigma_2}
\delta_{{\bf B}}^0A_{\mu}^*(k^{(2)})
\widehat{V}_{gl}^{(-1)\mu}(\sigma_2;k^{(2)})
|x_0^i\rangle_m 
\otimes 
|B \rangle_{gh}
\otimes 
|B;\eta \rangle_{sgh}~,
\nonumber \\
\label{supercurrent two-gluon boundary integral}
\end{eqnarray}
where we have omitted the terms proportional to 
$\delta^0_{{\bf B}}{\cal B}$. 

\subsubsection{Action of the closed-string BRST operator}
The action of the closed-string BRST operator 
is the incorporation of the actions 
of the Virasoro generators and the supercurrents.  
In particular, 
the contact terms which appear in the actions 
(\ref{Virasoro two-gluon boundary integral}) 
and 
(\ref{supercurrent two-gluon boundary integral}) 
are combined by taking account of the ghosts $c, \bar{c}$ and 
the superghosts $\gamma, \bar{\gamma}$. 
The gluon vertex operators of their expansions 
(\ref{expansion of two-gluon contact term 1})
and 
(\ref{expansion of two-gluon contact term 2})
are incorporated into the vertex operator 
$I_{\mu}\widehat{{\bf V}}_{gl}^{*\mu}$. 
The action of the closed-string BRST operator 
on the two-gluon boundary state is obtained from the actions 
(\ref{Virasoro two-gluon boundary integral}) 
and 
(\ref{supercurrent two-gluon boundary integral})  
as follows: 
\begin{eqnarray}
&&
Q_c \times 
\frac{i^4}{2}
C
\int_{{\cal I}_2(\epsilon)}
\bigl[ d^2S \bigr]
~g_{\eta}
\widehat{{\bf V}}_{A,\epsilon}
({\bf X}(S_1))
\widehat{{\bf V}}_{A,\epsilon}
({\bf X}(S_2))~
|x_0^i\rangle_m 
\otimes 
|B \rangle_{gh}
\otimes 
|B;\eta \rangle_{sgh}
\nonumber \\
&&
=
-iC
\int_{{\cal I}_1}
\!\! dS 
~g_{\eta}
\int \frac{d^{p+1}k}{(2\pi)^{\frac{p+1}{2}}}
\epsilon^{\alpha'k\cdot k}
\Biggl\{
\delta^1_{{\bf B}}A_{\mu}^*(k)
\widehat{{\bf V}}_{gl}^{* \mu}(S;k)+{\cal O}(\epsilon)
\Biggr\}
|x_0^i\rangle_m 
\otimes 
|B \rangle_{gh}
\otimes 
|B;\eta \rangle_{sgh}
\nonumber \\
&& ~~
+
\frac{i^4}{2}
C
\int_{{\cal I}_2(\epsilon)}
\!\! 
\bigl[d^2S \bigr]
~g_{\eta}
\int \frac{d^{p+1}k}{(2\pi)^{\frac{p+1}{2}}}
\epsilon^{\alpha'k\cdot k}
\Biggl\{
\delta_{{\bf B}}^0
A_{\mu}^*(k)
\widehat{{\bf V}}_{gl}^{* \mu}(S_1;k)
+
\delta^0_{{\bf B}}
{\cal B}(k)
\widehat{{\bf V}}_{a.g}(S_1;k)
\Biggr\}
\nonumber \\*
&&
~~~~
\times 
\widehat{{\bf V}}_{A,\epsilon}
({\bf X}(S_2))~
|x_0^i\rangle_m 
\otimes 
|B \rangle_{gh}
\otimes 
|B;\eta \rangle_{sgh}
\nonumber \\
&& ~~
-
\frac{i^4}{2}
C
\int_{{\cal I}_2(\epsilon)}
\!\! \bigl[d^2S \bigr]
~g_{\eta}
\widehat{{\bf V}}_{A,\epsilon}
({\bf X}(S_1))
\nonumber \\*
&&
~~~~
\times
\int \frac{d^{p+1}k}{(2\pi)^{\frac{p+1}{2}}}
\epsilon^{\alpha'k\cdot k}
\Biggl\{
\delta_{{\bf B}}^0
A_{\mu}^*(k)
\widehat{{\bf V}}_{gl}^{* \mu}(S_2;k)
+
\delta^0_{{\bf B}}
{\cal B}(k)
\widehat{{\bf V}}_{a.g}(S_2;k)
\Biggr\}
|x_0^i\rangle_m 
\otimes 
|B \rangle_{gh}
\otimes 
|B;\eta \rangle_{sgh}~, 
\nonumber\\
&&
\label{Qc on two-gluon BS}
\end{eqnarray} 
where the first term comes from the first terms of 
(\ref{Virasoro two-gluon boundary integral}) 
and 
(\ref{supercurrent two-gluon boundary integral}). 
In the above we have plugged the expansions 
(\ref{expansion of two-gluon contact term 1})
and 
(\ref{expansion of two-gluon contact term 2}) 
into the contact terms, 
and arranged the vertex operators 
$I_{\mu}\widehat{{\bf V}}_{gl}^{* \mu}$ into the form 
$\delta_{{\bf B}}^1A_{\mu}^*\widehat{{\bf V}}_{gl}^{* \mu}$. 
Here $\delta_{{\bf B}}^1A_{\mu}^*$ is defined by 
\begin{eqnarray}
\delta_{{\bf B}}^{1}A_{\mu}^*(k)
=
-\int 
\prod_{r=1,2}
\!
\frac{d^{p+1}k^{(r)}}{(2\pi)^{\frac{p+1}{2}}}
(2\pi)^{\frac{p+1}{2}}
\delta^{p+1}(k^{(1)}+k^{(2)}-k)
I_{\mu}(k^{(1)},k^{(2)})~.
\label{1st correction of BRS}
\end{eqnarray}

We can regard 
the first term of (\ref{Qc on two-gluon BS})
as a correction to the action 
of the BRST operator on the single gluon boundary state. 
It can be compared with the following part of the action 
on the single gluon boundary state (\ref{Qc on one-gluon BS}):  
\begin{eqnarray}
-iC
\int_{{\cal I}_1}
\!\! dS 
~g_{\eta} 
\int 
\frac{d^{p+1}k}{(2\pi)^{\frac{p+1}{2}}}
\epsilon^{\alpha' k \cdot k}
\delta^0_{{\bf B}}A_{\mu}^*(k)
\widehat{{\bf V}}_{gl}^{* \mu}(S;k)
|x_0^i\rangle_m 
\otimes 
|B \rangle_{gh}
\otimes 
|B;\eta \rangle_{sgh}~. 
\label{1 for 2 Qc}
\end{eqnarray}
From the comparison with the first term 
of (\ref{Qc on two-gluon BS}) we conclude that  
the modification is achieved by 
changing the BRST transformation 
$\delta^0_{{\bf B}}A_{\mu}^*$ 
into 
$\bigl(\delta^0_{{\bf B}}+\delta^1_{{\bf B}}
\bigr)A_{\mu}^*$ 
in the action on the single gluon boundary state. 
We can argue that the action of the closed-string BRST operator 
on the single gluon boundary state of the Wilson loop 
is modified by the actions on 
the multi-gluon boundary states and that the corrections come from 
the contact terms obtained at the boundaries of 
the configuration spaces as demonstrated by (\ref{Qc on two-gluon BS}). 
The modification up to the above correction  
can be written down by using the Moyal $\star$-product: 
\begin{eqnarray}
&&
\left(
\delta_{{\bf B}}^0+\delta_{{\bf B}}^1
\right)
A^*_{\mu}(k)
=
\alpha'
\int 
\!
\frac{d^{p+1}x}{(2\pi)^{\frac{p+1}{2}}}
e^{-ik_{\mu}x^{\mu}}
G^{\nu \rho}
\Bigl(
\partial_{\nu}\partial_{\rho}A_{\mu}
+
i\partial_{\mu}A_{\nu}\star A_{\rho}
\nonumber \\
&&
~~~~~~~~~~~~~~~~~~~~~~~~~~~~~~
-
iA_{\nu}\star\partial_{\mu}A_{\rho}
+
2iA_{\nu}\star\partial_{\rho}A_{\mu}
-
2i
\partial_{\nu}A_{\mu}\star A_{\rho}
\Bigr)(x)~,
\label{1st order BRST}
\end{eqnarray}
where  
$
A_{\mu}(x)= 
\int 
\frac{d^{p+1}k}{(2\pi)^{\frac{p+1}{2}}}
e^{i k_{\mu}x^{\mu}}
A_{\mu}(k)
$ 
is the gauge field on the non-commutative world-volume. 
The combination of the gauge fields in (\ref{1st order BRST}) 
is equal to $G^{\nu \rho}\nabla_{\nu}F_{\rho \mu}$ 
in the Lorentz gauge, up to the ${\cal O}(A^2)$ terms.  
Here  $\nabla_{\mu}=\partial_{\mu}+iA_{\mu}$ 
is the covariant derivative of the non-commutative gauge theory and 
$F_{\mu \nu}=\partial_{\mu}A_{\nu}-\partial_{\nu}A_{\mu}
+iA_{\mu}\star A_{\nu}-iA_{\nu}\star A_{\mu}$ is the field strength. 

\subsection{Contribution from three gluons}
We examine the three-gluon boundary state 
of the Wilson loop (\ref{SWilson}). 
The integration of the anti-commuting coordinates 
$\theta_1,\ \theta_2$ and $\theta_3$ gives 
\begin{eqnarray}
&&
\frac{i^9}{2}C
\int_{{\cal I}_3(\epsilon)}
\bigl[ d^3S \bigr]
~g_{\eta}
\widehat{{\bf V}}_{A,\epsilon}
({\bf X}(S_1))
\widehat{{\bf V}}_{A,\epsilon}
({\bf X}(S_2))
\widehat{{\bf V}}_{A,\epsilon}
({\bf X}(S_3))~
|x_0^i\rangle_m 
\otimes 
|B \rangle_{gh}
\otimes 
|B;\eta \rangle_{sgh}
\nonumber \\
&&
=\frac{i^3}{3}C
\int 
\!
\prod_{r=1}^{3}
\frac{d^{p+1}k^{(r)}}{(2\pi)^{\frac{p+1}{2}}}
~
g_{\eta}
\int_0^{2\pi}
\!\!
d\sigma_1
\widehat{V}^{(0)}_{A,\epsilon}(\sigma_1;k^{(1)}) 
\int_{\sigma_1+\epsilon}^{\sigma_1+2\pi-2\epsilon}
\!\!\!\!
d\sigma_2
\widehat{V}^{(0)}_{A,\epsilon}(\sigma_2;k^{(2)}) 
\nonumber \\*
&&
~~~~~~~~~~~~~~~~~~~~~~~~~~~~~~
\times 
\int_{\sigma_2+\epsilon}^{\sigma_1+2\pi-\epsilon}
\!\!\!\!
d\sigma_3
%
%
\widehat{V}^{(0)}_{A,\epsilon}(\sigma_3;k^{(3)}) 
|x_0^i\rangle_m 
\otimes 
|B \rangle_{gh}
\otimes 
|B;\eta \rangle_{sgh}
\nonumber \\
&&
~~+
\frac{i^2}{3}C
\int 
\prod_{r=1}^{3}
\frac{d^{p+1}k^{(r)}}{(2\pi)^{\frac{p+1}{2}}}
~
g_{\eta}
\int_0^{2\pi}
\!\!
d\sigma_1
%
%
\widehat{V}^{(-1)}_{A,\epsilon}(\sigma_1;k^{(1)}) 
\widehat{V}^{(-1)}_{A,\epsilon}(\sigma_1+\epsilon;k^{(2)})
\nonumber \\*
&&
~~~~~~~~~~~~~~~~~~~~~~~~~~~~~
\times 
\int_{\sigma_1+2\epsilon}^{\sigma_1+2\pi-\epsilon}
\!\!\!\!
d\sigma_2 
\widehat{V}^{(0)}_{A,\epsilon}(\sigma_2;k^{(3)}) 
|x_0^i\rangle_m 
\otimes 
|B \rangle_{gh}
\otimes 
|B;\eta \rangle_{sgh}
\nonumber \\
&&
~~
+\frac{i^2}{3}C
\int 
\prod_{r=1}^{3}
\frac{d^{p+1}k^{(r)}}{(2\pi)^{\frac{p+1}{2}}}
~
g_{\eta}
\int_0^{2\pi}
\!\!
d\sigma_1
%
%
\widehat{V}^{(0)}_{A,\epsilon}(\sigma_1;k^{(1)}) 
\nonumber \\*
&&
~~~~~~~~~~~~~~
\times 
\int_{\sigma_1+\epsilon}^{\sigma_1+2\pi-2\epsilon}
\!\!\!\!
d\sigma_2 
\widehat{V}^{(-1)}_{A,\epsilon}(\sigma_2;k^{(2)})
\widehat{V}^{(-1)}_{A,\epsilon}(\sigma_2+\epsilon;k^{(3)}) 
|x_0^i\rangle_m 
\otimes 
|B \rangle_{gh}
\otimes 
|B;\eta \rangle_{sgh}
\nonumber \\
&&
~~
+\frac{i^2}{3}C
\int 
\prod_{r=1}^{3}
\frac{d^{p+1}k^{(r)}}{(2\pi)^{\frac{p+1}{2}}}
~
g_{\eta}
\int_0^{2\pi}
\!\!
d\sigma_1
%
%
\widehat{V}^{(-1)}_{A,\epsilon}(\sigma_1-\epsilon;k^{(3)}) 
\widehat{V}^{(-1)}_{A,\epsilon}(\sigma_1;k^{(1)})
\nonumber \\*
&&
~~~~~~~~~~~~~~~~~~~~~~~~~~~~~
\times 
\int_{\sigma_1+\epsilon}^{\sigma_1+2\pi-2\epsilon}
\!\!\!\!
d\sigma_2 
\widehat{V}^{(0)}_{A,\epsilon}(\sigma_2;k^{(2)}) 
|x_0^i\rangle_m 
\otimes 
|B \rangle_{gh}
\otimes 
|B;\eta \rangle_{sgh}~.
\label{three-gluon BS}
\end{eqnarray} 
Here the first term is the state (\ref{test 1}) with $N=3$. 
The supersymmetric path-ordering provides the other three 
terms. These include the contact interactions of the 
$-1$ picture operators. 

\subsubsection{Action of the supercurrents}
Let us compute the action of the superconformal generators 
on the three-gluon boundary state (\ref{three-gluon BS}). 
We start with the action of the supercurrents. 
The first term of (\ref{three-gluon BS}) 
consists of the triple integral of 
the $0$ picture operators,  
$\frac{i^3}{3}
\int_{{\cal M}_{3,\epsilon}} \widehat{V}_{A,\epsilon}^{(0)}
\widehat{V}_{A,\epsilon}^{(0)} \widehat{V}_{A,\epsilon}^{(0)}$. 
When the supercurrent acts on this integral, 
by using the commutation relation 
$\bigl[ {\cal G}_r,~\widehat{V}_{A,\epsilon}^{(0)} \bigr]
\sim -i\partial_{\sigma}
\left\{ e^{ir\sigma}\widehat{V}_{A,\epsilon}^{(-1)} \right\}$, 
it gives rise to the boundary integrals 
over $\partial {\cal M}_{3,\epsilon}$. 
They emerge as the double integrals of the form 
$\frac{i^2}{3}\int 
\left\{e^{ir \sigma}\widehat{V}_{A,\epsilon}^{(-1)}\right.
\widehat{V}_{A,\epsilon}^{(0)}
-e^{ir(\sigma + \epsilon)}\widehat{V}_{A,\epsilon}^{(0)}
\left.\widehat{V}_{A,\epsilon}^{(-1)}\right\}
\int \widehat{V}_{A,\epsilon}^{(0)}$.   
Another kind of double integrals 
appear from the actions on the other three 
terms of (\ref{three-gluon BS}). 
For instance, let us consider the second term 
of (\ref{three-gluon BS}). 
This term includes the integral  
$\frac{i^2}{3}\int 
\widehat{V}_{A,\epsilon}^{(-1)}
\widehat{V}_{A,\epsilon}^{(-1)}
\int \widehat{V}_{A,\epsilon}^{(0)}$. 
We compute the action of the supercurrent 
on the $-1$ picture operators in this integral. 
By using the commutation relation 
$\bigl\{ {\cal G}_r,~\widehat{V}_{A,\epsilon}^{(-1)} \bigr\}
\sim e^{ir\sigma}\widehat{V}_{A,\epsilon}^{(0)}$,
we obtain the integral
$\frac{i^2}{3}\int 
\left\{e^{ir\sigma}\widehat{V}_{A,\epsilon}^{(0)}\right. 
\widehat{V}_{A,\epsilon}^{(-1)}
-e^{ir(\sigma +\epsilon)}\widehat{V}_{A,\epsilon}^{(-1)}
\left.\widehat{V}_{A,\epsilon}^{(0)}\right\}
\int \widehat{V}_{A,\epsilon}^{(0)}$. 
The similar integrals are obtainable 
from the remaining two terms. 
These two kinds of double integrals 
are put together in the form   
$\frac{i^2}{3}\int e^{ir\sigma}(1-e^{ir\epsilon})
\Delta_{A,\epsilon}^{(-1)}
\int \widehat{V}_{A,\epsilon}^{(0)}$.  
Single integrals appear from the last three terms of 
(\ref{three-gluon BS}). 
For instance, 
when the supercurrent acts on the $0$ picture operator 
in the second term of (\ref{three-gluon BS}), 
we obtain  
$\frac{i}{3}\int$
$\widehat{V}_{A,\epsilon}^{(-1)}$
$\widehat{V}_{A,\epsilon}^{(-1)}$
$\int d \left\{ e^{ir\sigma} 
\widehat{V}_{A,\epsilon}^{(-1)} \right\}$. 
This becomes single integrals of the product 
$\widehat{V}_{A,\epsilon}^{(-1)}$
$\widehat{V}_{A,\epsilon}^{(-1)}$
$\widehat{V}_{A,\epsilon}^{(-1)}$. 
These single integrals turn out to be collected in the form 
$i\int e^{ir\sigma}
(e^{-ir\epsilon}-e^{ir\epsilon})\widehat{V}_{A,\epsilon}^{(-1)}$
$\widehat{V}_{A,\epsilon}^{(-1)}$
$\widehat{V}_{A,\epsilon}^{(-1)}$.  
This leads us to introduce another contact term, 
which we call $\Upsilon_{A,\epsilon}^{(-1)}$, 
as follows:
\begin{eqnarray}
\Upsilon_{A,\epsilon}^{(-1)}(\sigma;k^{(1)},k^{(2)},k^{(3)})
=
\widehat{V}_{A,\epsilon}^{(-1)}(\sigma-\epsilon;k^{(1)})
\widehat{V}_{A,\epsilon}^{(-1)}(\sigma;k^{(2)})
\widehat{V}_{A,\epsilon}^{(-1)}(\sigma+\epsilon;k^{(3)}) ~.
\label{three-gluon contact term 1}
\end{eqnarray}

Finally we piece together all these computations 
and obtain the following expression of 
the action of the supercurrent:   
\begin{eqnarray}
&&
(G_r+i\eta \bar{G}_{-r})\times
\frac{i^9}{2}C
\int_{{\cal I}_3(\epsilon)}
\bigl[ d^3S \bigr]
~g_{\eta}
\widehat{{\bf V}}_{A,\epsilon}
({\bf X}(S_1))
\widehat{{\bf V}}_{A,\epsilon}
({\bf X}(S_2))
\widehat{{\bf V}}_{A,\epsilon}
({\bf X}(S_3))
\nonumber \\*
&&
~~~~~~~~~~~~~~~~~~~~~~~~~~~~~~~~~~~~
\times 
|x_0^i\rangle_m 
\otimes 
|B \rangle_{gh}
\otimes 
|B;\eta \rangle_{sgh}
\nonumber \\
&&
=
iC
\int 
\!\!
\prod_{r=1}^3
\frac{d^{p+1}k^{(r)}}{(2\pi)^{\frac{p+1}{2}}}
~g_{\eta}
\int_0^{2\pi}
\!\!d\sigma   
e^{ir\sigma}(e^{-ir\epsilon}-e^{ir\epsilon})
\Upsilon_{A,\epsilon}^{(-1)}(\sigma;k^{(1)},k^{(2)},k^{(3)}) 
\nonumber \\*
&&
~~~~~~~~~~~~~~~~~~~~~~~~~~~~~~~~~~~~
\times
|x_0^i\rangle_m 
\otimes 
|B \rangle_{gh}
\otimes 
|B;\eta \rangle_{sgh} 
\nonumber \\
&&
~~
+
\frac{i^2}{2}C
\int 
\!\!
\prod_{r=1}^3
\frac{d^{p+1}k^{(r)}}{(2\pi)^{\frac{p+1}{2}}}
~g_{\eta}
\int_{0}^{2\pi}
\!\!
d\sigma_1
%
%
e^{ir\sigma_1}(1-e^{ir\epsilon})
\Delta_{A,\epsilon}^{(-1)}(\sigma_1;k^{(1)},k^{(2)})
\nonumber \\*
&&
~~~~~~~~~~~~~~~~~~~~~~~~~~~~~~~~~~~~
\times 
\int_{\sigma_1+2\epsilon}^{\sigma_1+2\pi-\epsilon}
\!\!\!\!\!
d\sigma_2
\widehat{V}_{A,\epsilon}^{(0)}(\sigma_2;k^{(3)})
|x_0^i\rangle_m 
\otimes 
|B \rangle_{gh}
\otimes 
|B;\eta \rangle_{sgh} 
\nonumber \\
&&
+
~~
\frac{i^2}{2}
C
\int 
\!\!
\prod_{r=1}^3
\frac{d^{p+1}k^{(r)}}{(2\pi)^{\frac{p+1}{2}}}
~g_{\eta}
\int_{0}^{2\pi}
\!\!
d\sigma_1
%
%
\widehat{V}_{A,\epsilon}^{(0)}(\sigma_2;k^{(1)})
\nonumber \\*
&&
~~~~~~~~~~
\times 
\int_{\sigma_1+\epsilon}^{\sigma_1+2\pi-2\epsilon}
\!\!\!\!
d\sigma_2
e^{ir\sigma_2}(1-e^{ir\epsilon})
\Delta_{A,\epsilon}^{(-1)}(\sigma_2;k^{(2)},k^{(3)})
|x_0^i\rangle_m 
\otimes 
|B \rangle_{gh}
\otimes 
|B;\eta \rangle_{sgh}
\nonumber \\ 
&&
+ \cdots ~, 
\label{supercurrent three-gluon boundary integral}
\end{eqnarray}
where the terms proportional to 
$\delta_{{\bf B}}^0A_{\mu}^*$ and $\delta_{{\bf B}}^0{\cal B}$ 
are omitted.

Let us examine the role of the contact terms 
$\Delta_{A,\epsilon}^{(-1)}$ which appear 
in the second and the third terms of  
(\ref{supercurrent three-gluon boundary integral}).  
When the $0$ picture operator is apart from 
the contact term in the integrals, 
the expansion 
(\ref{expansion of two-gluon contact term 2}) 
can be used for the contact term. 
By this substitution, also 
taking account of the definition 
(\ref{1st correction of BRS}), 
the second term of 
(\ref{supercurrent three-gluon boundary integral}) 
is expressed as 
\begin{eqnarray}
&&
-\frac{i^2}{2}C
\int 
\!\!
\prod_{r=1,2}
\frac{d^{p+1}k^{(r)}}{(2\pi)^{\frac{p+1}{2}}}
\epsilon^{\alpha'k^{(1)}\cdot k^{(1)}}
~g_{\eta}
\int_{0}^{2\pi}
\!\!
d\sigma_1
~2re^{ir\sigma_1}
\left\{ 
\delta_{{\bf B}}^{1}A_{\mu}^*(k^{(1)})
\widehat{V}_{gl}^{(-1)\mu}(\sigma_1;k^{(1)})+
{\cal O}(\epsilon) 
\right\}
\nonumber \\*
&&
~~~~~~~~~~~~~~~~~~~~~~~~~~
\times 
\int_{\sigma_1+2\epsilon}^{\sigma_1+2\pi-\epsilon}
\!\!\!\!\!
d\sigma_2
\widehat{V}_{A,\epsilon}^{(0)}(\sigma_2;k^{(3)})
|x_0^i\rangle_m 
\otimes 
|B \rangle_{gh}
\otimes 
|B;\eta \rangle_{sgh}~. 
\label{from 3 to 2 for supercurrent}
\end{eqnarray} 
The similar expression is obtainable from the third term 
of (\ref{supercurrent three-gluon boundary integral}). 
These may be compared with the action of the 
supercurrent on the two-gluon boundary state 
(\ref{supercurrent two-gluon boundary integral}). 
The relevant terms of the action 
(\ref{supercurrent two-gluon boundary integral})  
are those proportional to $\delta_{{\bf B}}^0A^*_{\mu}$. 
Correspondingly to (\ref{from 3 to 2 for supercurrent}), 
we can find the following one: 
\begin{eqnarray}
&&
-\frac{i^2}{2}C
\int 
\!\!
\prod_{r=1,2}
\frac{d^{p+1}k^{(r)}}{(2\pi)^{\frac{p+1}{2}}}
\epsilon^{\alpha'k^{(1)}\cdot k^{(1)}}
~g_{\eta}
\int_{0}^{2\pi}
\!\!
d\sigma_1
~2re^{ir\sigma_1} 
\delta_{{\bf B}}^{0}A_{\mu}^*(k^{(1)})
\widehat{V}_{gl}^{(-1)\mu}(\sigma_1;k^{(1)}) 
\nonumber \\*
&&
~~~~~~~~~~~~~~~~~~~~~~~~~~
\times 
\int_{\sigma_1+\epsilon}^{\sigma_1+2\pi-\epsilon}
\!\!\!
d\sigma_2
\widehat{V}_{A,\epsilon}^{(0)}(\sigma_2;k^{(2)})
|x_0^i\rangle_m 
\otimes 
|B \rangle_{gh}
\otimes 
|B;\eta \rangle_{sgh}~. 
\label{2 from 3 for supercurrent}
\end{eqnarray} 
From the comparison between 
(\ref{from 3 to 2 for supercurrent}) 
and 
(\ref{2 from 3 for supercurrent})
we see that the second and the third terms of 
(\ref{supercurrent three-gluon boundary integral}) 
modify the action of the supercurrent on the 
{\it two-}gluon boundary state such that 
the $-1$ picture operators 
$\delta_{{\bf B}}^0A_{\mu}^*\widehat{V}_{gl}^{(-1)\mu}$ in 
the action (\ref{supercurrent two-gluon boundary integral}) 
change into 
$\bigl( \delta_{{\bf B}}^0 + \delta_{{\bf B}}^1 \bigr)
A_{\mu}^*\widehat{V}_{gl}^{(-1)\mu}$.  
The correction 
$\delta_{{\bf B}}^1A_{\mu}^*\widehat{V}_{gl}^{(-1)\mu}$ 
comes from the contact term $\Delta_{A,\epsilon}^{(-1)}$ 
in (\ref{supercurrent three-gluon boundary integral}).

At this stage it is convenient to recall 
how the supersymmetric path-ordering controls 
the divergences which originate in the propagations  
of unphysical open-string particle. 
In the channels where the $0$ picture gluon vertex operators 
are sufficiently close to one another, 
the unphysical open-string particles propagate. 
These propagations are cancelled out with 
the contact interactions between the $-1$ picture operators. 
The supersymmetric path-ordering gives rise to these interactions. 
In the first term of the three-gluon boundary state (\ref{three-gluon BS})     
we can find the channels which suffer 
the propagations of unphysical open-string particle.  
Their propagations will be deleted by the contact 
interactions of the $-1$ picture operators 
which are present in the other terms 
of (\ref{three-gluon BS}).

The cancellation should be maintained even in 
(\ref{supercurrent three-gluon boundary integral}). 
In the second and the third terms of 
(\ref{supercurrent three-gluon boundary integral}), 
we can find the channels where the $0$ picture operator gets closer to 
the contact term $\Delta_{A,\epsilon}^{(-1)}$.  
In such channels the contact term is just the product
$\widehat{V}_{A,\epsilon}^{(0)}\widehat{V}_{A,\epsilon}^{(-1)}$ and 
thereby the unphysical open-string particles propagate. 
The contributions from their propagations  
must be cancelled out with a suitable part of the first 
term of (\ref{supercurrent three-gluon boundary integral}). 
We need to extract it from the contact term 
$\Upsilon_{A,\epsilon}^{(-1)}$ in a proper manner.

For this purpose, let us remark that any successive two  operators 
in $\Upsilon_{A,\epsilon}^{(-1)}$ come from the contact interactions 
of  the three-gluon boundary state (\ref{three-gluon BS}). 
This indicates that the RHS of (\ref{three-gluon contact term 1}) 
should be understood as 
\begin{eqnarray}
&&
\Upsilon_{A,\epsilon}^{(-1)}(\sigma;k^{(1)},k^{(2)},k^{(3)})
\nonumber \\
&&
~~~~~~~~
=
\frac{1}{2}
\Biggl\{
\widehat{V}_{A,\epsilon}^{(-1)}(\sigma-\epsilon;k^{(1)})
\widehat{V}_{A,\epsilon}^{(-1)}(\sigma;k^{(2)})
\Biggr\}
\lim_{\sigma_3 \rightarrow \sigma+\epsilon}
\widehat{V}_{A,\epsilon}^{(-1)}(\sigma_3;k^{(3)})
\nonumber \\
&&
~~~~~~~~~~
+
\frac{1}{2}
\lim_{\sigma_1 \rightarrow \sigma-\epsilon}
\widehat{V}_{A,\epsilon}^{(-1)}(\sigma_1;k^{(1)})
\Biggl\{
\widehat{V}_{A,\epsilon}^{(-1)}(\sigma;k^{(2)})
\widehat{V}_{A,\epsilon}^{(-1)}(\sigma+\epsilon;k^{(3)})
\Biggr\}
~,
\label{def of three-gluon contact term 1}
\end{eqnarray}
where the products in the parentheses 
are the contact interactions which originally exist 
in the three-gluon boundary state (\ref{three-gluon BS}). 
We will expand $\Upsilon_{A,\epsilon}^{(-1)}$ 
into the power series of $\epsilon$ according to 
(\ref{def of three-gluon contact term 1}). 
Namely, we first expand the products in the parentheses 
up to the next-to-leading order. 
(This turns out to be  necessary since the next-to-leading 
terms provide the tensor $J_{\mu}$ in the below.)
Each coefficient in the expansions 
is paired with the $-1$ picture operator of 
(\ref{def of three-gluon contact term 1}). 
The expansion of $\Upsilon_{A,\epsilon}^{(-1)}$ 
is obtained from the expansions of these pairs. 
The expansion of the first 
term of (\ref{def of three-gluon contact term 1}) becomes 
as follows: 
\begin{eqnarray}
&&
\frac{1}{2}
\Biggl\{
\widehat{V}_{A,\epsilon}^{(-1)}(\sigma-\epsilon;k^{(1)})
\widehat{V}_{A,\epsilon}^{(-1)}(\sigma;k^{(2)})
\Biggr\}
\lim_{\sigma_3 \rightarrow \sigma+\epsilon}
\widehat{V}_{A,\epsilon}^{(-1)}(\sigma_3;k^{(3)})
\nonumber \\
&&
=
i
\epsilon^{\alpha'(\sum_{r}k^{(r)})^2-1}
e^{-\frac{i}{2}\sum_{r<s}k^{(r)}\times k^{(s)}}
\nonumber \\*
&&
~~~~
\times 
\alpha' 
\Biggl[
\biggl\{
A(k^{(2)})\!\cdot \!A(k^{(3)})A_{\mu}(k^{(1)})
-
A(k^{(1)})\!\cdot \!A(k^{(3)})A_{\mu}(k^{(2)})
\biggr\}
+
A(k^{(1)})\!\cdot \!A(k^{(2)})A_{\mu}(k^{(3)})
\Biggr]
\nonumber \\*
&& 
~~~~
\times 
\widehat{V}_{gl}^{(-1)\mu}(\sigma;\sum_{r}k^{(r)})
\nonumber \\
&&
~~
+
{\cal O}(\epsilon^{\alpha'(\sum_{r}k^{(r)})^2})~.
\label{pre-expansion three-gluon contact term 1}
\end{eqnarray}
The expansion of the second  
term of (\ref{def of three-gluon contact term 1}) 
turns out to be that obtained from 
(\ref{pre-expansion three-gluon contact term 1}) 
just by exchanging $A(k^{(1)}) \leftrightarrow A(k^{(3)})$: 
\begin{eqnarray}
&&
\frac{1}{2}
\lim_{\sigma_1 \rightarrow \sigma-\epsilon}
\widehat{V}_{A,\epsilon}^{(-1)}(\sigma_1;k^{(1)})
\Biggl\{
\widehat{V}_{A,\epsilon}^{(-1)}(\sigma;k^{(2)})
\widehat{V}_{A,\epsilon}^{(-1)}(\sigma+\epsilon;k^{(3)})
\Biggr\}
\nonumber \\
&&
=
i
\epsilon^{\alpha'(\sum_{r}k^{(r)})^2-1}
e^{-\frac{i}{2}\sum_{r<s}k^{(r)}\times k^{(s)}}
\nonumber \\*
&&
~~~~
\times 
\alpha' 
\Biggl[
\biggl\{
A(k^{(1)})\!\cdot \!A(k^{(2)})A_{\mu}(k^{(3)})
-
A(k^{(1)})\!\cdot \!A(k^{(3)})A_{\mu}(k^{(2)})
\biggr\}
+
A(k^{(2)})\!\cdot \!A(k^{(3)})A_{\mu}(k^{(1)})
\Biggr]
\nonumber \\*
&& 
~~~~
\times 
\widehat{V}_{gl}^{(-1)\mu}(\sigma;\sum_{r}k^{(r)})
\nonumber \\
&&
~~
+
{\cal O}(\epsilon^{\alpha'(\sum_{r}k^{(r)})^2})~.
\label{pre-expansion three-gluon contact term 2}
\end{eqnarray}
From these expansions we can read 
the part of $\Upsilon_{A,\epsilon}^{(-1)}$ 
which works to maintain the cancellation in  
(\ref{supercurrent three-gluon boundary integral}).  
For the first term of (\ref{def of three-gluon contact term 1}), 
it is the term proportional to 
$A(k^{(1)})\!\cdot \!A(k^{(2)})A_{\mu}(k^{(3)})$
$\widehat{V}_{gl}^{(-1)\mu}$ in the expansion 
(\ref{pre-expansion three-gluon contact term 1}). For the second 
term, it is given by the above replacement.  
The expansion of $\Upsilon_{A,\epsilon}^{(-1)}$ 
is the collection of 
(\ref{pre-expansion three-gluon contact term 1})
and 
(\ref{pre-expansion three-gluon contact term 2}). 
Let us write it in the following manner: 
\begin{eqnarray}
&&
\Upsilon_{A,\epsilon}^{(-1)}(\sigma;k^{(1)},k^{(2)},k^{(3)}) 
\nonumber \\ 
&& 
~~
=
i\epsilon^{\alpha' \left(\sum_{r}k^{(r)}\right)^2-1}
\Biggl\{
J_{\mu}(k^{(1)},k^{(2)},k^{(3)})
+
K_{\mu}(k^{(1)},k^{(2)},k^{(3)}) 
\Biggr\} 
\widehat{V}_{gl}^{(-1)\mu}(\sigma;\sum_{r}k^{(r)})
\nonumber \\
&& 
~~~~
+
{\cal O}(\epsilon^{\alpha' \left(\sum_{r}k^{(r)}\right)^2})~,
\label{expansion three-gluon contact term 1}
\end{eqnarray}
Here $K_{\mu}\widehat{V}_{gl}^{(-1)\mu}$ is the $-1$ picture 
operator which is responsible 
to the cancellation of the unphysical open-string particle.  
The tensor $K_{\mu}$ becomes  
\begin{eqnarray}
K_{\mu}(k^{(1)},k^{(2)},k^{(3)})
=
\alpha' 
e^{-\frac{i}{2}\sum_{r<s}k^{(r)}\times k^{(s)}}
\Biggl\{
A(k^{(1)})\!\cdot \!A(k^{(2)})A_{\mu}(k^{(3)})
+~1 \leftrightarrow 3~
\Biggr\}~.
\label{K}
\end{eqnarray} 
The $-1$ picture operator which is left in the expansion 
after the subtraction of $K_{\mu}\widehat{V}_{gl}^{(-1)\mu}$ 
is denoted by 
$J_{\mu}\widehat{V}_{gl}^{(-1)\mu}$. 
The tensor $J_{\mu}$ becomes 
\begin{eqnarray} 
&& 
J_{\mu}(k^{(1)},k^{(2)},k^{(3)}) 
\nonumber \\*
&&
~
=
\alpha' 
e^{-\frac{i}{2}\sum_{r<s}k^{(r)}\times k^{(s)}} 
\nonumber \\*
&&
~~
\times 
\Biggl \{
A(k^{(1)})\!\cdot \!A(k^{(2)})A_{\mu}(k^{(3)})
+  
A(k^{(2)})\!\cdot \!A(k^{(3)})A_{\mu}(k^{(1)})
-
2A(k^{(1)})\!\cdot \!A(k^{(3)})A_{\mu}(k^{(2)})
\Biggr\}~.
\nonumber \\
\label{J}
\end{eqnarray}

It is plausible that 
the tensor $J_{\mu}$ brings about  
another collection $\delta_{{\bf B}}^2A_{\mu}^*$ 
to the BRST transformation of the antifield $A^*_{\mu}$. 
This may be verified by the comparison  
of the first term of 
(\ref{supercurrent three-gluon boundary integral}) 
with  
the action of the supercurrent 
on the {\it single} gluon boundary state 
(\ref{one-gluon supercurrent 1}).
By using the expression 
(\ref{expansion three-gluon contact term 1}) 
we can write down the first term of 
(\ref{supercurrent three-gluon boundary integral}) 
as follows: 
\begin{eqnarray}
&&
iC
\int 
\!\!
\prod_{r=1}^3
\frac{d^{p+1}k^{(r)}}{(2\pi)^{\frac{p+1}{2}}}
~g_{\eta}
\int_0^{2\pi}
\!\!d\sigma   
e^{ir\sigma}(e^{-ir\epsilon}-e^{ir\epsilon})
\Upsilon_{A,\epsilon}^{(-1)}(\sigma;k^{(1)},k^{(2)},k^{(3)}) 
|x_0^i\rangle_m 
\otimes 
|B \rangle_{gh}
\otimes 
|B;\eta \rangle_{sgh} 
\nonumber \\
&&
=
iC 
\int
\!\!
\frac{d^{p+1}k}{(2\pi)^{\frac{p+1}{2}}}
\epsilon^{\alpha'k\cdot k}
~g_{\eta}
\int_0^{2\pi}
\!\!d\sigma   
2re^{ir\sigma} 
\nonumber \\*
&&
~~
\times 
\left\{ 
\int 
\!\!
\prod_{r=1}^3
\frac{d^{p+1}k^{(r)}}{(2\pi)^{\frac{p+1}{2}}}
(2\pi)^{\frac{p+1}{2}}
\delta^{p+1} 
\bigl(\sum_rk^{(r)}-k \bigr)
J_{\mu}(k^{(1)},k^{(2)},k^{(3)})
\right\}
\widehat{V}_{gl}^{(-1)\mu}(\sigma;k)
\nonumber \\*
&& 
~~~~~~
\times
|x_0^i\rangle_m \otimes |B \rangle_{gh} 
\otimes |B;\eta\rangle_{sgh}
\nonumber \\ 
&&
~~+ \cdots~.
\label{from 3 to 1 for supercurrent}
\end{eqnarray}
The relevant term in the action 
(\ref{one-gluon supercurrent 1})
is that proportional to $\delta_{{\bf B}}^0A^*_{\mu}$. 
It has the following form: 
\begin{eqnarray}
-iC 
\int
\!\!
\frac{d^{p+1}k}{(2\pi)^{\frac{p+1}{2}}}
\epsilon^{\alpha'k\cdot k}
~g_{\eta}
\int_0^{2\pi}
\!\!d\sigma   
~2re^{ir\sigma} 
\delta_{{\bf B}}^0A^*(k)
\widehat{V}_{gl}^{(-1)\mu}(\sigma;k)
|x_0^i\rangle_m \otimes |B \rangle_{gh} 
\otimes |B;\eta\rangle_{sgh}~.
\label{1 from 3 for supercurrent}
\end{eqnarray}
By comparing these two we can see that 
the action on the single gluon boundary state 
(\ref{one-gluon supercurrent 1}) is modified 
effectively by (\ref{from 3 to 1 for supercurrent}).  
The modification is exactly interpreted as 
the correction $\delta_{{\bf B}}^2A_{\mu}^*$
to the BRST transformation of the antifield
$A^*_{\mu}$. It can be read as 
\begin{eqnarray}
\delta_{{\bf B}}^2A_{\mu}^*(k)
=
-\int 
\!\!
\prod_{r=1}^3
\frac{d^{p+1}k^{(r)}}{(2\pi)^{\frac{p+1}{2}}}
(2\pi)^{\frac{p+1}{2}}
\delta^{p+1} 
\bigl(\sum_rk^{(r)}-k \bigr)
J_{\mu}(k^{(1)},k^{(2)},k^{(3)} \bigr)~.
\label{2nd correction of BRS}
\end{eqnarray}
This is also written in terms of 
the Moyal $\star$-product as 
\begin{eqnarray} 
&&
\delta_{{\bf B}}^2A_{\mu}^*(k)
\nonumber \\*
&&
=
\alpha'
\int 
\!
\frac{d^{p+1}x}{(2\pi)^{\frac{p+1}{2}}}
e^{-ik_{\mu}x^{\mu}}
G^{\nu \rho}
\Bigl(
-A_{\mu}\star A_{\nu} \star A_{\rho}
-A_{\nu}\star A_{\rho} \star A_{\mu}
+2A_{\nu}\star A_{\mu} \star A_{\rho}
\Bigr)(x)~.
\nonumber \\
\label{2nd order BRST}
\end{eqnarray}

So far, 
we have obtained the two corrections 
$\delta_{{\bf B}}^1A^*_{\mu}$ (\ref{1st correction of BRS}) 
and 
$\delta_{{\bf B}}^2A^*_{\mu}$ (\ref{2nd correction of BRS}) 
to the BRST transformation of 
the antifield $A^*_{\mu}$.  
The first correction 
$\delta_{{\bf B}}^1A^*_{\mu}$ 
is brought about from the two-gluon boundary state 
by the BRST operator. 
The second correction 
$\delta_{{\bf B}}^2A^*_{\mu}$ (\ref{2nd correction of BRS}) 
is obtained from the three-gluon boundary state by the supercurrent. 
We will show subsequently that  
the Virasoro action on the three-gluon boundary state 
is consistent with the action of the supercurrent. 
These actions of the superconformal generators 
are incorporated into the action of the BRST operator 
as in the case of the two-gluon boundary state.  
The two corrections  
$\delta_{{\bf B}}^1 A_{\mu}^*$ 
and
$\delta_{{\bf B}}^2 A_{\mu}^*$ 
make the BRST transformation of the antifield 
$A^*_{\mu}$ non-linear with the following form:  
\begin{eqnarray}
\Bigl(\delta_{{\bf B}}^0+\delta_{{\bf B}}^1
+\delta_{{\bf B}}^2 \Bigr)A_{\mu}^*(k)
=
\alpha'
\int 
\!
\frac{d^{p+1}x}{(2\pi)^{\frac{p+1}{2}}}
e^{-ik_{\mu}x^{\mu}}
G^{\nu \rho}
\nabla_{\nu}F_{\rho \mu}\Bigr|_{\partial \cdot A=0}(x)~. 
\label{total BRST}
\end{eqnarray}
The BRST invariance condition of the single 
gluon boundary state of the Wilson loop is modified non-linearly and 
becomes the equation of motion for the non-commutative gauge 
theory in the Lorentz gauge.   

\subsubsection{Action of the Virasoro generators}
Computation of the Virasoro action on the three-gluon boundary state 
(\ref{three-gluon BS}) 
becomes similar to the previous computation of the supercurrent. 
When the Virasoro generator acts on 
the triple integral 
$\int_{{\cal M}_{3,\epsilon}} 
\widehat{V}_{A,\epsilon}^{(0)} \widehat{V}_{A,\epsilon}^{(0)} 
\widehat{V}_{A,\epsilon}^{(0)}$ 
in the first term of (\ref{three-gluon BS}),  
it brings about the double integrals of the form 
$\frac{i^2}{3}\int 
e^{in \sigma}(1-e^{in\epsilon})
\widehat{V}_{A,\epsilon}^{(0)}\widehat{V}_{A,\epsilon}^{(0)} 
\int \widehat{V}_{A,\epsilon}^{(0)}$.   
Another kind of double integrals appear from 
the other three terms of (\ref{three-gluon BS}). 
For instance, let us consider the second term 
of (\ref{three-gluon BS}).
This term includes the integral  
$\frac{i^2}{3}\int 
\widehat{V}_{A,\epsilon}^{(-1)}
\widehat{V}_{A,\epsilon}^{(-1)}$
$\int \widehat{V}_{A,\epsilon}^{(0)}$. 
When the Virasoro generator acts on 
the $-1$ picture operator in this integral, 
it brings about the two integrals  
$\frac{i}{3}\int e^{in\sigma}
\bigl(\partial_{\sigma}-\frac{in}{2}\bigr) 
\widehat{V}_{A,\epsilon}^{(-1)}
\widehat{V}_{A,\epsilon}^{(-1)}
\int \widehat{V}_{A,\epsilon}^{(0)}$ 
and 
$\frac{i}{3}\int 
\widehat{V}_{A,\epsilon}^{(-1)} 
e^{in(\sigma +\epsilon)}
\bigl(\partial_{\sigma}-\frac{in}{2}\bigr) 
\widehat{V}_{A,\epsilon}^{(-1)}
\int \widehat{V}_{A,\epsilon}^{(0)}$. 
These two are collected in the double integral 
$\frac{i^2}{3}\int e^{in\sigma}(1-e^{in\epsilon})
\frac{i}{2}\Bigl\{
\widehat{V}_{A,\epsilon}^{(-1)}
\partial_{\sigma}\widehat{V}_{A,\epsilon}^{(-1)}
-
\partial_{\sigma}\widehat{V}_{A,\epsilon}^{(-1)}
\widehat{V}_{A,\epsilon}^{(-1)}
\Bigr\} 
\int \widehat{V}_{A,\epsilon}^{(0)}$ by the partial integrations 
as in (\ref{two-gluon integral 4}). 
Similar integrals are obtainable from 
the remaining two terms of (\ref{three-gluon BS}). 
These two kinds of double integrals are put together 
in the form   
$\frac{i^2}{3}\int e^{in\sigma}(1-e^{in\epsilon})
\Delta_{A,\epsilon}^{(0)}
\int \widehat{V}_{A,\epsilon}^{(0)}$. 
Single integrals of the product 
$\widehat{V}_{A,\epsilon}^{(0)}
\widehat{V}_{A,\epsilon}^{(-1)}
\widehat{V}_{A,\epsilon}^{(-1)}$ 
also appear from the last three 
terms of (\ref{three-gluon BS}).  
All these single integrals are incorporated 
in a suitable manner. 
They turn out to be expressed as the collection 
of two kinds of integrals. One consists 
of the integrals written by means of 
$\Delta_{A,\epsilon}^{(-1)}$
in the form 
$\frac{i}{3}\int e^{in\sigma}(1-e^{in\epsilon})
\Delta_{A,\epsilon}^{(-1)}
\widehat{V}_{A,\epsilon}^{(-1)}$. 
The other consists of the integrals 
of the form 
$\frac{i}{2}
\int e^{in\sigma}(e^{-in\epsilon}-e^{in\epsilon})
\widehat{V}_{A,\epsilon}^{(0)}
\widehat{V}_{A,\epsilon}^{(-1)}
\widehat{V}_{A,\epsilon}^{(-1)}$. 
The latter single integrals lead us to 
introduce another contact term, 
which we call $\Upsilon_{A,\epsilon}^{(0)}$: 
\begin{eqnarray}
&&
\Upsilon_{A,\epsilon}^{(0)}(\sigma;k^{(1)},k^{(2)},k^{(3)})
\nonumber \\
&&
~~
=
\frac{1}{2} 
\Biggl\{
\widehat{V}_{A,\epsilon}^{(0)}(\sigma-\epsilon;k^{(1)})
\widehat{V}_{A,\epsilon}^{(-1)}(\sigma;k^{(2)})
\widehat{V}_{A,\epsilon}^{(-1)}(\sigma+\epsilon;k^{(3)}) 
\nonumber \\
&&
~~~~~~~~~~~
-
\widehat{V}_{A,\epsilon}^{(-1)}(\sigma-\epsilon;k^{(1)})
\widehat{V}_{A,\epsilon}^{(0)}(\sigma;k^{(2)})
\widehat{V}_{A,\epsilon}^{(-1)}(\sigma+\epsilon;k^{(3)})
\nonumber \\
&&
~~~~~~~~~~~~~~~~~~~~
+
\widehat{V}_{A,\epsilon}^{(-1)}(\sigma-\epsilon;k^{(1)})
\widehat{V}_{A,\epsilon}^{(-1)}(\sigma;k^{(2)})
\widehat{V}_{A,\epsilon}^{(0)}(\sigma+\epsilon;k^{(3)})
\Biggr\}~.
\label{three-gluon contact term 2}
\end{eqnarray}

Finally we piece together all the above computations and 
obtain the following expression of the Virasoro action: 
\begin{eqnarray}
&&
(L_n-\bar{L}_{-n})\times
\frac{i^9}{2}C
\int_{{\cal I}_3(\epsilon)}
\bigl[ d^3S \bigr]
~g_{\eta}
\widehat{{\bf V}}_{A,\epsilon}
({\bf X}(S_1))
\widehat{{\bf V}}_{A,\epsilon}
({\bf X}(S_2))
\widehat{{\bf V}}_{A,\epsilon}
({\bf X}(S_3))
\nonumber \\*
&&
~~~~~~~~~~~~~~~~~~~~~~~~~~~~~~~~~~~~
\times 
|x_0^i\rangle_m 
\otimes 
|B \rangle_{gh}
\otimes 
|B;\eta \rangle_{sgh}
\nonumber \\
&&
=iC
\int 
\!
\prod_{r=1}^3
\frac{d^{p+1}k^{(r)}}{(2\pi)^{\frac{p+1}{2}}}
~g_{\eta}
\int_0^{2\pi}
\!\!d\sigma   
e^{in\sigma}(e^{-in\epsilon}-e^{in\epsilon})
\Upsilon_{A,\epsilon}^{(0)}(\sigma;k^{(1)},k^{(2)},k^{(3)}) 
\nonumber \\*
&&
~~~~~~~~~~~~~~~~~~~~~~~~~~~~~~~~~~~~
\times
|x_0^i\rangle_m 
\otimes 
|B \rangle_{gh}
\otimes 
|B;\eta \rangle_{sgh} 
\nonumber \\
&&
~+
\frac{i^2}{2}C
\int 
\!
\prod_{r=1}^3
\frac{d^{p+1}k^{(r)}}{(2\pi)^{\frac{p+1}{2}}}
~g_{\eta}
\int_{0}^{2\pi}
\!\!
d\sigma_1
%
%
e^{in\sigma_1}(1-e^{in\epsilon})
\Delta_{A,\epsilon}^{(0)}(\sigma_1;k^{(1)},k^{(2)})
\nonumber \\*
&&
~~~~~~~~~~~~~~~~~~~~~~~~~~~~~~~~~~~~
\times 
\int_{\sigma_1+2\epsilon}^{\sigma_1+2\pi-\epsilon}
\!\!\!\!\!
d\sigma_2
\widehat{V}_{A,\epsilon}^{(0)}(\sigma_2;k^{(3)})
|x_0^i\rangle_m 
\otimes 
|B \rangle_{gh}
\otimes 
|B;\eta \rangle_{sgh} 
\nonumber \\
&&
~+
\frac{i}{2}C
\int 
\!
\prod_{r=1}^3
\frac{d^{p+1}k^{(r)}}{(2\pi)^{\frac{p+1}{2}}}
~g_{\eta}
\int_{0}^{2\pi}
\!\!
d\sigma
e^{in\sigma}(1-e^{in\epsilon})
\Delta_{A,\epsilon}^{(-1)}(\sigma;k^{(1)},k^{(2)})
\nonumber \\*
&&
~~~~~~~~~~~~~~~~~~~~~~~~~~~~~~~~~~~~
\times 
\widehat{V}_{A,\epsilon}^{(-1)}(\sigma+\epsilon;k^{(3)})
|x_0^{i}\rangle_m
\otimes 
|B \rangle_{gh}
\otimes 
|B;\eta \rangle_{sgh} 
\nonumber \\
&&
~+
\frac{i^2}{2}C
\int 
\!
\prod_{r=1}^3
\frac{d^{p+1}k^{(r)}}{(2\pi)^{\frac{p+1}{2}}}
~g_{\eta}
\int_{0}^{2\pi}
\!\!
d\sigma_1
%
%
\widehat{V}_{A,\epsilon}^{(0)}(\sigma_2;k^{(1)})
\nonumber \\*
&&
~~~~~~~~~~~~
\times 
\int_{\sigma_1+\epsilon}^{\sigma_1+2\pi-2\epsilon}
\!\!\!\!\!
d\sigma_2
e^{in\sigma_2}(1-e^{in\epsilon})
\Delta_{A,\epsilon}^{(0)}(\sigma_2;k^{(2)},k^{(3)})
|x_0^i\rangle_m 
\otimes 
|B \rangle_{gh}
\otimes 
|B;\eta \rangle_{sgh} 
\nonumber \\
&&
~+
\frac{i}{2}C
\int 
\!
\prod_{r=1}^3
\frac{d^{p+1}k^{(r)}}{(2\pi)^{\frac{p+1}{2}}}
~g_{\eta}
\int_{0}^{2\pi}
\!\!
d\sigma
\widehat{V}_{A,\epsilon}^{(-1)}(\sigma;k^{(1)})
\nonumber \\*
&&
~~~~~~~~~~~~
\times 
e^{in(\sigma+\epsilon)}(1-e^{in\epsilon})
\Delta_{A,\epsilon}^{(-1)}(\sigma+\epsilon;k^{(2)},k^{(3)})
|x_0^{i}\rangle_m
\otimes 
|B \rangle_{gh}
\otimes 
|B;\eta \rangle_{sgh} 
\nonumber \\
&&
~+ \cdots ~,
\label{Virasoro three-gluon boundary integral}
\end{eqnarray}
where the terms proportional to 
$\delta_{{\bf B}}^0A_{\mu}^*$ and $\delta_{{\bf B}}^0{\cal B}$ 
are omitted.

Let us examine the role of the contact terms 
$\Delta_{A,\epsilon}^{(0)}$ which appear 
in the second and the fourth terms of 
(\ref{Virasoro three-gluon boundary integral}). 
By plugging the expansion 
(\ref{expansion of two-gluon contact term 1}) 
into the contact term and also taking 
account of (\ref{1st correction of BRS}), 
the second term of 
(\ref{Virasoro three-gluon boundary integral}) 
can be expressed as 
\begin{eqnarray}
&&
-\frac{i^2}{2}C
\int 
\!\!
\prod_{r=1,2}
\frac{d^{p+1}k^{(r)}}{(2\pi)^{\frac{p+1}{2}}}
\epsilon^{\alpha'k^{(1)}\cdot k^{(1)}}
g_{\eta}
\int_{0}^{2\pi}
\!\!
d\sigma_1
~ne^{in\sigma_1}
\left\{ 
\delta_{{\bf B}}^{1}A_{\mu}^*(k^{(1)})
\widehat{V}_{gl}^{(0)\mu}(\sigma_1;k^{(1)})+
{\cal O}(\epsilon) 
\right\}
\nonumber \\*
&&
~~~~~~~~~~~~~~~~~~~~~~~~~~
\times 
\int_{\sigma_1+2\epsilon}^{\sigma_1+2\pi-\epsilon}
\!\!\!\!\!
d\sigma_2
\widehat{V}_{A,\epsilon}^{(0)}(\sigma_2;k^{(2)})
|x_0^i\rangle_m 
\otimes 
|B \rangle_{gh}
\otimes 
|B;\eta \rangle_{sgh}~. 
\label{from 3 to 2 Virasoro}
\end{eqnarray} 
The similar expression is obtainable from the fourth term of 
(\ref{Virasoro three-gluon boundary integral}).
These modify the Virasoro action 
on the two-gluon boundary state 
such that the $0$ picture operators 
$\delta^0_{{\bf B}}A_{\mu}^*\widehat{V}_{gl}^{(0)\mu}$ 
in the action 
(\ref{Virasoro two-gluon boundary integral})  
change into 
$\bigl(\delta^0_{{\bf B}}+\delta_{{\bf B}}^1\bigr)
A_{\mu}^*\widehat{V}_{gl}^{(0)\mu}$. 
For instance, the second term of 
(\ref{Virasoro three-gluon boundary integral}) 
becomes the correction to the following term of 
(\ref{Virasoro two-gluon boundary integral}): 
\begin{eqnarray}
&&
-\frac{i^2}{2}C
\int 
\!\!
\prod_{r=1,2}
\frac{d^{p+1}k^{(r)}}{(2\pi)^{\frac{p+1}{2}}}
\epsilon^{\alpha'k^{(1)}\cdot k^{(1)}}
~g_{\eta}
\int_{0}^{2\pi}
\!\!
d\sigma_1
~ne^{in\sigma_1} 
\delta_{{\bf B}}^{0}A_{\mu}^*(k^{(1)})
\widehat{V}_{gl}^{(0)\mu}(\sigma_1;k^{(1)}) 
\nonumber \\*
&&
~~~~~~~~~~~~~~~~~~~~~~~~~~
\times 
\int_{\sigma_1+\epsilon}^{\sigma_1+2\pi-\epsilon}
\!\!\!
d\sigma_2
\widehat{V}_{A,\epsilon}^{(0)}(\sigma_2;k^{(2)})
|x_0^i\rangle_m 
\otimes 
|B \rangle_{gh}
\otimes 
|B;\eta \rangle_{sgh}~. 
\label{2 for 3 Virasoro}
\end{eqnarray}

It should be noticed that the above term 
(\ref{2 for 3 Virasoro}) comes from 
the first term of the two-gluon boundary state 
(\ref{two-gluon BS}) 
by the action of the Virasoro generator. 
The supersymmetric path-ordering 
requires that the first term is paired with the other 
two terms of (\ref{two-gluon BS}). 
It is still supported in the Virasoro action.  
For instance, correspondingly to 
(\ref{2 for 3 Virasoro}), 
the following term is present in 
the Virasoro action on the two-gluon boundary state 
(\ref{Virasoro two-gluon boundary integral}): 
\begin{eqnarray}
&&
-iC
\int 
\!\!
\prod_{r=1,2}
\frac{d^{p+1}k^{(r)}}{(2\pi)^{\frac{p+1}{2}}}
\epsilon^{\alpha'k^{(1)}\cdot k^{(1)}}
~g_{\eta}
\int_{0}^{2\pi}
\!\!
d\sigma
~ne^{in\sigma_1} 
\delta_{{\bf B}}^{0}A_{\mu}^*(k^{(1)})
\widehat{V}_{gl}^{(-1)\mu}(\sigma;k^{(1)}) 
\nonumber \\*
&&
~~~~~~~~~~~~~~~~~~~~~~~
\times 
\widehat{V}_{A,\epsilon}^{(-1)}(\sigma+\epsilon;k^{(2)})
|x_0^i\rangle_m 
\otimes 
|B \rangle_{gh}
\otimes 
|B;\eta \rangle_{sgh}~. 
\label{2 for 3 Virasoro 2}
\end{eqnarray}

The consistent corrections to such terms are obtained 
from the third and the fifth terms of 
(\ref{Virasoro three-gluon boundary integral}). 
The contact terms $\Delta_{A,\epsilon}^{(-1)}$ 
provide such corrections. 
By plugging the expansion 
(\ref{expansion of two-gluon contact term 2}) 
into the contact term, 
the third term of 
(\ref{Virasoro three-gluon boundary integral}) 
can be expressed as 
\begin{eqnarray}
&&
-iC
\int 
\!\!
\prod_{r=1,2}
\frac{d^{p+1}k^{(r)}}{(2\pi)^{\frac{p+1}{2}}}
\epsilon^{\alpha'k^{(1)}\cdot k^{(1)}}
~g_{\eta}
\int_{0}^{2\pi}
\!\!
d\sigma
~ne^{in\sigma_1} 
\left\{
\delta_{{\bf B}}^{1}A_{\mu}^*(k^{(1)})
\widehat{V}_{gl}^{(-1)\mu}(\sigma;k^{(1)})
+
{\cal O}(\epsilon)
\right\} 
\nonumber \\*
&&
~~~~~~~~~~~~~~~~~~~~~~~
\times 
\widehat{V}_{A,\epsilon}^{(-1)}(\sigma+\epsilon;k^{(2)})
|x_0^i\rangle_m 
\otimes 
|B \rangle_{gh}
\otimes 
|B;\eta \rangle_{sgh}~. 
\label{2 from 3 Virasoro 2}
\end{eqnarray} 
The similar expression is obtainable from the fifth term of 
(\ref{Virasoro three-gluon boundary integral}). 
From the comparison between (\ref{2 for 3 Virasoro 2}) and
(\ref{2 from 3 Virasoro 2}), 
we see that the $-1$ picture operators 
$\delta^0_{{\bf B}}A_{\mu}^*\widehat{V}_{gl}^{(-1)\mu}$ 
in the action on the two-gluon boundary state 
(\ref{Virasoro two-gluon boundary integral}) 
change into 
$\bigl(\delta^0_{{\bf B}}+\delta_{{\bf B}}^1\bigr)
A_{\mu}^*\widehat{V}_{gl}^{(-1)\mu}$ 
by the effect of these terms.

The above discussions show that 
the contact terms $\Delta_{A,\epsilon}^{(-1)}$ and 
$\Delta_{A,\epsilon}^{(0)}$ in 
(\ref{Virasoro three-gluon boundary integral}) 
modify the Virasoro action on the two-gluon boundary state 
(\ref{Virasoro two-gluon boundary integral}) 
such that the operators 
$\delta_{{\bf B}}^0A_{\mu}^*\widehat{V}_{gl}^{(-1,0) \mu}$ 
of the both pictures change into 
$\bigl(\delta^0_{{\bf B}}+\delta_{{\bf B}}^1\bigr)A_{\mu}^*
\widehat{V}_{gl}^{(-1,0) \mu}$. 
It is consistent with the previous correction  
to the action of the supercurrent on the two-gluon boundary state. 
This means that the following correction emerges  
from the action of the closed-string BRST operator on 
the three-gluon boundary state  
\begin{eqnarray}
&&
\frac{i^4}{2}
C
\int_{{\cal I}_2(\epsilon)}
\!\! 
\bigl[d^2S \bigr]
~g_{\eta}
\int \frac{d^{p+1}k}{(2\pi)^{\frac{p+1}{2}}}
\epsilon^{\alpha'k\cdot k}
\Biggl\{
\delta_{{\bf B}}^1
A_{\mu}^*(k)
\widehat{{\bf V}}_{gl}^{* \mu}(S_1;k)
+
{\cal O}(\epsilon)
\Biggr\}
\nonumber \\*
&&
~~~~
\times 
\widehat{{\bf V}}_{A,\epsilon}
({\bf X}(S_2))~
|x_0^i\rangle_m 
\otimes 
|B \rangle_{gh}
\otimes 
|B;\eta \rangle_{sgh}
\nonumber \\
&&
+
\frac{i^4}{2}
C
\int_{{\cal I}_2(\epsilon)}
\!\! \bigl[d^2S \bigr]
~g_{\eta}
\widehat{{\bf V}}_{A,\epsilon}
({\bf X}(S_1))
\nonumber \\*
&&
~~~~
\times
\int \frac{d^{p+1}k}{(2\pi)^{\frac{p+1}{2}}}
\epsilon^{\alpha'k\cdot k}
\Biggl\{
\delta_{{\bf B}}^1
A_{\mu}^*(k)
\widehat{{\bf V}}_{gl}^{* \mu}(S_2;k)
+
{\cal O}(\epsilon)
\Biggr\}
|x_0^i\rangle_m 
\otimes 
|B \rangle_{gh}
\otimes 
|B;\eta \rangle_{sgh}~. 
\nonumber\\
\label{from 3 to 2 Qc}
\end{eqnarray} 
The above correction modifies the action 
of the BRST operator on the two-gluon boundary state 
(\ref{Qc on two-gluon BS})
by changing the BRST transformation 
$\delta_{{\bf B}}^0A_{\mu}^*$ 
into 
$\bigl(\delta^0_{{\bf B}}+\delta_{{\bf B}}^1\bigr)A_{\mu}^*$.

Nextly we consider the role of the contact term 
$\Upsilon_{A,\epsilon}^{(0)}$. 
Instead of giving a detailed analysis 
such as provided on $\Upsilon_{A,\epsilon}^{(-1)}$,  
we take another route which respects the world-sheet 
supersymmetry on the boundary. 
Let us recall that the contact terms 
$\Delta_{A,\epsilon}^{(-1)}$ and 
$\Delta_{A,\epsilon}^{(0)}$ emerge from the two-gluon boundary state 
by the actions of the supercurrents and the Virasoro generators.  
These contact terms give rise to the gluon vertex 
operators $\delta_{{\bf B}}^1A_{\mu}^*\widehat{V}_{gl}^{(-1)\mu}$ 
and $\delta_{{\bf B}}^1A_{\mu}^*\widehat{V}_{gl}^{(0)\mu}$. 
The pictures of these operators are consistent 
with the world-sheet superconformal algebra and 
they become the supermultiplet on the boundary.  
We expect that the contact terms $\Upsilon_{A,\epsilon}^{(-1)}$ 
and $\Upsilon_{A,\epsilon}^{(0)}$ also become 
a supermultiplet on the boundary. 
In particular, this makes it possible to read the expansion of 
$\Upsilon_{A,\epsilon}^{(0)}$ from that of 
$\Upsilon_{A,\epsilon}^{(-1)}$ as 
\begin{eqnarray}
&&
\Upsilon_{A,\epsilon}^{(0)}(\sigma;k^{(1)},k^{(2)},k^{(3)}) 
\nonumber \\ 
&& 
~~
=
\frac{i}{2}
\epsilon^{\alpha' \left(\sum_{r}k^{(r)}\right)^2-1}
\Biggl\{
J_{\mu}(k^{(1)},k^{(2)},k^{(3)})
+
K_{\mu}(k^{(1)},k^{(2)},k^{(3)}) 
\Biggr\} 
\widehat{V}_{gl}^{(0)\mu}(\sigma;\sum_{r}k^{(r)})
\nonumber \\
&& 
~~~~
+
{\cal O}(\epsilon^{\alpha' \left(\sum_{r}k^{(r)}\right)^2})~,
\label{expansion three-gluon contact term 2}
\end{eqnarray}
where the tensors $J_{\mu}$ and $K_{\mu}$ are those in  
(\ref{K}) and (\ref{J}). 
Similarly to the case of the supercurrent, 
the $0$ picture operator $J_{\mu}\widehat{V}_{gl}^{(0)\mu}$ 
in the above expansion gives a correction to the Virasoro 
action on the single gluon boundary state. 
We can write the first term of 
(\ref{Virasoro three-gluon boundary integral}) as follows: 
\begin{eqnarray}
&&
iC
\int 
\!\!
\prod_{r=1}^3
\frac{d^{p+1}k^{(r)}}{(2\pi)^{\frac{p+1}{2}}}
~g_{\eta}
\int_0^{2\pi}
\!\!d\sigma   
e^{in\sigma}(e^{-in\epsilon}-e^{in\epsilon})
\Upsilon_{A,\epsilon}^{(0)}(\sigma;k^{(1)},k^{(2)},k^{(3)}) 
|x_0^i\rangle_m 
\otimes 
|B \rangle_{gh}
\otimes 
|B;\eta \rangle_{sgh} 
\nonumber \\
&&
=
-iC 
\int
\!\!
\frac{d^{p+1}k}{(2\pi)^{\frac{p+1}{2}}}
\epsilon^{\alpha'k\cdot k}
~g_{\eta}
\int_0^{2\pi}
\!\!d\sigma   
~ne^{in\sigma} 
\Biggl\{
\delta_{{\bf B}}^2A_{\mu}^*
\widehat{V}_{gl}^{(0)\mu}(\sigma;k)
+\mathcal{O}(\epsilon)\Biggr\}
|x_0^i\rangle_m \otimes |B \rangle_{gh} 
\otimes |B;\eta\rangle_{sgh}
\nonumber \\ 
&&
~~+ \cdots~.
\label{from 3 to 1 Virasoro}
\end{eqnarray}
This shows that the first term of 
(\ref{Virasoro three-gluon boundary integral}) 
modifies the Virasoro action on the single gluon boundary state 
(\ref{one-gluon Virasoro 1}) 
such that the $0$ picture operator 
$\delta_{{\bf B}}^0A_{\mu}^*\widehat{V}_{gl}^{(0)\mu}$ 
changes into 
$\bigl(\delta_{{\bf B}}^0+\delta_{{\bf B}}^2\bigr)
A_{\mu}^*\widehat{V}_{gl}^{(0)\mu}$. 
This is consistent with the previous correction 
to the action of the supercurrrent on the single gluon boundary state. 
Thus we obtain the following correction from the action of the 
closed-string BRST operator on the three-gluon boundary state: 
\begin{eqnarray}
-iC
\int_{{\cal I}_1}
\!\! dS
~g_{\eta}
\int \frac{d^{p+1}k}{(2\pi)^{\frac{p+1}{2}}}
\epsilon^{\alpha'k\cdot k}
\Biggl\{
\delta_{{\bf B}}^2
A_{\mu}^*(k)
\widehat{{\bf V}}_{gl}^{* \mu}(S_1;k)
+
{\cal O}(\epsilon)
\Biggr\}
|x_0^i\rangle_m \otimes |B \rangle_{gh} 
\otimes |B;\eta\rangle_{sgh}~. 
\label{from 3 to 1 Qc}
\end{eqnarray}
Therefore,  
putting together with the correction from the two-gluon boundary state,  
the action of the BRST operator on the single gluon 
state is modified by 
changing the BRST transformation 
$\delta_{{\bf B}}^0A_{\mu}^*$ into 
$\bigl(\delta_{{\bf B}}^0+\delta_{{\bf B}}^1+\delta_{{\bf B}}^2
\bigr)A_{\mu}^*$. 

\subsection{Action of the closed-string BRST operator on the Wilson loop}
We comment briefly on the contributions from the $N$-gluons boundary states 
($N \geq 4$). These cases are regarded as generalizations of the previous 
case of the three-gluon boundary state. 
In particular, the actions of the closed-string 
BRST operator on the multi-gluon boundary states     
will give rise to another corrections 
$\delta_{{\bf B}}^LA_{\mu}^*$ $(L \geq 3)$ other than 
$\delta_{{\bf B}}^{1,2}A_{\mu}^*$.  
The correction $\delta_{{\bf B}}^LA_{\mu}^*$ is a polynomial 
of $A_{\mu}$ with the degree equal to $L+1$. 
This means that these corrections with $L \geq 3$ contribute 
as the higher order terms of the $\alpha'$-expansion of 
the BRST transformation of the antifield $A_{\mu}^*$. 
At the $\alpha'$-order, 
the non-linear transformation of $A_{\mu}^*$ is given by 
(\ref{total BRST}).
The collection of the actions of the closed-string 
BRST operator on the multi-gluon boundary states 
can be written as follows: 
\begin{eqnarray}
&&
Q_c \times 
\sum_{N=0}^{\infty}
\frac{i^{N^2}}{N} C
\int_{{\cal I}_{N}(\epsilon)}
\bigl[ d^N S \bigr]
g_{\eta}~
\widehat{{\bf V}}_{A,\epsilon}({\bf X}(S_1))
\cdots 
\widehat{{\bf V}}_{A,\epsilon}({\bf X}(S_N))
|x_0^i\rangle_m \otimes |B \rangle_{gh}
\otimes |B;\eta \rangle_{sgh}
\nonumber \\
&&
=
Q_c \times 
iC
\int_{{\cal I}_1}
dS 
g_{\eta}~
\widehat{{\bf V}}_{A,\epsilon}({\bf X}(S))
|x_0^i\rangle_m \otimes |B \rangle_{gh}
\otimes |B;\eta \rangle_{sgh}  
\nonumber \\
&&
~~+
Q_c \times 
\frac{i^{4}}{2}C
\int_{{\cal I}_{2}(\epsilon)}
\bigl[ d^2 S \bigr]
g_{\eta}~
\widehat{{\bf V}}_{A,\epsilon}({\bf X}(S_1))
\widehat{{\bf V}}_{A,\epsilon}({\bf X}(S_2))
|x_0^i\rangle_m \otimes |B \rangle_{gh}
\otimes |B;\eta \rangle_{sgh}
\nonumber \\
&&
~~+\cdots 
\nonumber \\
&&
=
-iC
\int_{{\cal I}_1}
\!\! dS 
~g_{\eta}
\int \frac{d^{p+1}k}{(2\pi)^{\frac{p+1}{2}}}
\epsilon^{\alpha'k\cdot k}
\left\{
\sum_{L=0}^{\infty}
\delta_{{\bf B}}^L
A_{\mu}^*(k)
\widehat{{\bf V}}_{gl}^{* \mu}(S;k)
+
\delta^0_{{\bf B}}
{\cal B}(k)
\widehat{{\bf V}}_{a.g}(S;k)
+
{\cal O}(\epsilon)
\right\}
\nonumber \\*
&&
~~~~~~~~~~~~~~~~~~~~~~~~~~~~~~~~~~~~~
\times 
|x_0^i\rangle_m 
\otimes 
|B \rangle_{gh}
\otimes 
|B;\eta \rangle_{sgh}~.
\nonumber \\
&&
~~+\frac{i^4}{2}C
\int_{{\cal I}_2(\epsilon)}
\!\! 
\bigl[d^2S \bigr]
~g_{\eta}
\int \frac{d^{p+1}k}{(2\pi)^{\frac{p+1}{2}}}
\epsilon^{\alpha'k\cdot k}
\Biggl\{
\sum_{L=0}^{\infty}
\delta_{{\bf B}}^L
A_{\mu}^*(k)
\widehat{{\bf V}}_{gl}^{* \mu}(S_1;k)
+
\delta^0_{{\bf B}}
{\cal B}(k)
\widehat{{\bf V}}_{a.g}(S_1;k)
+
{\cal O}(\epsilon)
\Biggr\}
\nonumber \\*
&&
~~~~~~~~~~~~~~~~~~~~~~~~~
\times 
\widehat{{\bf V}}_{A,\epsilon}
({\bf X}(S_2))~
|x_0^i\rangle_m 
\otimes 
|B \rangle_{gh}
\otimes 
|B;\eta \rangle_{sgh}
\nonumber \\
&&
~~-\frac{i^4}{2}
C
\int_{{\cal I}_2(\epsilon)}
\!\! \bigl[d^2S \bigr]
~g_{\eta}
\widehat{{\bf V}}_{A,\epsilon}
({\bf X}(S_1))
\nonumber \\*
&&
~~~~~~~~~~~\times
\int \frac{d^{p+1}k}{(2\pi)^{\frac{p+1}{2}}}
\epsilon^{\alpha'k\cdot k}
\Biggl\{
\sum_{L=0}^{\infty}
\delta_{{\bf B}}^L
A_{\mu}^*(k)
\widehat{{\bf V}}_{gl}^{* \mu}(S_2;k)
+
\delta^0_{{\bf B}}
{\cal B}(k)
\widehat{{\bf V}}_{a.g}(S_2;k)
+{\cal O}(\epsilon)
\Biggr\}
\nonumber \\*
&&
~~~~~~~~~~~~~~~~~~~~~~~~~~~~~~~~
\times 
|x_0^i\rangle_m 
\otimes 
|B \rangle_{gh}
\otimes 
|B;\eta \rangle_{sgh}
\nonumber\\
&&
~
+\cdots ~~~~~~.
\label{pre-result}
\end{eqnarray}
Here ${\cal O}(\epsilon)$ terms represent the massive modes of 
the open-string and become irrelevant at the short distance 
$\epsilon \sim 0$. 
Eq.(\ref{pre-result}) is translated into the action 
on the Wilson loop (\ref{Qc on Wilsonloop}).

\section{BRST Transformation and World-sheet RG}
\label{section6}

The non-linear BRST transformation $\delta_{{\bf B}}$ 
is expected \cite{Nakatsu} 
to generate the beta functions of the boundary interactions. 
In this section we study the Wilson loop (\ref{SWilson}) 
from the perspective of the world-sheet renormalization group. 
We will show that the BRST transformations 
$\delta_{{\bf B}}A_{\mu}^*(k)$ (\ref{non-linear A*}) 
are precisely the beta functions of $A_{\mu}(k)$.  
Here the beta functions  
$\beta_{A_{\mu}(k)}\left(=\epsilon 
\frac{\partial}{\partial \epsilon}
A_{\mu}(k;\epsilon)\right)$
are defined by the following condition 
on the Wilson loop: 
\begin{eqnarray}
\left\{
\epsilon \frac{\partial}{\partial \epsilon}
+\int 
\!d^{p+1}k
~\beta_{A_{\mu}(k)}
\frac{\delta}{\delta A_{\mu}(k)}
\right\}  
\Bigl| W_{\epsilon} [A];\eta \Bigr\rangle_{tot}
=0~.
\label{def of beta}
\end{eqnarray}

Let us compute the effect of infinitesimal deformation 
of the cut-off parameter $\epsilon$ on the Wilson loop: 
\begin{eqnarray}
\epsilon \frac{\partial}{\partial \epsilon} 
\Bigl| W_{\epsilon} [A];\eta \Bigr\rangle_{tot}
~.
\label{epsilon WL}
\end{eqnarray}
This can be obtained from the following variations 
of the multi-gluon boundary states: 
\begin{eqnarray}
\epsilon 
\frac{\partial }{\partial \epsilon}
\left\{
\frac{i^{N^2}}{N}C 
\int_{{\cal I}_{N}(\epsilon)}
\bigl[ d^NS \bigr]
~g_{\eta} 
\widehat{{\bf V}}_{A,\epsilon}({\bf X}(S_1))
\cdots 
\widehat{{\bf V}}_{A,\epsilon}({\bf X}(S_N)) 
\right\}
|x_0^i \rangle_m \otimes 
|B \rangle_{gh} \otimes 
|B;\eta \rangle_{sgh}~, 
\label{epsilon WL 1}
\end{eqnarray}
where the superspace notation 
(\ref{N-gluon BS}) 
is used to express the gluon boundary states. 
We describe computations of the variations (\ref{epsilon WL 1}) 
briefly. We first remark that the infinitesimal transformation of 
the gluon vertex operator becomes  
\begin{eqnarray}
\epsilon \frac{\partial}{ \partial \epsilon}
\widehat{{\bf V}}_{A,\epsilon}(S;k)
&=&
\alpha'k \cdot k
\widehat{{\bf V}}_{A,\epsilon}(S;k) 
\nonumber \\
&=&
-\epsilon^{\alpha' k \cdot k}
\delta_{{\bf B}}^0 
A_{\mu}^*(k)
\widehat{{\bf V}}_{gl}^{\mu}(S;k)~, 
\label{epsilon V}
\end{eqnarray}
where the canonical dimension of 
the coupling constant $A_{\mu}(k)$, $\alpha'k\cdot k$,  
appears due to the rescaling (\ref{rvertex}). 
The variation of the single gluon boundary state follows from 
the transformation (\ref{epsilon V}) as 
\begin{eqnarray}
&&
\epsilon \frac{\partial}{ \partial \epsilon}
\Biggl\{
iC 
\int_{{\cal I}_1}
\! dS 
~g_{\eta}
\widehat{{\bf V}}_{A,\epsilon}({\bf X}(S))
\Biggr\}
|x_0^i\rangle_m 
\otimes 
|B \rangle_{gh} 
\otimes 
|B;\eta \rangle_{sgh}
\nonumber \\
&&
~
=-iC 
\int_{{\cal I}_1} 
\!dS
~g_{\eta}
\int \! 
\frac{d^{p+1}k}{(2\pi)^{\frac{p+1}{2}}}
\epsilon^{\alpha' k \cdot k}
\delta^0_{{\bf B}}A^*_{\mu}(k)
\widehat{{\bf V}}^{\mu}_{gl}(S;k)
|x_0^i\rangle_m 
\otimes 
|B \rangle_{gh} 
\otimes 
|B;\eta \rangle_{sgh}~. 
\label{epsilon one-gluon 1}
\end{eqnarray}
The transformation of the gluon vertex operator 
(\ref{epsilon V}) also appears in the variations 
of the multi-gluon boundary states. 
In the cases of the multi-gluon boundary states, 
the $\epsilon$-dependence of the supermoduli 
${\cal I}_{N}(\epsilon)$ must be taken into account. 
The supermoduli is introduced 
in (\ref{super multiple-ordered integral}) and 
the $\epsilon$-dependence originates 
in the product of the regularized supersymmetric step functions 
$\prod_{\alpha=1}^N\Theta(S_{\alpha}+\epsilon,S_{\alpha+1})$. 
We compute these effects 
by using the expressions  
(\ref{two-gluon BS}) and (\ref{three-gluon BS}) 
of the gluon boundary states. 
Although the deformation of $\epsilon$ is not generated  
by the Virasoro algebra  $L_n-\bar{L}_{-n}$ or ${\cal L}_n$, 
the computations themselves become analogous to the computations 
of the Virasoro actions in section \ref{section5}. 
For the two-gluon boundary state, we obtain
\footnote{Cf. the Virasoro action 
(\ref{Virasoro two-gluon boundary integral}).}: 
\begin{eqnarray}
&&
\epsilon \frac{\partial}{\partial \epsilon}
\Biggl \{
\frac{i^4}{2}
C
\int_{{\cal I}_2(\epsilon)}
\bigl[ d^2S \bigr]
~g_{\eta}
\widehat{{\bf V}}_{A,\epsilon}
({\bf X}(S_1))
\widehat{{\bf V}}_{A,\epsilon}
({\bf X}(S_2))
\Biggr \}
|x_0^i\rangle_m 
\otimes 
|B \rangle_{gh}
\otimes 
|B;\eta \rangle_{sgh}
\nonumber \\
&&
=
C\int \!
\prod_{r=1,2}
\frac{d^{p+1}k^{(r)}}{(2\pi)^{\frac{p+1}{2}}}
g_{\eta}
\int_0^{2\pi}
\!\!d\sigma 
~\epsilon
\Delta_{A,\epsilon}^{(0)}(\sigma;k^{(1)},k^{(2)}) 
|x_0^i\rangle_m 
\otimes 
|B \rangle_{gh}
\otimes 
|B;\eta \rangle_{sgh}
\nonumber \\
&&
-\frac{i^4}{2}C 
\int_{{\cal I}_2(\epsilon)}\bigl[ d^2S \bigr]
~g_{\eta}
\int 
\!
\frac{d^{p+1}k}{(2\pi)^{\frac{p+1}{2}}}
\epsilon^{\alpha'k \cdot k}
\delta_{{\bf B}}^0A_{\mu}^*(k)
\widehat{{\bf V}}_{gl}^{\mu}(S_1;k)
\widehat{{\bf V}}_{A,\epsilon}({\bf X}(S_2))
|x_0^i\rangle_m 
\otimes 
|B \rangle_{gh}
\otimes 
|B;\eta \rangle_{sgh}
\nonumber \\
&&
-\frac{i^4}{2}C 
\int_{{\cal I}_2(\epsilon)}
\bigl[ d^2S \bigr]
~g_{\eta}
\widehat{{\bf V}}_{A,\epsilon}({\bf X}(S_1))
\int 
\!
\frac{d^{p+1}k}{(2\pi)^{\frac{p+1}{2}}}
\epsilon^{\alpha' k\cdot k}
\delta_{{\bf B}}^0A_{\mu}^*(k)
\widehat{{\bf V}}_{gl}^{\mu}(S_2;k)
|x_0^i\rangle_m 
\otimes 
|B \rangle_{gh}
\otimes 
|B;\eta \rangle_{sgh}~.
\nonumber \\
\label{epsilon two-gluon 1}
\end{eqnarray}
In the above, the first term is brought about by the deformation 
of the supermoduli ${\cal I}_2(\epsilon)$. 
This term includes the contact term 
$\Delta_{A,\epsilon}^{(0)}(\sigma)$. 
By plugging the expansion 
(\ref{expansion of two-gluon contact term 1}) into 
the contact term, we can write the first term of 
(\ref{epsilon two-gluon 1}) as follows: 
\begin{eqnarray}
&&
C\int \!
\prod_{r=1,2}
\frac{d^{p+1}k^{(r)}}{(2\pi)^{\frac{p+1}{2}}}
g_{\eta}
\int_0^{2\pi}
\!\!d\sigma 
~\epsilon
\Delta_{A,\epsilon}^{(0)}(\sigma;k^{(1)},k^{(2)}) 
|x_0^i\rangle_m 
\otimes 
|B \rangle_{gh}
\otimes 
|B;\eta \rangle_{sgh}
\nonumber \\
&&
=
-iC
\int_{{\cal I}_1}
\!\! dS 
~g_{\eta}
\int \frac{d^{p+1}k}{(2\pi)^{\frac{p+1}{2}}}
\epsilon^{\alpha'k\cdot k}
\Biggl\{
\delta^1_{{\bf B}}A_{\mu}^*(k)
\widehat{{\bf V}}_{gl}^{\mu}(S;k)+{\cal O}(\epsilon)
\Biggr\}
|x_0^i\rangle_m 
\otimes 
|B \rangle_{gh}
\otimes 
|B;\eta \rangle_{sgh}~, 
\nonumber \\*
\label{epsilon two-gluon contact}
\end{eqnarray}
where $\delta_{{\bf B}}^1A_{\mu}^*$ is the first correction 
(\ref{1st correction of BRS}) 
to the BRST transformation. 
 ${\cal O}(\epsilon)$ term of (\ref{epsilon two-gluon contact}) 
represents the massive modes of the open-string. 
These massive modes become irrelevant at the short distance  
$\epsilon \sim 0$. 
The variation (\ref{epsilon two-gluon 1}) clarifies the possible 
UV divergence of the two-gluon boundary state. 
Eq.(\ref{epsilon two-gluon contact}) shows that 
the two-gluon boundary state suffers only 
from the logarithmic divergence. 
This divergence is caused by the massless pole of gluon. 
There is no $1/\epsilon$ divergence in the two-gluon boundary state. 
It is convenient to recall the expression 
(\ref{two-gluon BS}) of the two-gluon 
boundary state.  The first term of (\ref{two-gluon BS}) 
diverges $\sim 1/\epsilon$ due to the unphysical mode 
in the product of the $0$ picture gluon vertex operators 
(\ref{OPE between 0 gluon vertices}). 
The absence of $1/\epsilon$ divergence means that 
it cancels out precisely with the $1/\epsilon$ 
divergences of the other two terms of (\ref{two-gluon BS}).

We move on to the case of the three-gluon boundary state. 
It becomes \footnote{Cf. the Virasoro action 
(\ref{Virasoro three-gluon boundary integral}).}: 
\begin{eqnarray}
&&
\epsilon 
\frac{\partial}{\partial \epsilon}
\left\{
\frac{i^9}{2}C
\int_{{\cal I}_3(\epsilon)}
\bigl[ d^3S \bigr]
~g_{\eta}
\widehat{{\bf V}}_{A,\epsilon}
({\bf X}(S_1))
\widehat{{\bf V}}_{A,\epsilon}
({\bf X}(S_2))
\widehat{{\bf V}}_{A,\epsilon}
({\bf X}(S_3))
\right\}
|x_0^i\rangle_m 
\otimes 
|B \rangle_{gh}
\otimes 
|B;\eta \rangle_{sgh}
\nonumber \\
&&
=C
\int 
\!
\prod_{r=1}^3
\frac{d^{p+1}k^{(r)}}{(2\pi)^{\frac{p+1}{2}}}
~g_{\eta}
\int_0^{2\pi}
\!\!d\sigma   
~2\epsilon
\Upsilon_{A,\epsilon}^{(0)}(\sigma;k^{(1)},k^{(2)},k^{(3)}) 
|x_0^i\rangle_m 
\otimes 
|B \rangle_{gh}
\otimes 
|B;\eta \rangle_{sgh} 
\nonumber \\
&&
~+
\frac{i}{2}C
\int 
\!
\prod_{r=1}^3
\frac{d^{p+1}k^{(r)}}{(2\pi)^{\frac{p+1}{2}}}
~g_{\eta}
\int_{0}^{2\pi}
\!\!
d\sigma_1
~\epsilon
\Delta_{A,\epsilon}^{(0)}(\sigma_1;k^{(1)},k^{(2)})
\nonumber \\*
&&
~~~~~~~~~~~~~~~~~~~~~~~~~~~~~~~~~~~~~~~~~
\times 
\int_{\sigma_1+2\epsilon}^{\sigma_1+2\pi-\epsilon}
\!\!\!\!\!
d\sigma_2
\widehat{V}_{A,\epsilon}^{(0)}(\sigma_2;k^{(3)})
|x_0^i\rangle_m 
\otimes 
|B \rangle_{gh}
\otimes 
|B;\eta \rangle_{sgh} 
\nonumber \\
&&
~+
\frac{1}{2}C
\int 
\!
\prod_{r=1}^3
\frac{d^{p+1}k^{(r)}}{(2\pi)^{\frac{p+1}{2}}}
~g_{\eta}
\int_{0}^{2\pi}
\!\!
d\sigma
~\epsilon
\Delta_{A,\epsilon}^{(-1)}(\sigma;k^{(1)},k^{(2)})
\nonumber \\*
&&
~~~~~~~~~~~~~~~~~~~~~~~~~~~~~~~~~~~~~~~~~
\times 
\widehat{V}_{A,\epsilon}^{(-1)}(\sigma+\epsilon;k^{(3)})
|x_0^{i}\rangle_m
\otimes 
|B \rangle_{gh}
\otimes 
|B;\eta \rangle_{sgh} 
\nonumber \\
&&
~+
\frac{i}{2}C
\int 
\!
\prod_{r=1}^3
\frac{d^{p+1}k^{(r)}}{(2\pi)^{\frac{p+1}{2}}}
~g_{\eta}
\int_{0}^{2\pi}
\!\!
d\sigma_1
\widehat{V}_{A,\epsilon}^{(0)}(\sigma_2;k^{(1)})
\nonumber \\*
&&
~~~~~~~~~~~~~~~~~~~~~~~~~~
\times 
\int_{\sigma_1+\epsilon}^{\sigma_1+2\pi-2\epsilon}
\!\!\!\!\!
d\sigma_2
~\epsilon
\Delta_{A,\epsilon}^{(0)}(\sigma_2;k^{(2)},k^{(3)})
|x_0^i\rangle_m 
\otimes 
|B \rangle_{gh}
\otimes 
|B;\eta \rangle_{sgh} 
\nonumber \\
&&
~+
\frac{1}{2}C
\int 
\!
\prod_{r=1}^3
\frac{d^{p+1}k^{(r)}}{(2\pi)^{\frac{p+1}{2}}}
~g_{\eta}
\int_{0}^{2\pi}
\!\!
d\sigma
\widehat{V}_{A,\epsilon}^{(-1)}(\sigma;k^{(1)})
\nonumber \\*
&&
~~~~~~~~~~~~~~~~~~~~~~~~~~~
\times 
~\epsilon
\Delta_{A,\epsilon}^{(-1)}(\sigma+\epsilon;k^{(2)},k^{(3)})
|x_0^{i}\rangle_m
\otimes 
|B \rangle_{gh}
\otimes 
|B;\eta \rangle_{sgh} 
\nonumber \\
&&
~+ \cdots ~,
\label{epsilon three-gluon 1}
\end{eqnarray}
where the terms proportional to 
$\delta_{{\bf B}}^0A_{\mu}^*\widehat{{\bf V}}_{gl}^{\mu}$ 
are omitted. In the above, the terms which include the contact 
terms are all brought about by the deformation of the supermoduli 
${\cal I}_{3}(\epsilon)$. 
We plug the expansions 
(\ref{expansion of two-gluon contact term 1}) 
and 
(\ref{expansion of two-gluon contact term 2}) 
into the contact terms $\Delta_{A,\epsilon}^{(0)}(\sigma)$ and 
$\Delta_{A,\epsilon}^{(-1)}(\sigma)$. 
For the contact term $\Upsilon_{A,\epsilon}^{(0)}(\sigma)$, 
as discussed in 
the computation of the Virasoro action, 
we should use the part of the expansion 
(\ref{expansion three-gluon contact term 2}) 
which is proportional to the tensor $J_{\mu}$. 
After these substitutions for the contact terms,   
Eq.(\ref{epsilon three-gluon 1}) becomes as follows: 
\begin{eqnarray}
&&
\epsilon 
\frac{\partial}{\partial \epsilon}
\left\{
\frac{i^9}{2}C
\int_{{\cal I}_3(\epsilon)}
\bigl[ d^3S \bigr]
~g_{\eta}
\widehat{{\bf V}}_{A,\epsilon}
({\bf X}(S_1))
\widehat{{\bf V}}_{A,\epsilon}
({\bf X}(S_2))
\widehat{{\bf V}}_{A,\epsilon}
({\bf X}(S_3))
\right\}
|x_0^i\rangle_m 
\otimes 
|B \rangle_{gh}
\otimes 
|B;\eta \rangle_{sgh}
\nonumber \\
&&
~
=-iC 
\int_{{\cal I}_1} 
\!dS
~g_{\eta}
\int \! 
\frac{d^{p+1}k}{(2\pi)^{\frac{p+1}{2}}}
\epsilon^{\alpha' k \cdot k}
\Biggl\{
\delta^2_{{\bf B}}A^*_{\mu}(k)
\widehat{{\bf V}}^{\mu}_{gl}(S;k)
+{\cal O}(\epsilon) 
\Biggr\}
|x_0^i\rangle_m 
\otimes 
|B \rangle_{gh} 
\otimes 
|B;\eta \rangle_{sgh}
\nonumber \\
&&
~~~~
-\frac{i^4}{2}C 
\int_{{\cal I}_2(\epsilon)}\bigl[ d^2S \bigr]
~g_{\eta}
\int 
\!
\frac{d^{p+1}k}{(2\pi)^{\frac{p+1}{2}}}
\epsilon^{\alpha'k \cdot k}
\Biggl\{
\delta_{{\bf B}}^1A_{\mu}^*(k)
\widehat{{\bf V}}_{gl}^{\mu}(S_1;k)
+
{\cal O}(\epsilon)
\Biggr\}
\nonumber \\*
&&
~~~~~~~~~~~~~~~~~
\times 
\widehat{{\bf V}}_{A,\epsilon}({\bf X}(S_2))
|x_0^i\rangle_m 
\otimes 
|B \rangle_{gh}
\otimes 
|B;\eta \rangle_{sgh}
\nonumber \\
&&
~~~~
-\frac{i^4}{2}C 
\int_{{\cal I}_2(\epsilon)}
\bigl[ d^2S \bigr]
~g_{\eta}
\widehat{{\bf V}}_{A,\epsilon}({\bf X}(S_1))
\nonumber \\*
&&
~~~~~~~~~~~~~~~~~
\times 
\int 
\!
\frac{d^{p+1}k}{(2\pi)^{\frac{p+1}{2}}}
\epsilon^{\alpha' k\cdot k}
\Biggl\{
\delta_{{\bf B}}^1A_{\mu}^*(k)
\widehat{{\bf V}}_{gl}^{\mu}(S_2;k)
+
{\cal O}(\epsilon) 
\Biggr\}
|x_0^i\rangle_m 
\otimes 
|B \rangle_{gh}
\otimes 
|B;\eta \rangle_{sgh} 
\nonumber \\
&&
~~~~
+ \cdots~, 
\label{epsilon three-gluon 2}
\end{eqnarray}
where $\delta_{{\bf B}}^2A_{\mu}^*$ is the second correction 
(\ref{2nd correction of BRS}) to the BRST transformation. 
For the $N$-gluon boundary states ($N \geq 4$), 
the deformation of the supermoduli ${\cal I}_N(\epsilon)$ 
will give rise to another corrections $\delta_{{\bf B}}^LA_{\mu}^*$ 
$(L \geq 3)$ other than $\delta_{{\bf B}}^{1,2}A_{\mu}^*$. 
The corrections  $\delta_{{\bf B}}^LA_{\mu}^*$ $(L \geq 3)$
become the higher order terms in the $\alpha'$-expansion 
of the non-linear BRST transformation $\delta_{{\bf B}}A_{\mu}^*$.

The effect of the deformation on the Wilson loop (\ref{epsilon WL}) 
is given by the collection of the variations of the gluon 
boundary states. The multi-gluon boundary states 
bring about corrections to 
the single gluon boundary state. 
For instance, 
the $\delta_{{\bf B}}^1A_{\mu}^*$-term of (\ref{epsilon two-gluon 1}) 
and 
the $\delta_{{\bf B}}^2A_{\mu}^*$-term of (\ref{epsilon three-gluon 2}) 
contribute to the single gluon boundary state. 
These corrections are absorbed 
in the variation (\ref{epsilon one-gluon 1}) 
of the single gluon boundary state by changing  
the BRST transformation $\delta_{{\bf B}}^0A_{\mu}^*$ into 
$\bigl(\delta_{{\bf B}}^0+\delta_{{\bf B}}^1+\delta_{{\bf B}}^2
\bigr)A_{\mu}^*$. This is not limited to the single gluon 
boundary state. For instance, 
the $\delta_{{\bf B}}^1A_{\mu}^*$-terms 
of (\ref{epsilon three-gluon 2}) 
contribute to the two-gluon boundary state. 
They are 
absorbed in the variation (\ref{epsilon two-gluon 1})
of the two-gluon boundary state by changing 
the BRST transformation $\delta_{{\bf B}}^0A_{\mu}^*$ into 
$\bigl(\delta_{{\bf B}}^0+\delta_{{\bf B}}^1\bigr)A_{\mu}^*$.
These could be generalized to the other gluon boundary states 
so that the effect of the deformation of $\epsilon$ is described 
by replacing each coupling constant $A_{\mu}(k)$ with 
the non-linear BRST transformation $-\delta_{{\bf B}}A_{\mu}^*$. 
It is analogous to the action of the closed-string BRST operator. 
Hence, the effect of the deformation of $\epsilon$ 
on the Wilson loop becomes as follows: 
\begin{eqnarray}
&&
\epsilon 
\frac{\partial }{\partial \epsilon}
\Bigl| W_{\epsilon} [A];\eta \Bigr\rangle_{tot}
=
-\int 
\!d^{p+1}k
~\delta_{{\bf B}}A_{\mu}^*(k)
\frac{\delta}{\delta A_{\mu}(k)}
\Bigl| W_{\epsilon} [A];\eta \Bigr\rangle_{tot}
~.
\label{epsilon WL result}
\end{eqnarray}
By plugging (\ref{epsilon WL result}) into (\ref{def of beta}) 
we obtain 
\begin{eqnarray}
\beta_{A_{\mu}(k)}
&=&
\delta_{{B}}A_{\mu}^*(k)
\nonumber \\*
&=&
\alpha^{\prime}\int\frac{d^{p+1}x}{(2\pi)^{\frac{p+1}{2}}}
e^{-ik\cdot{x}}G^{\nu\rho}\nabla_{\nu}F_{\rho\mu}
\Bigm|_{\partial \cdot A =0}(x)
+
{\cal O}(\alpha'^2)~.
\label{beta}
\end{eqnarray}

\subsection*{Acknowledgements}
We would like to thank K. Murakami for his collaboration 
in the early stage of this work. 
We benefitted from several discussions with him. 
The work of T.N. is supported in part by Grant-in-Aid for 
Scientific Research No.15540273.


\end{document}